\documentclass[11pt]{article}

\usepackage[T1]{fontenc}
\usepackage[latin1]{inputenc}
\usepackage[nosort]{cite}
\usepackage{color}
\usepackage[bulletsep]{collref}
\usepackage{graphicx}
\usepackage{bbm}
\usepackage{amsmath}
\usepackage{amssymb}
\usepackage{subfig}
\usepackage{multirow}
\usepackage{tikz}
\usepackage{hyperref}

\makeatletter \@addtoreset{equation}{section} \makeatother

\makeatletter
\let\old@startsection=\@startsection
\let\oldl@section=\l@section
\renewcommand{\@startsection}[6]{\old@startsection{#1}{#2}{#3}{#4}{#5}{#6\mathversion{bold}}}
\renewcommand{\l@section}[2]{\oldl@section{\mathversion{bold}#1}{#2}}
\makeatother

\makeatletter
\let\old@makecaption=\@makecaption
\def\@makecaption{\small\old@makecaption}
\makeatother

\let\oldPhi=\Phi
\let\oldPsi=\Psi
\let\oldGamma=\Gamma
\let\oldDelta=\Delta
\let\oldSigma=\Sigma
\let\oldTheta=\Theta
\let\oldPi=\Pi
\let\oldUpsilon=\Upsilon
\renewcommand{\Phi}{\mathnormal{\oldPhi}}
\renewcommand{\Psi}{\mathnormal{\oldPsi}}
\renewcommand{\Gamma}{\mathnormal{\oldGamma}}
\renewcommand{\Sigma}{\mathnormal{\oldSigma}}
\renewcommand{\Delta}{\mathnormal{\oldDelta}}
\renewcommand{\Theta}{\mathnormal{\oldTheta}}
\renewcommand{\Pi}{\mathnormal{\oldPi}}
\renewcommand{\Upsilon}{\mathnormal{\oldUpsilon}}

\newcommand{\Action}{\mathcal{S}}

\newcommand{\tr}{\mathop{\mathrm{tr}}}

\newcommand{\diag}{\mathop{\mathrm{diag}}}

\newcommand{\imag}{\mathbbm{i}}

\newcommand{\order}{\mathcal{O}}

\newcommand{\trans}{{\scriptscriptstyle\mathrm{T}}}

\newcommand{\Hyp}{\mathbbm{H}}
\newcommand{\Sphere}{S}  
\newcommand{\AdS}{\mathrm{AdS}}

\ifx\genfrac\sdflkaj

\else

\fi
\newcommand{\sfrac}[2]{{\textstyle\frac{#1}{#2}}}
\newcommand{\half}{\sfrac{1}{2}}

\newcommand{\Half}{\frac{1}{2}}

\newcommand{\Quarter}{\frac{1}{4}}
\newcommand{\p}{\partial}

\newcommand{\matr}[2]{\left(\begin{array}{#1}#2\end{array}\right)}

\newcommand{\grp}[1]{\mathrm{#1}}

\newcommand{\grSL}{\grp{SL}}

\newcommand{\lrbrk}[1]{\left(#1\right)}
\newcommand{\bigbrk}[1]{\bigl(#1\bigr)}
\newcommand{\Bigbrk}[1]{\Bigl(#1\Bigr)}
\newcommand{\biggbrk}[1]{\biggl(#1\biggr)}

\newcommand{\lrsbrk}[1]{\left[#1\right]}

\newcommand{\Bigsbrk}[1]{\Bigl[#1\Bigr]}

\newcommand{\comm}[2]{[#1,#2]}

\newcommand{\abs}[1]{{|#1|}}

\newcommand{\bigeval}[1]{#1\big|}

\newcommand{\nn}{\nonumber}
\newcommand{\nln}{\nonumber\\}
\newcommand{\nl}[1][0pt]{\nonumber\\[#1]&\hspace{-4\arraycolsep}&\mathord{}}

\newcommand{\earel}[1]{\mathrel{}&\hspace{-2\arraycolsep}#1\hspace{-2\arraycolsep}&\mathrel{}}
\newcommand{\eq}{\earel{=}}

\def\[{\begin{equation}}
\def\]{\end{equation}}

\makeatletter
\def\mr@ignsp#1 {\ifx\:#1\@empty\else #1\expandafter\mr@ignsp\fi}%
\newcommand{\multiref}[1]{\begingroup
\xdef\mr@no@sparg{\expandafter\mr@ignsp#1 \: }%
\def\mr@comma{}%
\@for\mr@refs:=\mr@no@sparg\do{\mr@comma\def\mr@comma{,}\ref{\mr@refs}}%
\endgroup}
\makeatother

\newcommand{\hypref}[2]{\ifx\href\asklfhas #2\else\href{#1}{#2}\fi}

\newcommand{\secref}[1]{Sec.~\multiref{#1}}

\newcommand{\appref}[1]{App.~\multiref{#1}}

\newcommand{\figref}[1]{Fig.~\multiref{#1}}
\renewcommand{\eqref}[1]{(\multiref{#1})}

\ifx\href\asklfhas\newcommand{\href}[2]{#2}\fi

\newcommand{\comma}{\quad,\quad}

\newcommand{\eps}{\varepsilon}

\newcommand{\be}{\begin{eqnarray}}
\newcommand{\ee}{\end{eqnarray}}

\DeclareMathOperator{\sech}{sech}
\DeclareMathOperator{\csch}{csch}

\DeclareMathOperator{\arctanh}{arctanh}

\newcommand{\Elliptic}[1]{\mathbbm{#1}}
\newcommand{\EllipticE}{\Elliptic{E}}
\newcommand{\EllipticF}{\Elliptic{F}}
\newcommand{\EllipticK}{\Elliptic{K}}
\newcommand{\EllipticPi}{\oldPi}
\DeclareMathOperator{\JacobiAM}{am}

\DeclareMathOperator{\JacobiCN}{cn}
\DeclareMathOperator{\JacobiDN}{dn}
\DeclareMathOperator{\JacobiSN}{sn}

\DeclareMathOperator{\JacobiSD}{sd}
\DeclareMathOperator{\JacobiCD}{cd}

\DeclareMathOperator{\JacobiDS}{ds}
\DeclareMathOperator{\JacobiCS}{cs}
\DeclareMathOperator{\JacobiNS}{ns}

\begin{document}

\thispagestyle{empty}
\begin{flushright}\footnotesize
\texttt{UUITP-14/13}\\
\texttt{HU-EP-13/45}%
\end{flushright}
\vspace{1cm}

\begin{center}%
{\Large\textbf{\mathversion{bold}%
Correlation Function of Circular Wilson Loops at Strong Coupling
}\par}

\vspace{1.5cm}

\textrm{Amit Dekel$^1$ and Thomas Klose$^{1,2}$} \vspace{8mm} \\
\textit{%
$^1$ Department of Physics and Astronomy, Uppsala University \\
SE-75108 Uppsala, Sweden \\[6mm]
$^2$ Institut f\"{u}r Physik, Humboldt-Universit\"{a}t zu Berlin, \\
Newtonstra{\ss}e 15, 12489 Berlin, Germany
} \\

\texttt{\\ amit.dekel@physics.uu.se, thomas.klose@physics.uu.se}

\par\vspace{14mm}

\textbf{Abstract} \vspace{5mm}

\begin{minipage}{14cm}
We study the correlation function of two circular Wilson loops at strong coupling in $\mathcal{N}=4$ super Yang-Mills theory. Using the AdS/CFT correspondence, the problem maps to finding the minimal surface between two circles defined on the boundary of AdS, and the fluctuations around the classical solution in $\AdS_5\times \Sphere^5$. At the classical level, we derive the string solution in $\Hyp_3\times\Sphere^1$ explicitly, and focus on properties such as stability and phase transition. Furthermore, a computation of the associated algebraic curve is given. At the quantum level, the one-loop partition function is constructed by introducing quadratic bosonic and fermionic fluctuations around the classical solution, embedded in $\AdS_5\times \Sphere^5$. We find an analytic, formal expression for the partition function in terms of an infinite product by employing the Gel'fand-Yaglom method and supersymmetric regularization. We regulate the expression and evaluate the partition function numerically.

\end{minipage}

\end{center}

\newpage

\tableofcontents

\bigskip
\noindent\hrulefill
\bigskip

\section{Introduction}

Wilson loops are interesting and important non-local observables in gauge theories. For instance, they can be used as order parameters in the definition of the various phases in which the theory may be. In a weakly coupled gauge theory, their expectation values can be estimated using perturbation theory. If the gauge theory is very special and possesses a string theory dual, then Wilson loops may also be studied indirectly in the strong coupling regime by means of the AdS/CFT correspondence \cite{Maldacena:1997re,Witten:1998qj,Gubser:1998bc}. In this holographic description, the Wilson loop expectation value is calculated from the string path integral subject to the condition that the string world-sheet has to end at the boundary of AdS on the contour that is defined by the Wilson loop \cite{Rey:1998ik,Maldacena:1998im}. Moreover, in integrable theories such as $\mathcal{N}=4$ supersymmetric Yang-Mills theory (SYM), one may hope to go even further and study these observables at any value of the coupling constant; see \cite{Beisert:2010jr} for a review. In fact, exact results have been obtained using the idea of localization \cite{Pestun:2007rz}.

\bigskip

In this paper, we consider the correlator of two concentric circular Wilson loops at strong coupling in $\mathcal{N}=4$ SYM using holography. The special case where one of the loops shrinks to zero size is contained as a limit of this generic setup and describes the string theory dual of a correlator of a circular Wilson loop and a local operator. Likewise, the single circular Wilson loop can be obtained by taking a limit. All configurations that we consider are characterized by two charges, a dilatation charge $Q$ and an angular momentum $J$. At strong coupling, the Wilson loop expectation value is encoded in the string partition function. For the generic case, we compute the tree-level partition function and for vanishing angular momentum, we also compute its one-loop correction.

\bigskip

At leading order, we solve the problem of finding the analytical solution describing a general minimal surface of revolution in the Poincar\'{e} model of hyperbolic three-space $\Hyp_3$ times a circle $\Sphere^1$, see \secref{sec:classical}. These are precisely the surfaces that end on circular contours when extended to the boundary of $\Hyp_3 \subset \AdS_5$. It is the dependence on $\Sphere^1$ that imparts the angular momentum to the corresponding string, and consequently the Wilson loop. Thus, a configuration with vanishing angular momentum charge is confined to $\Hyp_3$. Throughout the paper, we use two parameters, $j_1$ and $j_2$, to characterize the different configurations, see \figref{fig:surface-plots}, their charges and the ratio of the Wilson loop radii.

\bigskip

Generally, given two concentric circles on the AdS boundary, there exist several string solutions. In the simpler case, where the angular momentum vanishes and the solution is confined to $\Hyp_3$, there can be two, one or zero solutions for a given boundary configuration. Then, in case there are two solutions, one is stable against perturbations while the other one is not. If there is only one solution, we call it a \emph{critical} solution. It is a saddle point of the string action and can be thought of as the degeneration of the stable and the unstable solution. These feature are shared, at least qualitatively, by the catenoid in flat space. Back to the $\Hyp_3\times\Sphere^1$ case, we similarly define critical configurations in terms of $j_1$ and $j_2$ as those with only one solution. These configurations define a curve in the $j_1$-$j_2$-plane, separating stable and unstable configurations, see \figref{fig:Phases}. The stability of a configuration is directly related to the partition function, where we study fluctuations around the classical solution. For stable configurations, all the eigenvalues are positive, while for unstable ones, at least one is negative. For the system studied here, unstable configurations have exactly one negative eigenvalue associated with the fluctuations in the normal directions in $\Hyp_3$. As the partition function is proportional to the square root of this determinant, the negative eigenvalue renders the partition function for unstable configurations imaginary.

\bigskip

The classical solution for this problem was discussed in \cite{Zarembo:1999bu,Olesen:2000ji} for two parallel circular Wilson loops with equal radii, and more extensively in \cite{Drukker:2005cu} with some generalizations which also include the sphere in the target space. In \cite{Drukker:2005cu}, it was also noticed that the configurations of \cite{Zarembo:1999bu} are related to the concentric Wilson loops by a conformal transformation. These papers also discuss the Gross-Ooguri phase transition of the system \cite{Gross:1998gk}. If the loops are separated by a distance much greater than their radii, the configuration can be approximated by two disconnected surfaces which interact by exchange of supergravity modes as shown in \cite{Berenstein:1998ij}. In this paper, we present the results for the classical solution in a slightly different fashion and extend the stability analysis.

\bigskip

To get the one-loop correction, we should consider fluctuations along all transverse directions in $\AdS_5\times \Sphere^5$ as well as fermionic fluctuations. We follow \cite{Drukker:2000ep}, where a prescription for computing the one-loop partition function in the Green-Schwarz formalism was given. The resulting partition function is given formally in terms of determinants of second order differential operators as we show in \secref{sec:partition-function}. In order to evaluate these determinants, one should take the product of all eigenvalues of the relevant operators. Finding the explicit analytic expressions for the eigenvalues is, however, a hard task since the operators turn out to be quite complicated. Although, we do present approximate formulas that are rather accurate for large eigenvalues, see \appref{app:frequencies}, there is a way to circumvent the problem of finding the individual eigenvalues altogether, namely the Gel'fand-Yaglom method (GY).

\bigskip

The GY method, which we use extensively in this paper (see \appref{app:GY} for details), allows one to evaluate at once the product of all eigenvalues of a one-dimensional second order differential operator by solving a homogeneous initial value problem. In general, this method is not applicable to the string world-sheet theory as the operators are two-dimensional, however, by virtue of the azimuthal symmetry of the classical solution, it is easy to trade the value of the two-dimensional determinant for an infinite product of one-dimensional determinants which we do know how to solve analytically using the GY method. Thus, we finally end up with a formal infinite product over know functions which represents the partition function as shown in \secref{sec:detsUsingGY}.

\bigskip

Here we follow \cite{Kruczenski:2008zk} which first used the GY method for computing the straight line and the circular Wilson loop partition functions. We also combine it with some ideas introduced in \cite{Chu:2009qt} and \cite{Forini:2010ek}, where the partition function of two infinite parallel lines (known also as the $q\bar q$-potential) was studied using the GY method (see also \cite{Drukker:2011za} for the generalized $q\bar q$-potential, and \cite{Beccaria:2010ry} for a similar treatment for computing the one-loop corrections to the energy of the folded string). In the latter papers, the approach is a bit different from the one used in \cite{Kruczenski:2008zk} as we shall explain in the text. A yet slightly different approach was taken in \cite{Kristjansen:2012nz} for the computation of the straight line and circular Wilson loop partition functions. The weak coupling counterpart of our computation was carried out some time ago to second order in perturbation theory in \cite{Plefka:2001bu,Arutyunov:2001hs}.

\bigskip

As usual \cite{Kruczenski:2008zk,Forini:2010ek}, the Wilson loop partition function is both UV and IR divergent, and one has to regulate the result. In \secref{sec:REG}, we give our regularization scheme, where we subtract a reference solution which regulates both UV and IR divergences and yields a finite result. This regularization procedure determines the result up to an overall constant, which is not important if one is interested in the effective action as a function of the ratio of the radii. For this reason, we modify the reference function so that the sum converges faster at the price of changing the (anyway) unknown overall constant.

\bigskip

As a consequence of the integrability of the theory, the classical string solutions in $\AdS_5\times \Sphere^5$ are characterized by an algebraic curve \cite{Kazakov:2004qf}. These algebraic curves are usually associated with closed string solutions where one can define a nontrivial monodromy. However, for Wilson loops, there are usually no nontrivial monodromies, and the algebraic curve introduced in \cite{Kazakov:2004qf} is not defined (see however \cite{Janik:2012ws}). For the case of our interest, there is a nontrivial monodromy, and we compute the associated algebraic curve using the method given in \cite{Dekel:2013dy}.

\section{Classical string solutions with axial symmetry}
\label{sec:classical}

The classical string solutions that we consider are minimal surfaces that can be embedded inside $\Hyp_3$ or $\Hyp_3\times\Sphere^1$. Moreover, we focus on solutions that are rotationally symmetric about the ``vertical'' axis of $\Hyp_3$, i.e.\ about the $z$-axis if the metric in the Poincar\'{e} coordinates is given by
\be \label{eqn:Metric-H3}
  ds^2 = \frac{d\vec{x}^{\,2}+dz^2}{z^2} \; .
\ee
In this section, we will find a one-parameter set of solutions in $\Hyp_3$ and their generalization to a two-parameter set of solutions in $\Hyp_3\times\Sphere^1$. A figure showing the types of surfaces that we are going to find is given on page \pageref{fig:surface-plots}.

\subsection{Action and equations of motion}

\paragraph{Surfaces in $\Hyp_3$.} Surfaces of revolution can be parametrized as
\be \label{eqn:surf-of-revolution}
  \vec{x} = \matr{c}{r(\tau)\cos\sigma \\ r(\tau)\sin\sigma} \comma z = z(\tau) \; ,
\ee
where $\sigma = 0..2\pi$. The target-space metric induces the metric
\be \label{eqn:induced-metric-H3}
  g_{\tau\tau} = \frac{\dot{r}^2 + \dot{z}^2}{z^2}
  \comma
  g_{\tau\sigma} = 0
  \comma
  g_{\sigma\sigma} = \frac{r^2}{z^2}
\ee
on the (euclidean) string world-sheet. Thus, the Nambu-Goto action for effective string tension $\sqrt{\lambda}$ is given by
\be \label{eqn:Nambu-Goto-H3}
  \Action = \sqrt{\lambda} \int d\tau\: \frac{r\sqrt{\dot{r}^2+\dot{z}^2}}{z^2} \; ,
\ee
where the trivial $\sigma$-integration has been performed. The equations of motion for $r$ and $z$ are satisfied if and only if
\be \label{eqn:eom-rz}
  \dot{r} \ddot{z} - \dot{z} \ddot{r} + \lrbrk{\frac{2\dot{r}}{z} + \frac{\dot{z}}{r} } \lrbrk{ \dot{z}^2+\dot{r}^2 } = 0	\; .
\ee
These equations are invariant under reparametrization of $\tau$. We choose to fix this gauge freedom by imposing conformal gauge, namely by setting
\be \label{eqn:conformal-gauge-condition-H3}
  g_{\tau\tau} - g_{\sigma\sigma} = \frac{\dot{r}^2 + \dot{z}^2 - r^2}{z^2} \stackrel{!}{=} 0 \; .
\ee
This is the usual (diagonal) Virasoro constraint.

The action is invariant under dilatations. The associated conserved charge, $Q = \sqrt{\lambda} \mathcal{Q}$, is
\be \label{eqn:dilatation-charge}
  \mathcal{Q} = \frac{r}{\sqrt{\dot{r}^2 + \dot{z}^2}} \frac{r\dot{r} + z\dot{z}}{z^2} = \frac{r\dot{r} + z\dot{z}}{z^2} \; .
\ee
where we used \eqref{eqn:conformal-gauge-condition-H3} in the second step. This equation is a first integral of \eqref{eqn:eom-rz} and thus replaces the equation of motion.

\paragraph{Surfaces in $\Hyp_3\times\Sphere^1$.} We generalize the setup slightly by adding a circle, $\Sphere^1$, to the target space. The coordinate on the circle will be called $\phi$ and the metric is now
\be \label{eqn:Metric-H3S1}
  ds^2 = \frac{d\vec{x}^{\,2}+dz^2}{z^2} \pm d\varphi^2 \; ,
\ee
where we leave the signature of the circle undetermined ($+$ for space-like, $-$ for time-like). We will assume that $\varphi$ only depends on $\tau$ and not on $\sigma$, which is nothing but in the spirit of studying surfaces of revolution. The Nambu-Goto action then reads
\be \label{eqn:Nambu-Goto-H3S1}
  \Action = \sqrt{\lambda} \int d\tau\: \frac{r\sqrt{\dot{r}^2 + \dot{z}^2 \pm z^2 \dot{\varphi}^2}}{z^2}
\ee
and the conformal gauge constraint is generalized to
\be \label{eqn:conformal-gauge-condition-H3S1}
  \frac{\dot{r}^2 + \dot{z}^2 - r^2}{z^2} \pm \dot{\varphi}^2 \stackrel{!}{=} 0 \; .
\ee
The dilatation charge simplifies to the same expression that we already had above
\be
  \mathcal{Q} = \frac{r}{\sqrt{\dot{r}^2 + \dot{z}^2 \pm z^2 \dot{\varphi}^2}} \frac{r\dot{r} + z\dot{z}}{z^2} = \frac{r\dot{r} + z\dot{z}}{z^2} \; .
\ee

In the following, we will assume a linear ``motion'' on the circle and set
\be \label{eqn:sol-varphi}
  \varphi(\tau) = \mathcal{J} \tau \; .
\ee
This is a solution to the $\varphi$-equation of motion in conformal gauge and the constant of proportionality is, in fact, the conserved charge, $J = \sqrt{\lambda} \mathcal{J}$, associated with translations along $\Sphere^1$. Then, dilatation charge conservation and the gauge constraint, which are equivalent to the equations of motion for $r$ and $z$ are given by
\be
  \frac{r\dot{r} + z\dot{z}}{z^2} = \mathcal{Q}
  \comma
  \frac{\dot{r}^2 + \dot{z}^2 - r^2}{z^2} = \mp \mathcal{J}^2
  \; .
\ee
It will be convenient to introduce two new parameters, $j_1$ and $j_2$, in place of $\mathcal{Q}$ and $\mathcal{J}$ such that both signatures can be discussed simultaneously. The equations we will study in the following are
\be \label{eqn:generalized-eom-rz}
  \frac{r\dot{r} + z\dot{z}}{z^2} = \frac{j_1+j_2}{2}
	\comma
  \frac{\dot r^2 + \dot z^2 - r^2}{z^2} = j_1 j_2 \; .
\ee
If $j_1$ or $j_2$ or both vanish, then $\mathcal{J} = 0$ and the surface is confined to $\Hyp_3$. If both are non-zero, then the sign of their product determines whether the circle is space- or time-like.

\subsection{General solution}

\begin{figure}%
\begin{center}
\includegraphics[width=60mm]{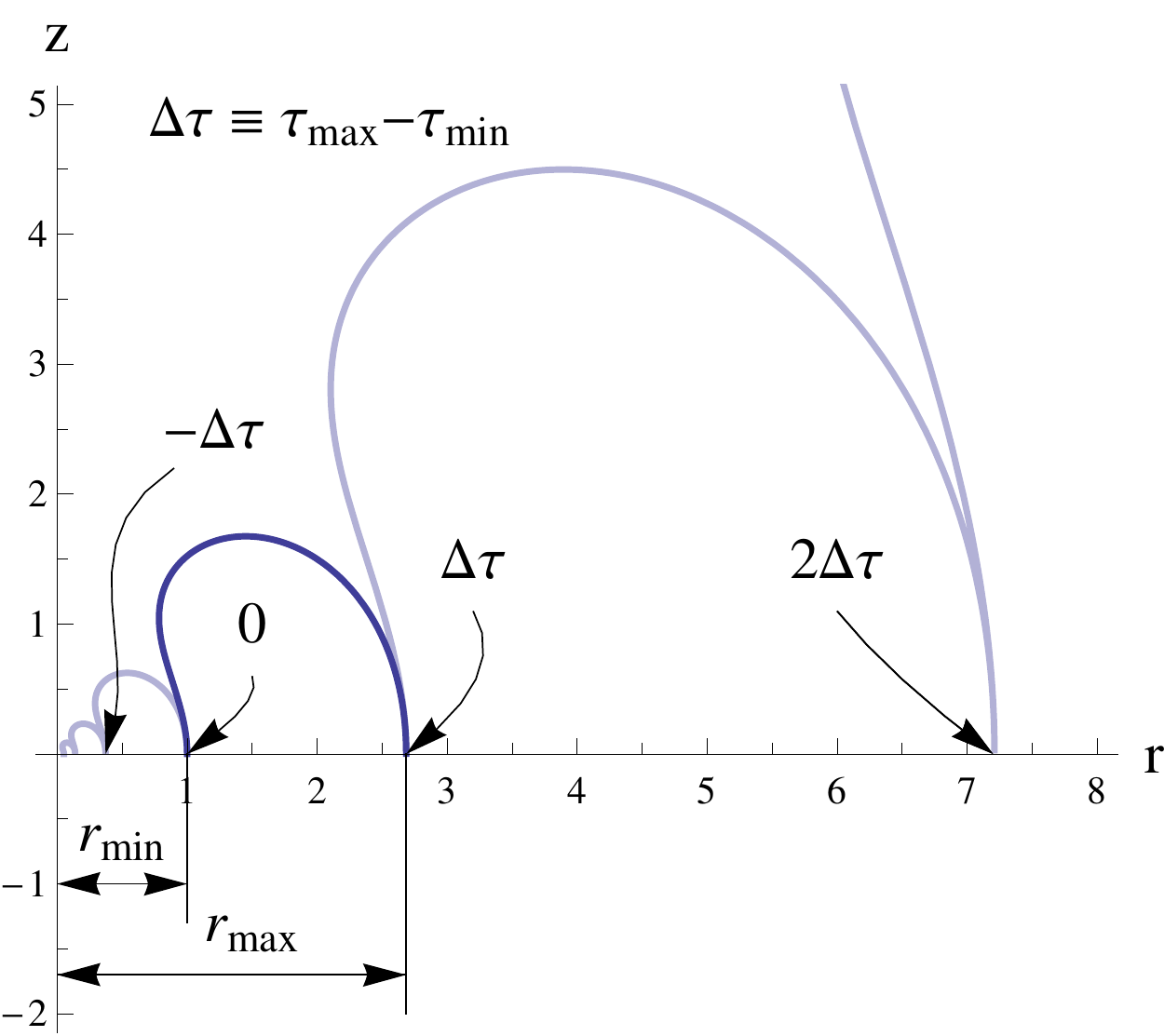}
\caption{\textbf{Branches of the solution.} The profile $\{r(\tau),z(\tau)\}$ of the surface is plotted for $j_1=1.5$ and $j_2=0$. The branch of the solution that we consider is shown in bold and corresponds to the $\tau$-interval \protect\eqref{eqn:tau-interval}. If the functions $r(\tau)$ and $z(\tau)$ are extended beyond this interval, one obtains larger and smaller copies of the principal branch.}%
\label{fig:solution}%
\end{center}
\end{figure}

For solving \eqref{eqn:generalized-eom-rz}, it is convenient to change variables from $r(\tau)$ and $z(\tau)$ to $h(\tau)$ and $f(\tau)$ via
\be \label{eqn:rz_to_hf}
  r = \sqrt{1-\frac{1}{h^2}} \, e^{f}
  \comma
  z = \frac{1}{h} \, e^{f} \; ,
\ee
where $h\ge1$ is required for all $\tau$ such that $r$ is real. Then, the two equations in \eqref{eqn:generalized-eom-rz} take the form
\be\label{eqn:generalized-eom-hf}
  2 h^2 \dot{f} = j_1+j_2
  \comma
  \frac{\dot h^2}{h^2 - 1} + h^2 \dot{f}^2 = h^2 - 1 + j_1 j_2 \; .
\ee
Eliminating $\dot{f}$, we end up with a first order equation for $h$, namely
\be \label{eqn:eom-h}
  \frac{\dot h^2}{h^2-1} + \frac{(j_1+j_2)^2}{4 h^2} = h^2 - 1 + j_1 j_2 \; .
\ee
This equation is solved by
\be \label{eqn:general-solution-h}
  h(\tau) = \sqrt{1+a\JacobiDS^2(\sqrt{a}\,\tau|m)} \; ,
\ee
where $\JacobiDS(u|m)$ is the Jacobi elliptic function\footnote{We collect the properties of various special functions that are relevant for our purposes in \appref{app:EllipticFunctions}.} and
\be \label{eqn:parameters-a-m}
  a = \sqrt{\left(1+j_1^2\right) \left(1+j_2^2\right)}
  \comma
  m = \frac{1}{2} \lrbrk{ 1 + \frac{1 + j_1 j_2}{\sqrt{\left(1+j_1^2\right) \left(1+j_2^2\right)}}}
  \; .
\ee
The ranges of these two parameters are $a\ge1$ and $0\le m\le1$ and they can be used as alternative labels for the solution instead of $j_1$ and $j_2$, or $\mathcal{Q}$ and $\mathcal{J}$. The integration constant that arises when solving \eqref{eqn:eom-h} corresponds to shifts in $\tau$. We chose it such that $z(\tau=0)=0$, i.e.\ that the solution starts at the boundary of the Poincar\'{e} patch. The next zero of $z$, or pole of $h$, is at $\tau = \frac{2}{\sqrt{a}} \EllipticK(m)$, where $\EllipticK(m)$ is the complete elliptic integral of the first kind. Thus, a solution in the range
\be \label{eqn:tau-interval}
  \tau_{\mathrm{min}} = 0  \;\le\; \tau \;\le\; \tau_{\mathrm{max}}= \frac{2}{\sqrt{a}} \EllipticK(m)
\ee
is one ``branch'' that reaches from boundary to boundary, see \figref{fig:solution}. Inserting the solution for $h$ into \eqref{eqn:generalized-eom-hf}, one can solve for $f$ and finds
\be \label{eqn:general-solution-f}
  f(\tau) = \frac{j_1+j_2}{2\sqrt{a}(am-1)} \Bigsbrk{ \EllipticPi\bigbrk{m-\tfrac{1}{a}, \JacobiAM(\sqrt{a} \, \tau|m) |m} - \sqrt{a} \, \tau } + f_0 \; ,
\ee
where $\JacobiAM(u|m)$ is the Jacobi amplitude and $\EllipticPi(n,\phi|m)$ is the incomplete elliptic integral of the third kind. The integration constant $f_0$ enters the solution \eqref{eqn:rz_to_hf} as an overall constant $e^{f_0}$ and determines the scale of the solution. In the limit $j_1=j_2=j$, the modulus becomes $m=1$ and the solutions simplify to
\be
  h(\tau) = \sqrt{1 + (1+j^2) \csch^2\bigbrk{\tau\sqrt{1+j^2}}}
	\comma
  f(\tau) = \arctanh\frac{j\tanh\bigbrk{\tau\sqrt{1+j^2}}}{\sqrt{1+j^2}} \; .
\ee

\begin{figure}%
\begin{center}
\includegraphics[width=160mm]{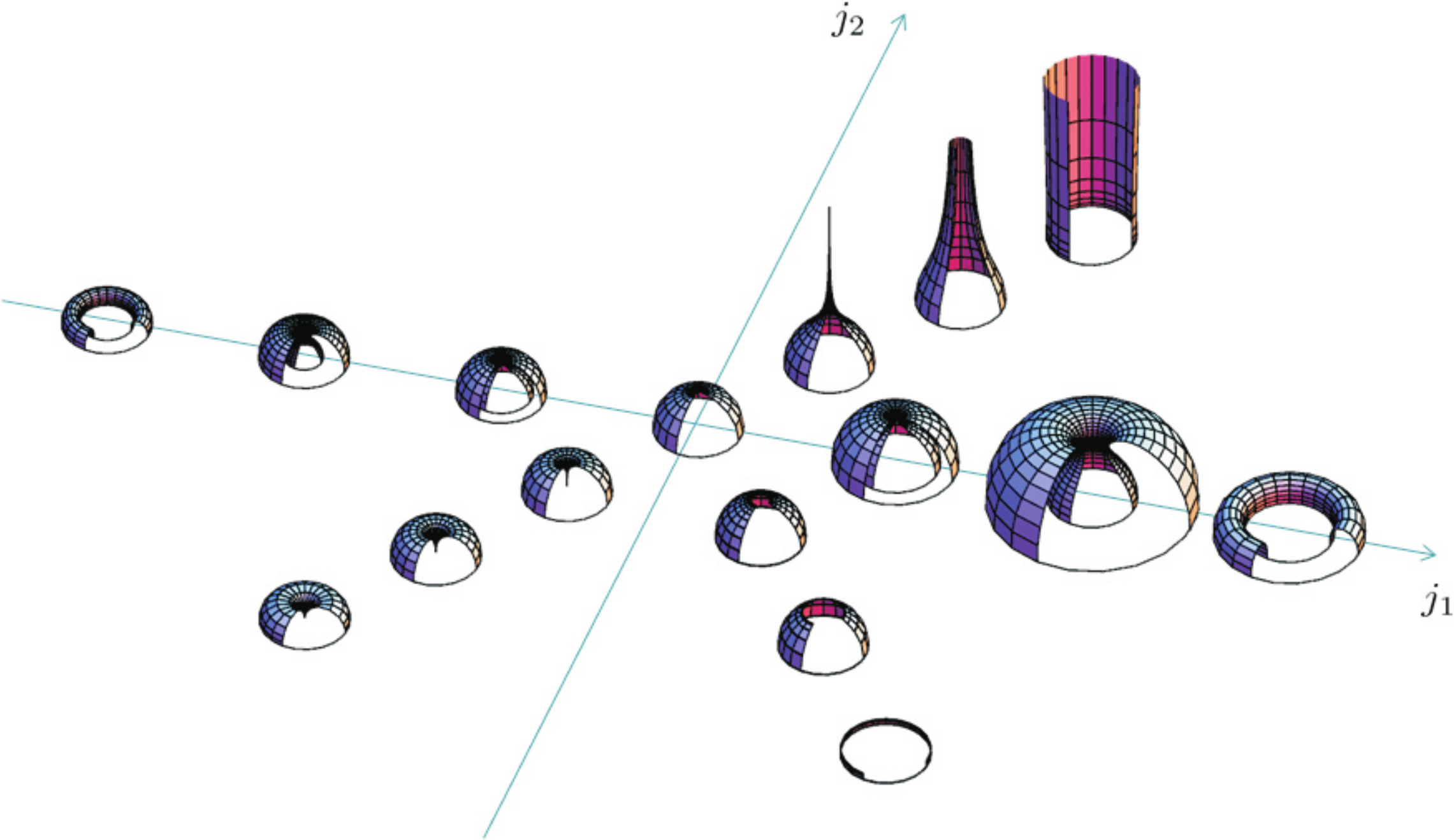}
\caption{\textbf{Parameter dependence.} This figure displays the qualitative difference of the surface as a function of the parameters $j_1$ and $j_2$. This ``phase diagram'' is symmetric under $j_1\leftrightarrow j_2$. We have used this fact to omit some figures. If $j_1$ or $j_2$ is zero, the surface is minimal in $\Hyp_3$. If either of the parameters is non-zero, then the surface ``moves'' along a great circle on the internal space.}%
\label{fig:surface-plots}%
\end{center}
\end{figure}

Being a surface of revolution implies that if the surface touches the boundary of the hyperbolic space then it has to do so along a circle. The gauge theory interpretation is thus that of a correlator of two circular Wilson loops. The radii of the circles at $\tau_{\mathrm{min}}$ and $\tau_{\mathrm{max}}$ are
\begin{align} \label{eqn:r-min-max}
  r_{\mathrm{min}} &\equiv r(\tau_{\mathrm{min}}) = e^{f_0} \; , \\
  r_{\mathrm{max}} &\equiv r(\tau_{\mathrm{max}}) = e^{f_0}
	 \exp\lrbrk{ \frac{j_1+j_2}{\sqrt{a}(am-1)} \Bigsbrk{ \EllipticPi\bigbrk{m-\tfrac{1}{a}|m} - \EllipticK(m) } } \; , \nn
\end{align}
respectively. Note that neither is $r_{\mathrm{min}}$ necessarily the smallest nor is $r_{\mathrm{max}}$ the biggest radius of the solution. The labels ``min'' and ``max'' just refer to the left and right ends of the interval \eqref{eqn:tau-interval}, see \figref{fig:solution}. If $j_1$ and $j_2$ are both positive or both negative, i.e. for a space-like $\Sphere^1$, then $0 < r_{\mathrm{min}} < r_{\mathrm{max}}$, otherwise, for a time-like $\Sphere^1$, we have $0 < r_{\mathrm{max}} < r_{\mathrm{min}}$. If both are zero, something special happens. Formally, $\tau_{\mathrm{max}} = \infty$ and $r_{\mathrm{min}} = r_{\mathrm{max}}$, however, the surface does not actually return back to the boundary but closes up and forms a hemisphere---it is the string solution for the circular Wilson loop. What mathematically happens in this limit is that it already takes infinite $\tau$ to reach the north pole and the way back to the equator lies inaccessibly behind this ``horizon.''  Representative surfaces for various values of the parameters are plotted in \figref{fig:surface-plots}.

\bigskip

\begin{figure}%
\begin{center}
\includegraphics[width=60mm]{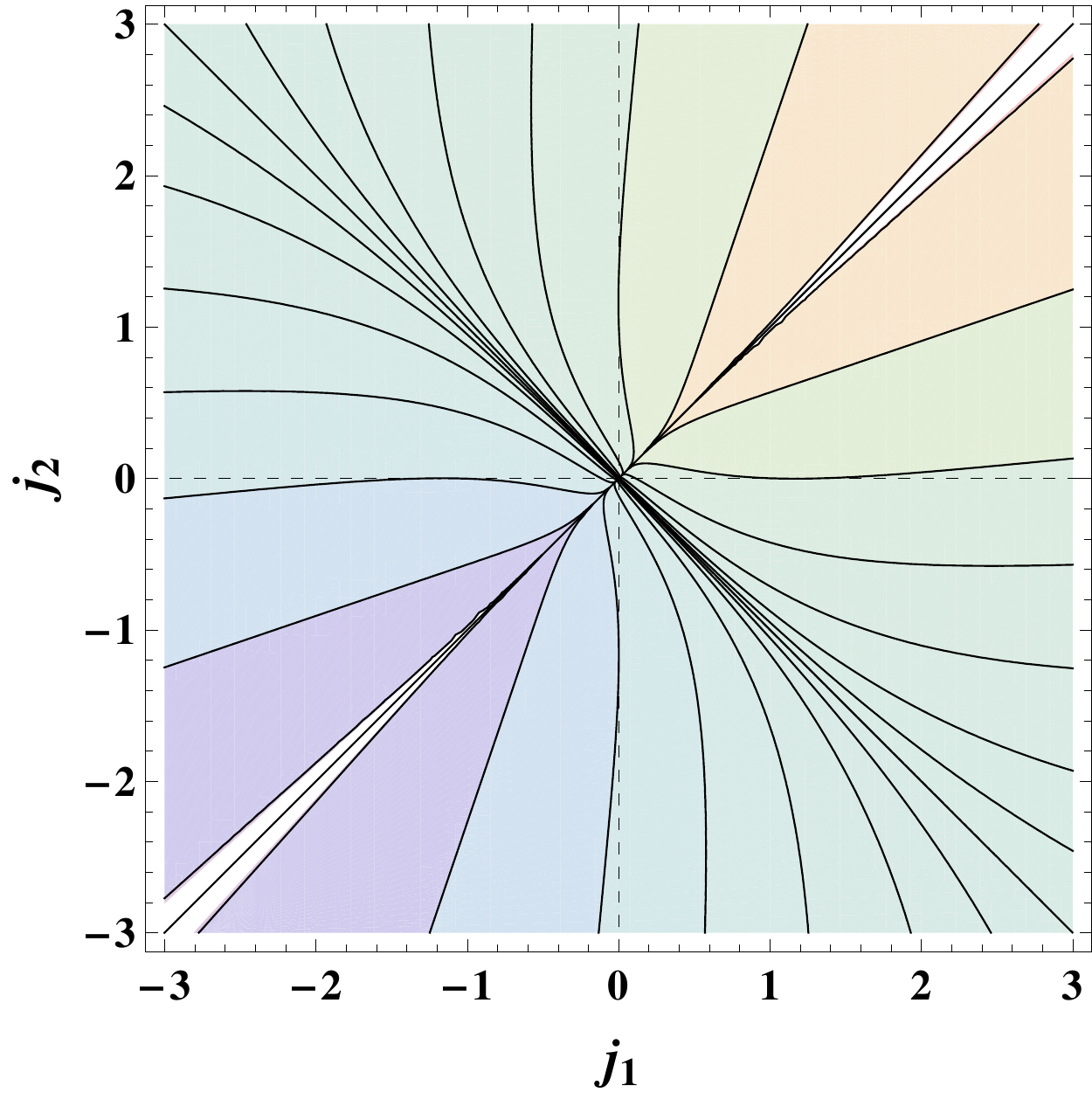}
\caption{\textbf{Ratios of boundary circles.} The string solution ends on two circles on the boundary of $\Hyp_3$. The ratio of their radii depends on the parameters $j_1$ and $j_2$. All points that lie on the same contour in this plot correspond to the same ratio. For $j_1 = j_2 < 0$, this ratio is zero, and for $j_1 = j_2 > 0$, this ratio is infinity. For $j_1 = j_2 = 0$ the surface ends on only one circle and thus a ratio cannot be defined.}%
\label{fig:ratioradii-2d}%
\end{center}
\end{figure}

The parameters $j_1$ and $j_2$ determine the ratio
\be
  \rho \equiv \frac{r_{\mathrm{max}}}{r_{\mathrm{min}}}
  = \exp\lrbrk{ \frac{j_1+j_2}{\sqrt{a}(am-1)} \Bigsbrk{ \EllipticPi\bigbrk{m-\tfrac{1}{a}|m} - \EllipticK(m) } }
\ee
of the radii at which the surface reaches the boundary. The overall scale is determined by the integration constant $f_0$, see \eqref{eqn:general-solution-f}, but because of scale invariance nothing is going to depend on $f_0$. A contour plot of the function $\rho(j_1,j_2)$ is provided in \figref{fig:ratioradii-2d}. For any non-negative value of $\rho$, there is a continuous set of solutions.

If we want to consider solutions that can be embedded in $\Hyp_3$, we have to set one of the $j$'s to zero and we are left with a one-parameter family of solutions labeled by, say,
\be \label{eqn:js-for-H3}
  j_1 = j \comma j_2 = 0 \; .
\ee
In this case, the parameters $a$ and $m$ are not independent anymore. The expression for the ratio simplifies to
\be
  \ln\rho = \pm \sqrt{\frac{4m(2m-1)}{1-m}} \Bigsbrk{ \EllipticPi(1-m|m) - \EllipticK(m) } \; ,
\ee
where we chose to write the logarithm for convenience. The positive sign of the square root applies for positive $j$ and the negative sign for negative $j$. Two useful alternative ways of writing this formula are
\be
  \ln\rho = \pm 2 \Bigsbrk{ \EllipticK(m) \EllipticE(\phi|m) - \EllipticE(m) \EllipticF(\phi|m) }
          = \pm 2 \EllipticK(m) \, Z(\alpha|m) \; ,
\ee
where $\phi = \arcsin\sqrt{\frac{1-m}{m}}$ and $\JacobiSN(\alpha|m) = \sqrt{\frac{1-m}{m}}$, respectively, and $Z(\alpha|m)$ is the Jacobi Zeta function.

\begin{figure}%
\begin{center}
\includegraphics[width=60mm]{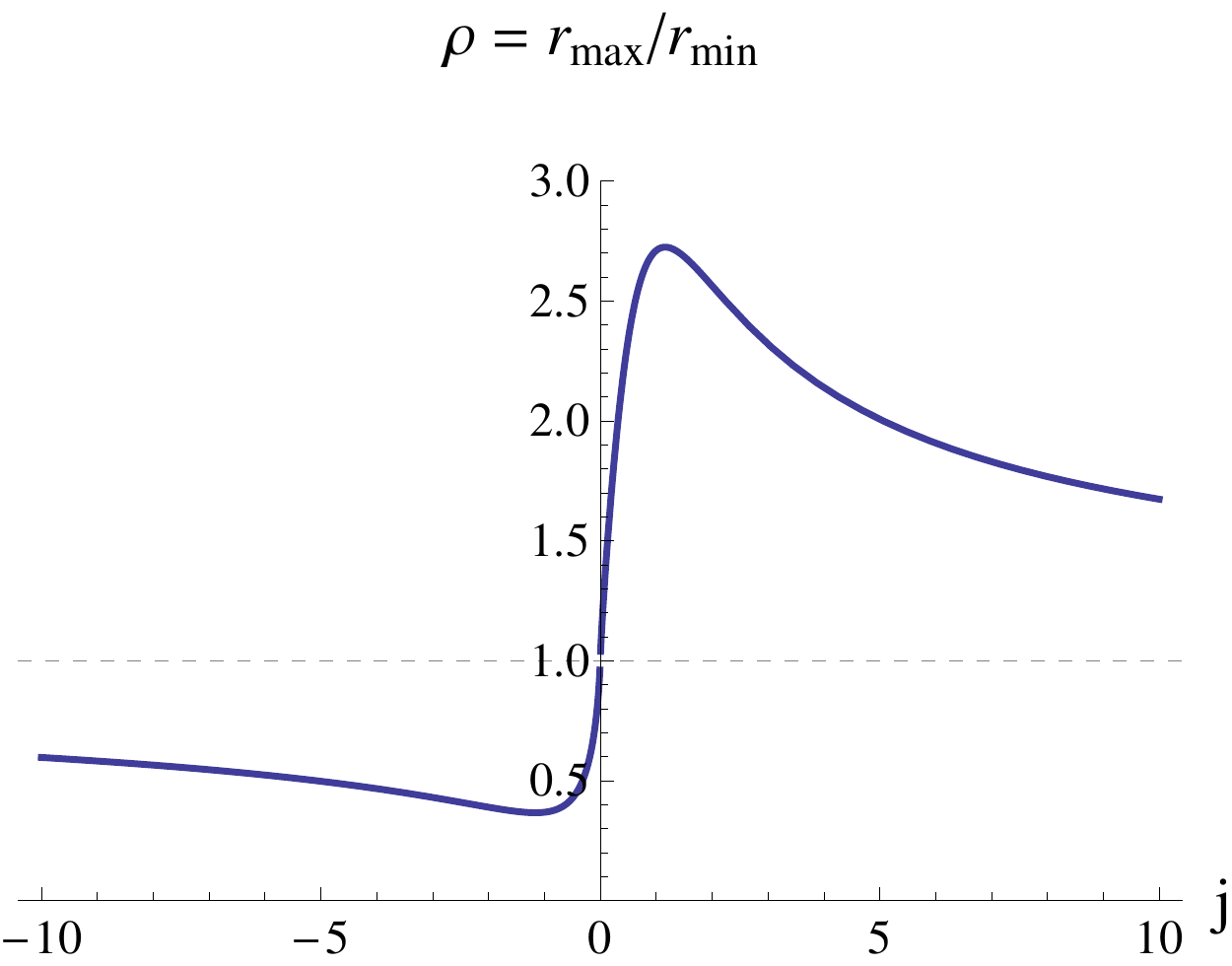}
\caption{\textbf{Ratio of radii.} The parameter $j$ determines the ratio of the radii $r_{\mathrm{min}}$ and $r_{\mathrm{max}}$.}%
\label{fig:ratioradii}%
\end{center}
\end{figure}

\begin{figure}%
\begin{center}
\subfloat[][$\rho=1.05$ \par \mbox{} \hspace{2mm} $j\approx0.008$ and $j\approx1200$]{
  \includegraphics[width=35mm]{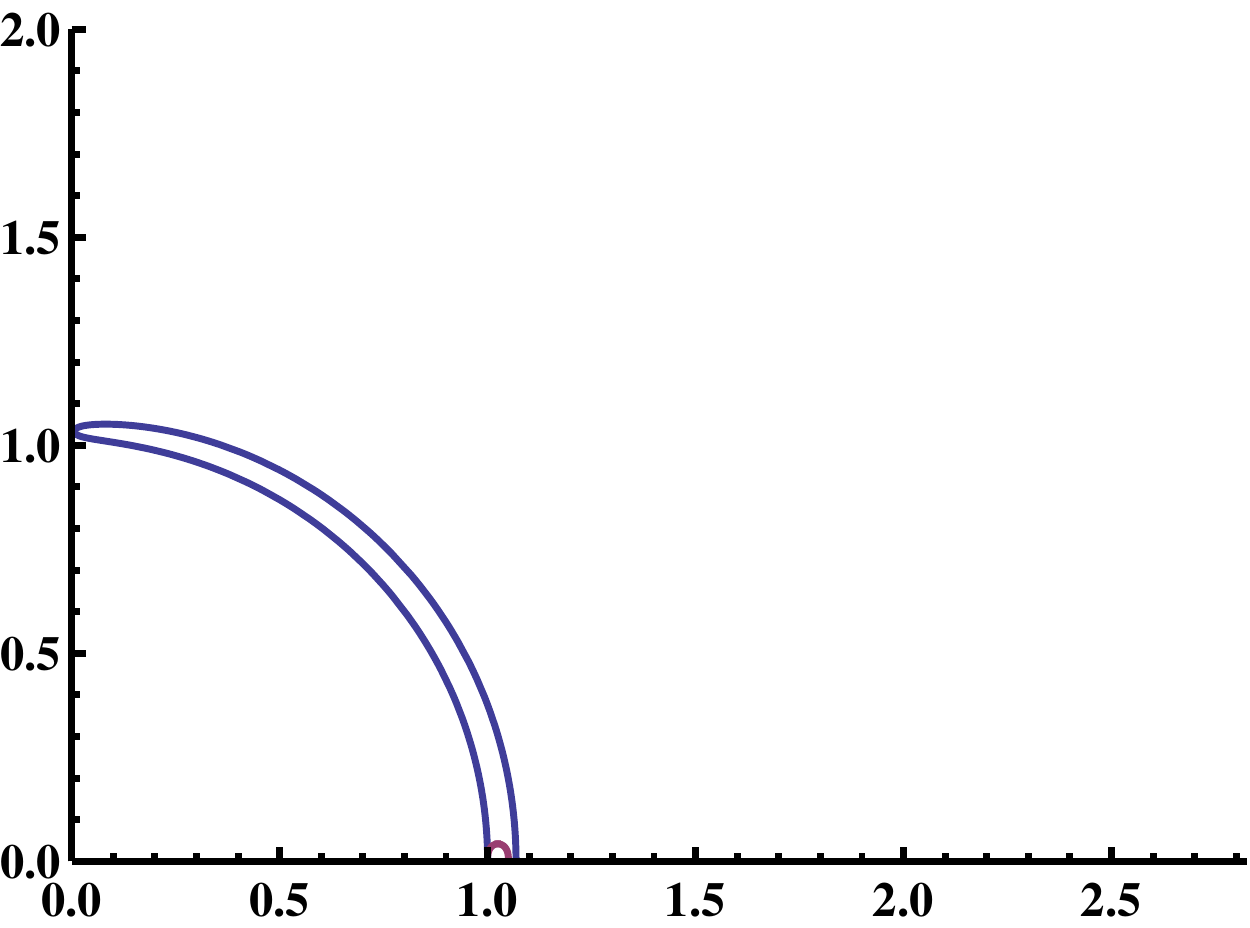}} \hspace{3mm}
\subfloat[][$\rho=1.6$ \par \mbox{} \hspace{3mm} $j\approx0.164$ and $j\approx12.2$]{
  \includegraphics[width=35mm]{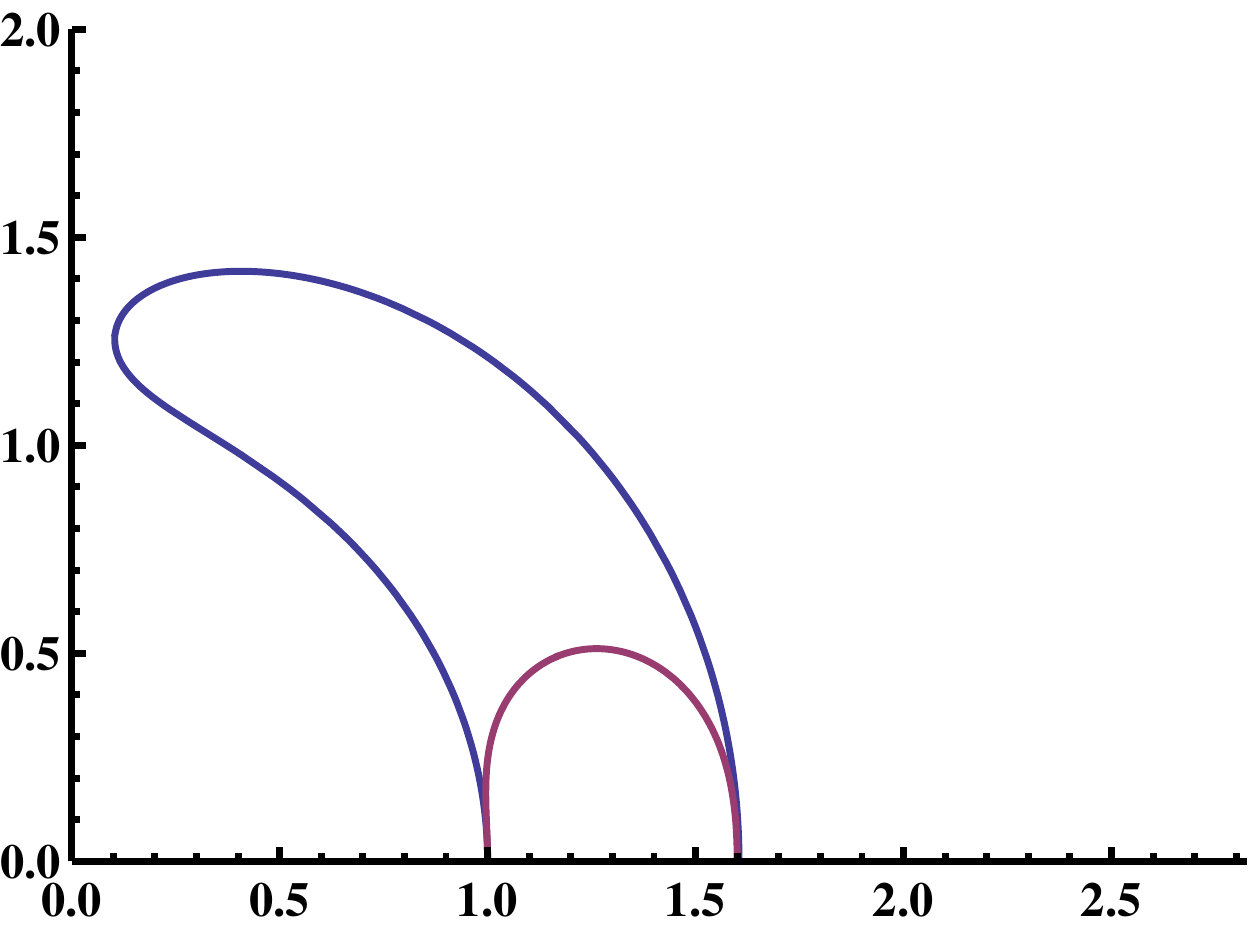}} \hspace{3mm}
\subfloat[][$\rho=2.1$ \par \mbox{} \hspace{3mm} $j\approx0.364$ and $j\approx4.28$]{
  \includegraphics[width=35mm]{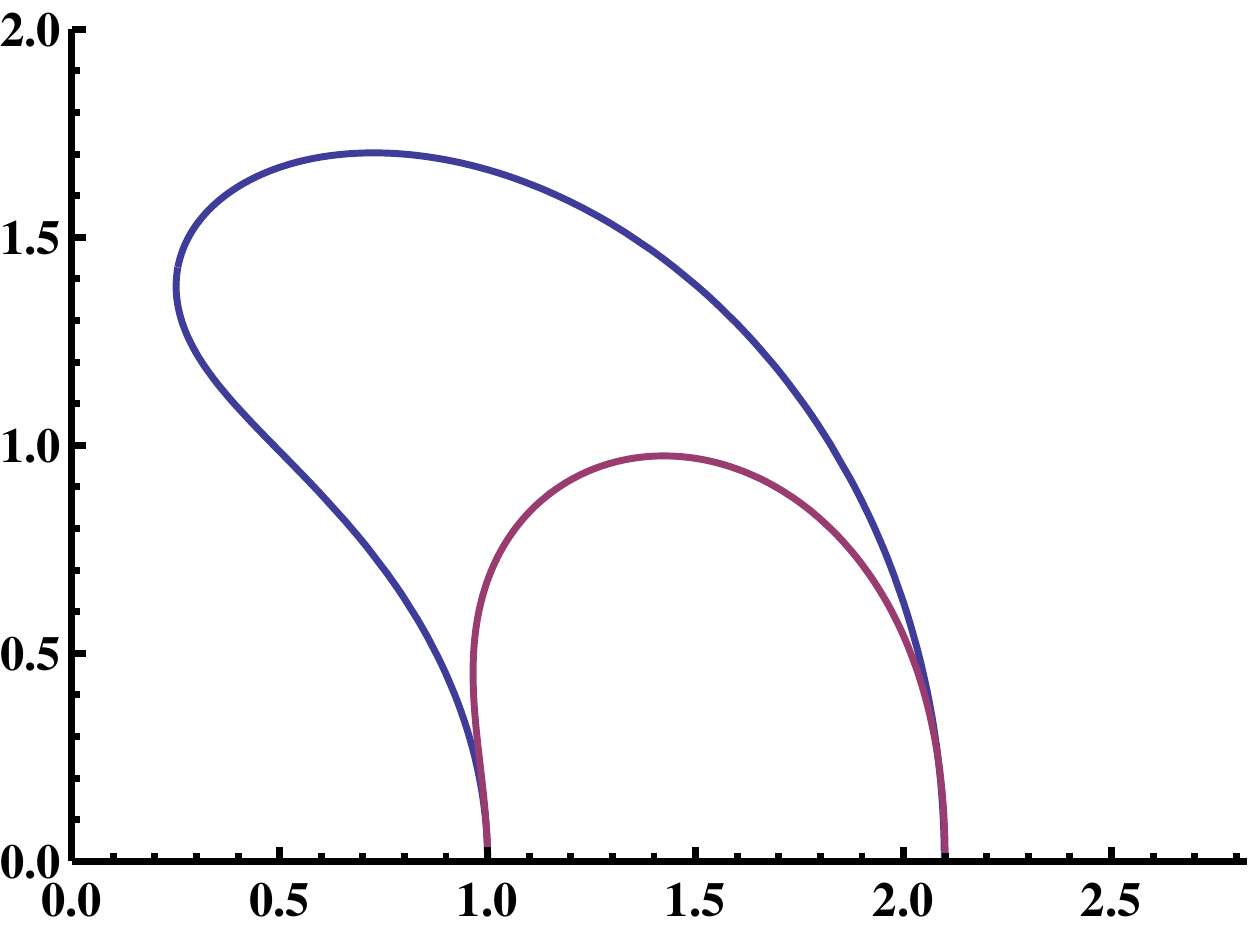}} \hspace{3mm}
\subfloat[][$\rho=\rho_c$ \par \mbox{} \hspace{3mm} $j=j_c\approx1.162$]{
  \includegraphics[width=35mm]{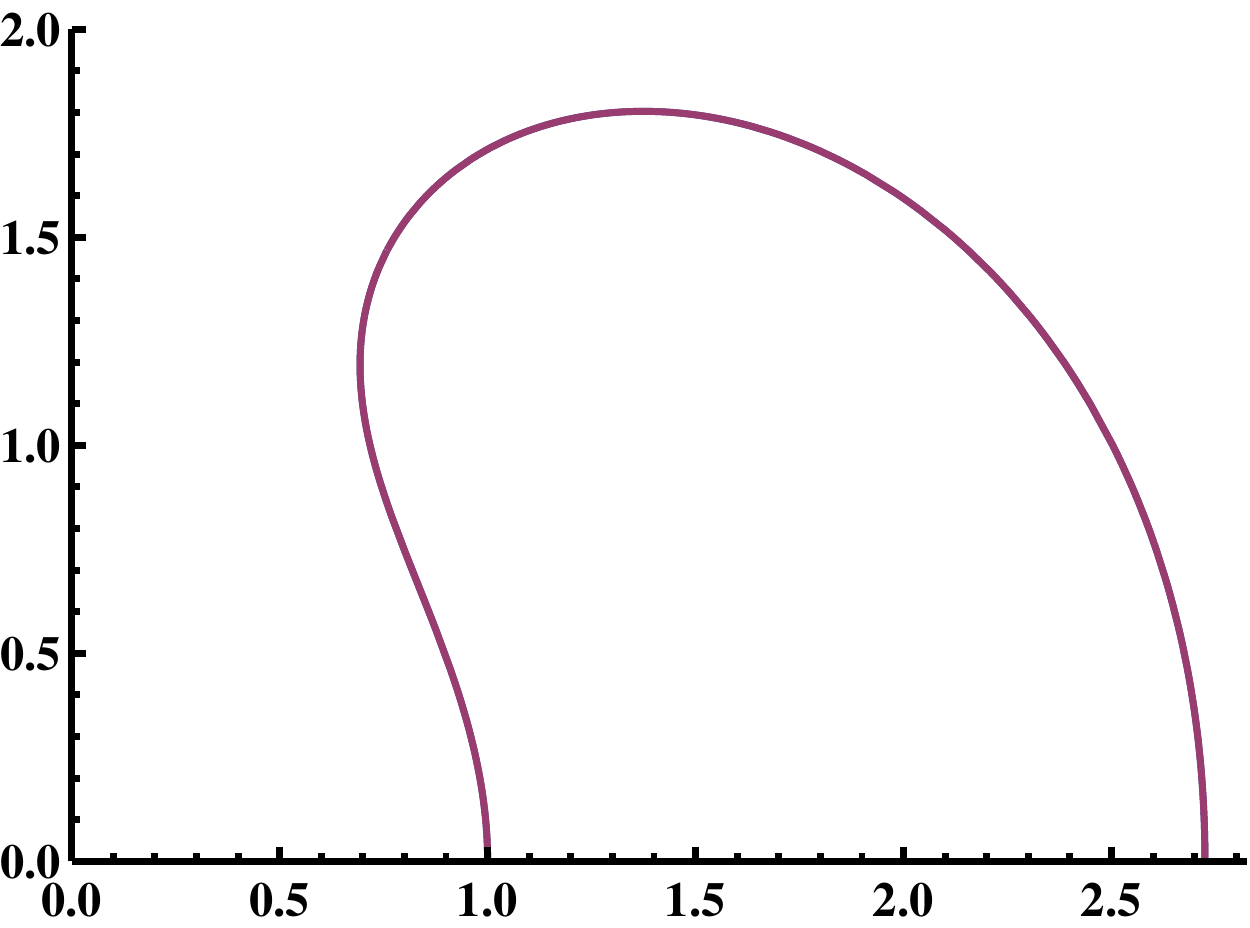}}
\caption{\textbf{Profiles of minimal surface ending on two concentric circles.}}%
\label{fig:WilsonCorr}%
\end{center}
\end{figure}

The function $\rho(j)$, which is the slice $\rho(j,0)$ of the above contour plot, is displayed in \figref{fig:ratioradii}. Note that for solutions on this slice, i.e.\ for surfaces entirely in $\Hyp_3$, there is a largest and a smallest possible value for $\rho$. Numerically, we can easily compute those ``critical'' values and find
\be
  \rho_{\mathrm{largest}} = \rho_c \approx 2.72450
  \comma
  \rho_{\mathrm{smallest}} = 1/\rho_c \approx 0.36704
  \; ,
\ee
which were already obtained in \cite{Olesen:2000ji}. These values are attained for $j = j_c \approx 1.16220$ and $j = -j_c$, respectively. In \secref{sec:partition-function}, we will derive a compact equation for $j_c$, see \eqref{eqn:eqn-for-jc}. \figref{fig:ratioradii} furthermore shows that for any ratio $\rho$ between those extremes, there are \emph{two} possible values for $j$, i.e. there are two solutions for the same boundary conditions. When $\rho$ is tuned to $\rho_{\mathrm{largest}}$ or $\rho_{\mathrm{smallest}}$, those two solutions degenerate into one. We plot some examples in \figref{fig:WilsonCorr}. Notice that the system is invariant under $\rho\to1/\rho$ which amounts to interchanging the two circles. This means that it is enough to consider $1\leq\rho\leq \rho_{\mathrm{max}}$ or, equivalently, $0\leq j< \infty$.

\subsection{Classical action and regularized area}

We evaluate the action \eqref{eqn:Nambu-Goto-H3S1} on the general two-parameter solution that we found above. As the solution satisfies the gauge condition \eqref{eqn:conformal-gauge-condition-H3S1}, the action simplifies to
\be \label{eqn:classical-action-h}
  \Action = \sqrt{\lambda} \int_{\tau_{\mathrm{min}}}^{\tau_{\mathrm{max}}} \!d\tau\: \frac{r^2}{z^2}
          = \sqrt{\lambda} \int_{\tau_{\mathrm{min}}}^{\tau_{\mathrm{max}}} \!d\tau\: (h^2-1)
          = \sqrt{\lambda} \, a \int_{\tau_{\mathrm{min}}}^{\tau_{\mathrm{max}}} \!d\tau\: \JacobiDS^2(\sqrt{a}\,\tau|m) \; ,
\ee
where we use \eqref{eqn:rz_to_hf} in the second and \eqref{eqn:general-solution-h} in the third step. Up to a factor, this action is the area of the corresponding world-sheet
\be
  A = \frac{2\pi}{\sqrt{\lambda}} \, \Action \; .
\ee
The integral in \eqref{eqn:classical-action-h} can be expressed in terms of elliptic functions as
\be
  A \eq - 2\pi \sqrt{a} \Bigsbrk{ \JacobiCS(\sqrt{a}\tau|m)\JacobiDN(\sqrt{a}\tau|m)  }_{\tau_{\mathrm{min}}}^{\tau_{\mathrm{max}}}
	            - 4\pi \sqrt{a} \Bigsbrk{ \EllipticE(m) - (1-m)\EllipticK(m) } \; ,
\ee
where the first term was not evaluated at the boundaries of \eqref{eqn:tau-interval} yet as it is divergent. We regularize it by introducing cutoffs $\eps_0$ and $\eps_1$ at the left and right ends of the interval, respectively, i.e. we replace
\be
  \tau_{\mathrm{min}} \to \tau_{\mathrm{min}} + \eps_0
  \qquad \text{and} \qquad
  \tau_{\mathrm{max}} \to \tau_{\mathrm{max}} - \eps_1
  \; .
\ee
Then, we can expand the divergent term in the area formula for small $\eps_0$ and $\eps_1$ and find
\be
  - 2\pi \sqrt{a} \Bigsbrk{ \JacobiCS(\sqrt{a}\tau|m)\JacobiDN(\sqrt{a}\tau|m)  }_{\tau_{\mathrm{min}} + \eps_0}^{\tau_{\mathrm{max}} - \eps_1}
	= 2\pi \lrsbrk{ \frac{1}{\eps_0} + \frac{1}{\eps_1} } + \order(\eps_0,\eps_1) \; .
\ee
Note that there are no terms of order $\eps_0^0$ or $\eps_1^0$ in these expansions. Now, we would like to relate these two world-sheet cutoffs to a single target-space cutoff $\eps$ through the relations
\be
   z(\tau_{\mathrm{min}} + \eps_0) = z(\tau_{\mathrm{max}} - \eps_1) = \eps \; .
\ee
We do this again by expanding the function $z$ for small $\eps_0$ and $\eps_1$ about $\tau_{\mathrm{min}}$ and $\tau_{\mathrm{max}}$, respectively. We find
\be
  \eps_0 = \frac{\eps}{r_{\mathrm{min}}} + \order(\eps^3)
  \comma
  \eps_1 = \frac{\eps}{r_{\mathrm{max}}} + \order(\eps^3)
\ee
or
\be
  \frac{1}{\eps_0} + \frac{1}{\eps_1} = \frac{r_{\mathrm{min}} + r_{\mathrm{max}}}{\eps} + \order(\eps) \; ,
\ee
again without constant term. Hence, we can write the area as
\be
  A \eq \frac{2\pi(r_{\mathrm{min}} + r_{\mathrm{max}})}{\eps}
	    - 4\pi \sqrt{a} \Bigsbrk{ \EllipticE(m) - (1-m)\EllipticK(m) }
	    + \order(\eps) \; .
\ee
The coefficient of the pole is recognized as the sum of the circumferences of the circles on which the world-sheet ends at the boundary, i.e.\ the total length of the boundary of the world-sheet. In fact, the entire first term, i.e. the coefficient times $1/\eps$, is the area of a surface that ends on the Wilson loop (or rather $\eps$ above it) and extends straight to $z=\infty$. This can be seen by computing the area of a cylinder of radius $R$ as this gives
\be
  A_{\mathrm{cylinder}} = 2\pi \int_\eps^\infty \!d\tau\: \frac{R}{\tau^2} = \frac{2\pi R}{\eps} \; .
\ee
It is customary to drop this divergence and call the second term in the expansion the regularized area
\be
  A_{\mathrm{reg}} = - 4\pi \sqrt{a} \Bigsbrk{ \EllipticE(m) - (1-m)\EllipticK(m) } \; .
\ee
Hence, we can say that the area of the circular Wilson loop correlator has been regularized by subtracting two cylinders. We point out, however, that the latter surface is not a solution of the string equation of motion. The solution that comes closest to a cylinder is a straight line in a target space with compact dimension.

Another surface, which ends on the same two circles and \emph{is} a solution to the string equations of motion is that of two disconnected hemispheres. The area of a hemisphere of radius $R$ that is cut off at $z=\eps$ is given by
\be
  A_{\mathrm{hemisphere}} = 2\pi \int_\eps^R \!d\tau\: \frac{R}{\tau^2} = \frac{2\pi R}{\eps} - 2\pi \; .
\ee
Two such surfaces of radius $r_{\mathrm{min}}$ and $r_{\mathrm{max}}$, respectively, have the same divergent contribution to the area as the connected surface or the two cylinders, and as, in fact, any surface that ends orthogonally on the same boundary \cite{Drukker:1999zq,Polyakov:2000ti}. The hemispheres can thus be used to regularize our surface just as well as the two cylinders. The only difference would be that the regularized area is rendered larger by $4\pi$. In passing, we also note that the area of a hemisphere can be regularized by a cylinder of the same radius yielding $-2\pi$.

\begin{figure}%
\begin{center}
\includegraphics[width=80mm]{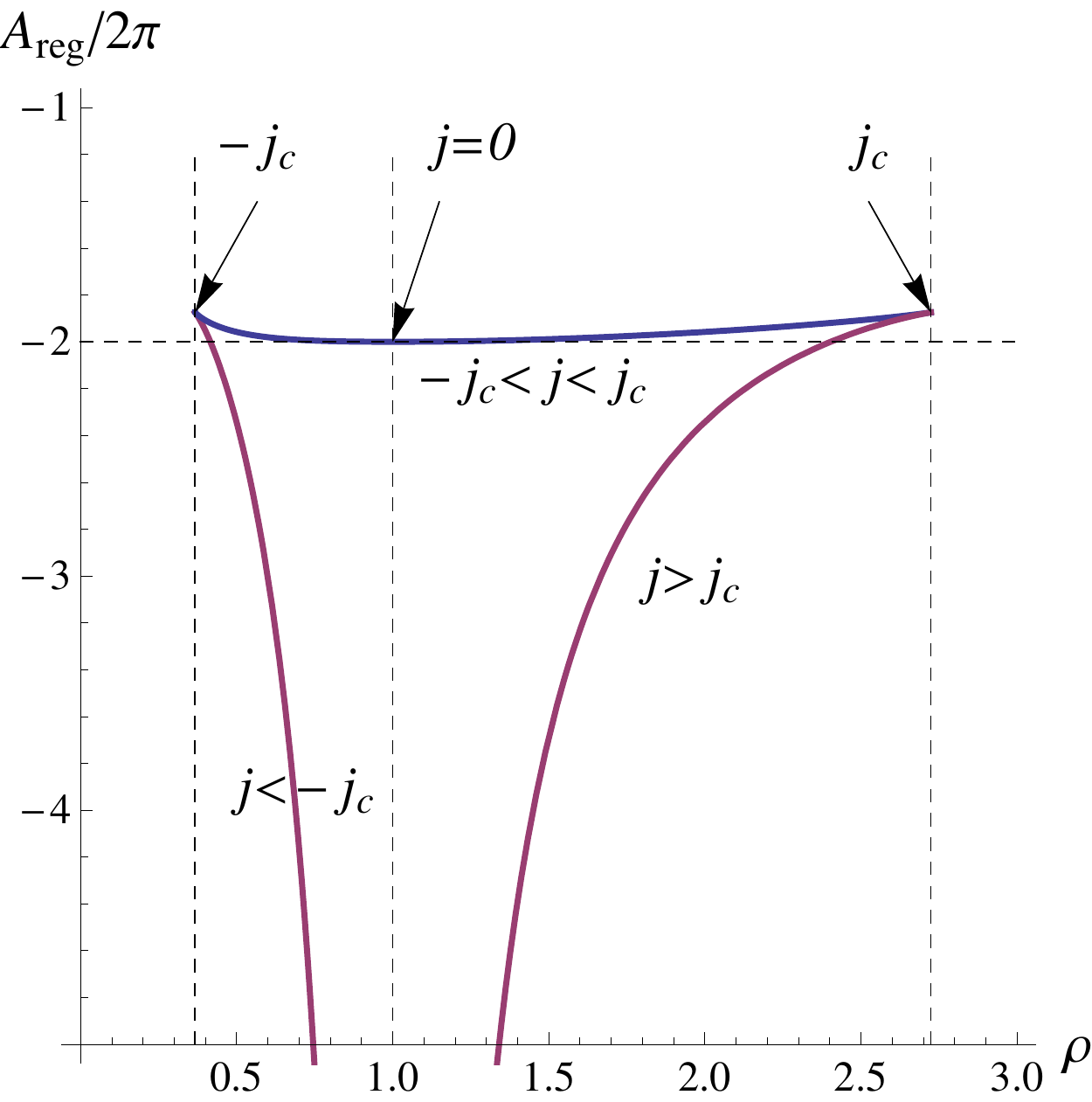}
\caption{\textbf{Regularized area.} The colors of the curves corresponds to the colors of the solutions in \protect\figref{fig:WilsonCorr}. The outer vertical dashed lines mark the range of possible $\rho$ values. The horizontal dashed line is the regularized area of two hemispheres that are attached to each Wilson loop separately.}%
\label{fig:RegArea}%
\end{center}
\end{figure}

The regularized area of the solution in $\Hyp_3$ as a function of $j$ is plotted in the $\rho$-$A_{\mathrm{reg}}$-plane in \figref{fig:RegArea}. We note that the two solutions that exist for the same boundary conditions have different areas; the one with $\abs{j}>j_c$ has smaller area and therefore contributes more significantly to the saddle-point approximation of the partition function
\be
  \mathcal{Z}_{\mathrm{saddle}} = e^{-\frac{\sqrt{\lambda}}{2\pi} \, A_{\mathrm{reg}}} \; .
\ee
The solution with larger area does not only have less importance, it is in fact unstable under perturbations \cite{Zarembo:1999bu}. We will encounter this instability explicitly in our fluctuation analysis in \secref{sec:fluctuations}. In this plot, we have also indicated the regularized area of the two hemispheres by a dashed line. For $\rho \gtrsim 2.4034$ or $\rho \lesssim 0.4161$, this solution has an area that is even less than the smaller area of the two connected solution. These points mark the Gross-Ooguri phase transition \cite{Gross:1998gk}.

When we compute the one-loop partition function in \secref{sec:partition-function}, we will again be faced with the question of regularization. In addition to the pole in $\eps$ (IR divergence), there will be a divergence in the sum over fluctuations (UV divergence). The IR divergence can be regularized in a similar way as done here at the classical level.

\subsection{Algebraic curve for surface of revolution in \texorpdfstring{$\Hyp_3\times \Sphere^1$}{H3S1}}

In this section, we determine the algebraic curve that describes the general two-parameter family of solutions found above. In principle, this involves the calculation of a path-ordered exponential along a non-contractible loop on the world-sheet. Such a computation can hardly ever be carried out explicitly. However, the special, ``factorized'' form of the solution allows us to employ the techniques put forward in \cite{Dekel:2013dy}. The calculation then boils down to fixing five constants in terms of the parameters $j_1$ and $j_2$.

We start by writing the general form of a surface of revolution, \eqref{eqn:surf-of-revolution}, in the $\grSL(2)$ group representation of $\Hyp_3$, namely
\be
  Y(\tau,\sigma) = \frac{1}{z} \matr{cc}{x_1^2 + x_2^2 + z^2 & x_1 + \imag x_2 \\ x_1 - \imag x_2 & 1}
                 = \frac{1}{z} \matr{cc}{r^2+z^2 & r e^{\imag\sigma} \\ r e^{-\imag\sigma} & 1} \; .
\ee
Note that $Y$ depends non-trivially on both $\tau$ and $\sigma$, but it can be written in the factorized form
\be
  Y(\tau,\sigma) = S^{-1}(\sigma) Y(\tau,0) S(\sigma)
  \qquad \text{with} \qquad
  S(\sigma) = e^{-\imag\sigma/2\,\sigma_3} = \matr{cc}{ e^{-\imag\sigma/2} & 0 \\ 0 & e^{\imag\sigma/2} } \; ,
\ee
which implies that the algebraic curve can be easily computed in terms of the flat connection. We define the Maurer-Cartan one-form $j = Y^{-1} d Y$, which satisfies the equation of motion and flatness condition
\be
  \partial_\tau j_\tau + \partial_\sigma j_\sigma = 0
  \comma
  \partial_\sigma j_\tau - \partial_\tau j_\sigma - \comm{j_\tau}{j_\sigma} = 0 \; ,
\ee
respectively. These two equations are combined into the flatness condition
\be
  \partial_\sigma A_\tau - \partial_\tau A_\sigma - \comm{A_\tau}{A_\sigma} = 0 \; ,
\ee
of the connection one-form $A(x)$ with components\footnote{This is the $\tau$\&$\sigma$-version of $A_z = j_z/(1-x)$, $A_{\bar{z}} = j_{\bar{z}}/(1+x)$ for $z = \tau + \imag \sigma$ and $j_z = (j_\tau - \imag j_\sigma)/2$.}
\be
  A_\tau = \frac{j_\tau - \imag x j_\sigma}{1-x^2}
 \comma
  A_\sigma = \frac{j_\sigma + \imag x j_\tau}{1-x^2}
\ee
for any value of the spectral parameter $x$. Importantly, the connection inherits the factorization property of $Y$ and we have
\be
  A_\alpha(\tau,\sigma) = S^{-1}(\sigma) A_\alpha(\tau,0) S(\sigma) \; .
\ee
In turn, we deduce that also the transfer matrix from $\sigma_0$ to $\sigma_1$ along a fixed-$\tau$ line,
\be
  \Omega(\tau,\sigma_1,\sigma_0,x) = \mathcal{P}\exp\lrbrk{-\int_{\sigma_0}^{\sigma_1} A_\sigma d\sigma} \; ,
\ee
is the similarity transform of a path-ordered exponential
\be
  \Omega(\tau,\sigma_1,\sigma_0,x) = S^{-1}(\sigma_1) \mathcal{P} \exp\lrbrk{- \int_{\sigma_0}^{\sigma_1} L(\tau,x) d\sigma } S(\sigma_0) \; ,
\ee
whose integrand
\be \label{eqn:monodromy-integrand}
  L(\tau,x) = A_\sigma(\tau,\sigma=0,x) - \partial_\sigma S S^{-1}
\ee
is indeed independent of $\sigma$ since $\partial_\sigma S S^{-1}=-\frac{\imag}{2}\sigma_3$. Hence, the integral is trivially computed
\be
  \Omega(\tau,\sigma_1,\sigma_0,x) = S(\sigma_1)^{-1} e^{-(\sigma_1 - \sigma_0) L(\tau,x) } S(\sigma_0) \; .
\ee
The quasi momenta $p_1(x)$ and $p_2(x)$, which in turn define the algebraic curve, are read off as the eigenvalues in the form $e^{\imag p_i(x)}$ of the monodromy matrix
\be
  \Omega(x) = \Omega(\tau,2\pi,0,x) = e^{\imag\pi \sigma_3} e^{-2\pi L(\tau,x)} \; .
\ee
As $L$ is a traceless $2\times2$ matrix, the two eigenvalues are the negatives of each other with one being given by
\be
  p(x) = 2\pi\sqrt{\det L(\tau,x)}+\pi \; .
\ee
Thus, computing the quasi-momenta has been reduced to computing the determinant of the matrix \eqref{eqn:monodromy-integrand}. This is a straight forward exercise. However, from the results of \cite{Dekel:2013dy}, we know that the most general form of this determinant is given by
\be
  \det L(\tau,x) = \frac{1}{4(1-x^2)^2} \sum_{i=0}^{4} c_i x^i \; ,
\ee
where the $c_i$ are some constants. For a surface of revolution, we find that $c_0 = c_4 = 1$ and $c_1=-c_3$, which is a consequence of the vanishing of the off-diagonal components of the energy momentum tensor. The diagonal components are related to $c_2$ through $T_{\tau\tau} \propto c_0-c_2+c_4$. Thus, the solution depends only on the two parameters $c_1$ and $c_2$, which are another way of encoding the information contained in $j_1$ and $j_2$. By an explicit computation of the determinant, we find
\be
  c_1 = -2(j_1 + j_2)
  \comma
  c_2 = 2(2 j_1 j_2-1),
\ee
such that
\be
  \det L(x,\tau) = \frac{(1-2 j_1 x-x^2)(1-2 j_2 x-x^2)}{4(1-x^2)^2}
\ee
or
\be \label{eqn:algebraic-curve}
  p(x) = \frac{\pi}{1-x^2}\sqrt{(1-2 j_1 x-x^2)(1-2 j_2 x-x^2)} + \pi \; .
\ee
As defined in \cite{Kazakov:2004qf}, the algebraic curve can be read off from the differential of the quasi-momentum
\be \label{eqn:algebraic-curve2}
 p'(x) =
 \frac{\pi  \left(1+x^2\right) \left((j_1+j_2) \left(x^2-1\right)+4 j_1 j_2 x\right)}{\left(x^2-1\right)^2 \sqrt{\left(1-2 j_1 x-x^2\right) \left(1-2 j_2 x-x^2\right)}}
 = \frac{\sum_{k = 1}^{5}a_k x^{k-1}}{(x^2-1)^2 \sqrt{y^2(x)}},
\ee
giving the curve equation
\be
y^2 = \left(1-2 j_1 x-x^2\right) \left(1-2 j_2 x-x^2\right).
\ee
This expression defines the two-cut algebraic curve for the general minimal surface of revolution in $\Hyp_3\times\Sphere^1$. It unifies various special cases in one formula which we can identify because we know the explicit string solution and the meaning of the parameters $j_1$ and $j_2$. The curve associated to the correlator of two circular Wilson loops without any dependence on the sphere is obtained by setting one of the parameters to zero while keeping the other non-zero. If, however, also the second parameter is set to zero, then we get the algebraic curve for the circular Wilson loop given in \cite{Janik:2012ws,Dekel:2013dy}. The algebraic curve associated to the quark anti-quark potential is contained in \eqref{eqn:algebraic-curve} as the limit $j_1=0$ and $j_2 \rightarrow \infty$ (see \secref{sec:qqbar}) and matches the results of \cite{Janik:2012ws}. The solutions for $j_1=j_2=j$ connect a circle with an isolated point, i.e.\ they are the holographic duals of correlators of a circular Wilson loop and a local operator with charge $j$, viz.\ $\tr Z^j$ as given in \cite{Janik:2012ws}. We would like to point out that in this case the quasi-momentum simply becomes $\frac{2\pi j x}{1-x^2}$, which is exactly the BMN quasi-momentum of the operator. As discussed in \cite{Janik:2012ws}, this is expected for any correlation function involving $\tr Z^j$. Here, we could demonstrate explicitly how the nontrivial information of the curve is lost by introducing the quasi-momenta. Finally, we stress that the curve underlying \eqref{eqn:algebraic-curve} above does not carry any information regarding the boundary conditions in the $\tau$ direction.

\section{Partition function}
\label{sec:partition-function}


In this section, we compute the partition function for a minimal surface of revolution embedded in $\Hyp_3 \subset \AdS_5$ which corresponds to circular Wilson loops correlator, following \cite{Drukker:2000ep}. First, we analyze the fluctuations around the classical solution using the Nambu-Goto (NG) action for the bosons and the Polyakov action for the fermions. After having obtained an expression for the partition function in terms of determinants of differential operators, we compute the determinants using the Gel'fand-Yaglom method and regularize the partition function to get a finite result.

\subsection{Constructing the partition function}

We construct the partition function by studying fluctuations around the classical solution given in \secref{sec:classical}.
In the bosonic sector, we use the NG action because it is not clear how to gauge fix the Polyakov action such that the contribution from the longitudinal modes and the ghosts are explicitly seen to cancel. This is similar to the $q\bar q$-potential case studied in \cite{Drukker:2000ep}. After finding the bosonic contribution, we move on to the fermionic sector, where we analyze it using the Polyakov action. Finally we put the bosonic and fermionic contribution together to get an expression for the partition function.

\subsubsection{The Nambu-Goto action}

Starting with the NG action we introduce fluctuations on top on the classical solution. We gauge fix the fluctuations such that there are no fluctuations along the $\phi=\sigma$ direction and that the fluctuations in the $z$-$r$-plane are normal to the surface. We denote these fluctuations by $\xi_R$. The remaining two fluctuations inside $\AdS_5$ and outside of $\Hyp_3$ are denoted by $\xi^v$ with $v=2,3$. After rescaling $\xi_R \rightarrow z \zeta_R$ and $\xi^v\rightarrow z\zeta^v$, we arrive at
\be \label{eqn:action-2B}
\Action_{2B}=\frac{1}{2}\int d^2\sigma \sqrt{g}\left(g^{ij}\partial_i\zeta^v\partial_j\zeta^v+2\zeta^v\zeta^v
+g^{ij}\partial_i \zeta_R\partial_j \zeta_R+(\mathcal{R}+4) \zeta_R \zeta_R
+g^{ij}\partial_i\xi^q\partial_j\xi^q\right),
\ee
where
\be \label{eqn:scalar-curvature}
\mathcal{R}=\frac{2}{r^4} \Bigsbrk{z^2 (\dot{r}^2 - r\ddot{r}) - r^2 (\dot{z}^2 - z\ddot{z})}
=-2-\frac{j^2}{2}\left(\frac{z}{r}\right)^4,
\ee
is the world-sheet Ricci scalar and
\be
g_{ij}=\frac{r^2}{z^2}\left(
                      \begin{array}{cc}
                        1 & 0 \\
                        0 & 1 \\
                      \end{array}
                    \right),\quad
\sqrt{g}=\frac{r^2}{z^2},\quad
\sqrt{g}g^{ij}=\left(
                      \begin{array}{cc}
                        1 & 0 \\
                        0 & 1 \\
                      \end{array}
                    \right),
\ee
upon using the Virasoro constraints \eqref{eqn:generalized-eom-rz}. Notice that this action has the same form as for fluctuations around the classical solution for parallel lines and circular Wilson loop (with the appropriate curvature) \cite{Drukker:2000ep}. We also notice that the natural norms of the fluctuations, before and after rescaling, are
\be \label{eqn:natural-norm-fluctuations}
\|\xi\|^2 = \int d^2\sigma \sqrt{g}(z^{-2}(\xi_R\xi_R + \xi^v\xi^v)+\xi^q\xi^q)
          = \int d^2\sigma \sqrt{g}( \zeta_R \zeta_R + \zeta^v\zeta^v+\xi^q\xi^q) \; ,
\ee
respectively.

\subsubsection{Fermionic sector}

Next, we use the quadratic part of the fermionic Lagrangian which is given by \cite{Drukker:2000ep}
\be\label{eq:L2F}
L_{2F}=-\imag(\sqrt{g}g^{ij}\delta^{IJ}-\epsilon^{ij} s^{IJ})\bar\theta^I\rho_i D_j \theta^J.
\ee
Using the convention $x^\mu=(r,\phi,x_1,x_2,z)$ and $\bar{x}^\mu$ to denote the classical solution, the $\rho$ matrices are given by
$\rho_i
=\Gamma_a E^a_\mu\partial_i \bar x^\mu
=\Gamma_a \eta^a_i$, where the vielbein of the AdS space is
\be
E^a_\mu=\mathrm{diag}\left(\frac{1}{z},\frac{r}{z},\frac{1}{z},\frac{1}{z},\frac{1}{z}\right) \; .
\ee
These satisfy $G_{\mu \nu}=\eta_{ab}E^a_\mu E^b_\nu$ with $\eta_{ab}=\mathrm{diag}\left(1,1,1,1,1\right)$. The projection of the vielbein onto the world-sheet, $\eta_i^a=\partial_i \bar x^\mu E^a_\mu$, and the mass matrix are given by
\be
\eta_\tau^a=\frac{1}{z}(\dot r,0,0,0,\dot z) \; , \quad
\eta_\sigma^a=\frac{1}{z}(0,r,0,0,0) \; ,\quad
X_{ab}=2 \delta_{ab}-g^{ij}\eta^a_i\eta^b_j \; .
\ee
Note that the eigenvalues of $X_{ab}$ are $(1,1,2,2,2)$ upon using the Virasoro constraints. Thus, the $\rho$ matrices take the explicit form
\be
\rho_\tau   = \frac{1}{z}\left(\dot r\Gamma_0+\dot z\Gamma_4\right) \; , \quad
\rho_\sigma = \frac{1}{z} \, r \Gamma_1 \; .
\ee
Next, the covariant derivatives which appear in \eqref{eq:L2F} are defined by $D_i \theta^I=(\delta^{IJ}\mathcal{D}_i-\frac{1}{2}\imag\epsilon^{IJ}\tilde\rho_i)\theta^J$, where $\mathcal{D}_i=\partial_i+\frac{1}{4}\partial_i \bar x^\mu\Omega_\mu^{a b}\Gamma_{a b}$ (in our case $\tilde\rho_i=\rho_i$ since the classical solution on the sphere is trivial), so that
\be
\mathcal{D}_\tau=\partial_\tau-\frac{1}{2}\frac{\dot r}{z}\Gamma_{0 4},\quad
\mathcal{D}_\sigma=\partial_\sigma+\frac{1}{2}(\Gamma_{1 0}-\frac{r}{z}\Gamma_{1 4}),
\ee
and
\begin{align}
D_0\theta^I&=\left(\delta^{IJ}(\partial_\tau-\frac{1}{2}\frac{\dot r}{z}\Gamma_{0 4})-\frac{\imag}{2}\epsilon^{IJ}\frac{1}{z}(\dot r\Gamma_0+\dot z\Gamma_4)\right)\theta^J,\nonumber\\
D_1\theta^I&=\left(\delta^{IJ}(\partial_\sigma+\frac{1}{2}(\Gamma_{1 0}-\frac{r}{z}\Gamma_{1 4}))-\frac{\imag}{2}\epsilon^{IJ}\frac{r}{z}\Gamma_1\right)\theta^J.
\end{align}
Notice also the we use the Minkowski version of the induced metric, so that
\be
e^0_\tau=e^1_\sigma=\frac{r}{z},\quad g_{ij}=\frac{r^2}{z^2}\mathrm{diag}(1,-1).
\ee

We would like to further simplify the quadratic fermionic action. We note that the $\rho$ matrices are related to the $\Gamma$ matrices by a local rotation
\be
\rho_\tau   = \frac{r}{z}(\Gamma_0 \cos\chi + \Gamma_4 \sin\chi)
            = e^\alpha_\tau S \Gamma_\alpha S^{-1} \; , \quad
\rho_\sigma = e^\alpha_\sigma S \Gamma_\alpha S^{-1} \; ,
\ee
with $S=\exp\left(-\frac{\chi}{2}\Gamma_{04}\right)$, $\sin\chi = \frac{\dot{z}}{r}$ and $\cos\chi = \frac{\dot{r}}{r}$.
Thus we have
\begin{align}
L_{2F}
=-\imag(\sqrt{g}g^{ij}\delta^{IJ}-\epsilon^{ij} s^{IJ})
(\bar\Psi^I\tau_i \hat\nabla_j\Psi^J
-\frac{i}{2}\epsilon^{JK}\bar\Psi^I\tau_i \tau_j \Psi^K) \; ,
\end{align}
where
\be
\Psi^I \eq S^{-1}\theta^I \; , \quad
\tau_i=e^\alpha_i\Gamma_\alpha = \frac{r}{z}(\Gamma_0,\Gamma_1) \; , \quad
\mathcal{D}_i=S \hat\nabla_i S^{-1} \; , \nln
\hat\nabla_\tau \eq \partial_\tau
+\frac{j}{4}\frac{z}{r}\Gamma_{04} \; , \quad
\hat\nabla_\sigma = \partial_\sigma
-\frac{1}{2}\left(\frac{\dot r}{r}-\frac{\dot z}{z}\right)\Gamma_{01}
-\frac{j}{4}\frac{z}{r}\Gamma_{14} \; .
\ee
We can arrive at $\hat\nabla_\tau=\partial_\tau$ if we add $\frac{j}{4}\frac{z}{r}\Gamma_{14}$ to the definition of $\hat\nabla_1$ (we assume $\Gamma_1^2=-\Gamma_0^2=-\Gamma_4^2$), that is
\be
\hat\nabla_\tau=\partial_\tau,\quad
\hat\nabla_\sigma
=\partial_\sigma
-\frac{1}{2}\left(\frac{\dot r}{r}-\frac{\dot z}{z}\right)\Gamma_{01}.
\ee
We gauge fix kappa symmetry such that $\theta^1=\theta^2$ so that $\Psi^1=\Psi^2$ and the Lagrangian simplifies to
\be
L_{2F}=-2\imag\sqrt{g}(\bar\Psi\tau^i\hat\nabla_i\Psi+\imag\hat\Psi\tau_3\Psi) \; ,
\ee
where $\tau_3 \equiv  \Gamma_0\Gamma_1$ with $\tau_3^2=1$.

From this Lagrangian, we can read off the Dirac operator $D_F = \imag\tau^i\hat\nabla_i-\tau_3$. Later, when analyzing the determinant of the Dirac operators, we will need its square which is given by
\begin{align}
D_F^2&
=-\left(\frac{z}{r}\Gamma^0 \hat\nabla_\tau+\frac{z}{r}\Gamma^1 \hat\nabla_\sigma\right)^2+1
=
-\left(\frac{z}{r}\right)^2(\hat\nabla_\tau^2-\hat\nabla_\sigma^2)
+\frac{1}{4}\left(2-\frac{j^2}{2}\frac{z^4}{r^4}\right)\nonumber\\
&=-\hat\nabla^2+\frac{1}{4}\mathcal{R}+1 \; ,
\end{align}
where $\hat\nabla^2=\frac{1}{\sqrt{g}}\hat\nabla_j(\sqrt{g}g^{ij}\hat\nabla_i)=\left(\frac{z}{r}\right)^2(\hat\nabla_\tau^2-\hat\nabla_\sigma^2)$.

\subsubsection{The partition function}

In the last two subsections, we constructed the bosonic and fermionic parts of the partition function, respectively.
Putting the bosonic and fermionic contributions together yields
\be
\mathcal{Z} = \frac{\det^{8}(-\imag\tau^\alpha \hat\nabla_\alpha + \tau_3)}
{
\det^{5/2}(-\nabla^2)
\det^{1/2}(-\nabla^2+\mathcal{R}+4)
\det(-\nabla^2+2)
},
\ee
with the different Laplacians and the Ricci scalar given in the previous section. In order to compute the partition function we need the explicit form of the operators inside the determinants which are
\begin{subequations}
\label{eq:operators}
\be
\label{eq:operator-Om}
-\nabla^2 + m^2
         \eq c^2 \lrbrk{ - \partial_\tau^2 - \partial_\sigma^2 + \frac{m^2}{c^2} }
         \equiv c^2 \mathcal{O}_m \; , \\
\label{eq:operator-OR}
-\nabla^2 + \mathcal{R} + 4
         \eq c^2 \left( - \partial_\tau^2 - \partial_\sigma^2 + \frac{2}{c^2} - \frac{j^2}{2} c^2 \right)
         \equiv c^2 \mathcal{O}_R \; , \\
\label{eq:operator-Opsi}
-\imag\tau^\alpha \hat\nabla_\alpha + \tau_3
         \eq c \left( - \imag \Gamma_0\left(\partial_\tau-\frac{\dot c}{2 c}\right) + \imag\Gamma_1\partial_\sigma + \frac{1}{c}\Gamma_{01}\right)
         \equiv c \, \tilde{\mathcal{O}}_\psi \; ,
\ee
\end{subequations}
where we defined
\be\label{eq:def-of-c}
c\equiv \frac{z}{r}=\frac{1}{\sqrt{a}}\JacobiSD\left(\sqrt{a}\tau|m\right).
\ee
On the right hand sides in \eqref{eq:operators}, we defined \emph{rescaled} operators $\mathcal{O}$. We would like to get rid of the $c$ factors in front of the rescaled operators as explained in \appref{app:GY}. However, in general the factor from the fermionic operator does not cancel the factors of the bosonic operators as shown in \eqref{eq:noncancelation-of-prefactors}. Therefore, we have to look at the square of the fermionic operator and pull out an overall $c^2$ factor which will then cancel the contribution from the bosonic operators,
\be
\bigbrk{ c \, \tilde{\mathcal{O}}_\psi}^2 =
\bigbrk{ -\imag\tau^\alpha \hat\nabla_\alpha + \tau_3 }^2
         \eq c^2 \biggbrk{-\partial_\tau^2 -\partial_\sigma^2 +\imag\frac{\dot c}{c}\Gamma_{01} \partial_\sigma +\frac{3}{4}\left(\frac{\dot c}{c}\right)^2 -\frac{1}{2} }
         \equiv c^2 {\mathcal{O}}^2_\psi \; .
\ee
Diagonalizing ${\mathcal{O}}_\psi^2$ such that
\be
  \det{\mathcal{O}}_\psi^2=\det{\mathcal{O}}_+\det{\mathcal{O}}_-
\ee
with
\be
\mathcal{O}_\pm =
-\partial_\tau^2
-\partial_\sigma^2
\pm \imag\frac{\dot c}{c}\partial_\sigma
+\frac{3}{4}\left(\frac{\dot c}{c}\right)^2
-\frac{1}{2} \; ,
\ee
we arrive at the following expressions for the partition function
\be\label{eqn:partition_functionCWL}
\mathcal{Z} =\frac{\det^{2}{\mathcal{O}}_+ \: \det^{2}{\mathcal{O}}_-}
                   { \det^{5/2}\mathcal{O}_0 \:  \det\mathcal{O}_2 \: \det^{1/2}\mathcal{O}_R} \; .
\ee

Unfortunately, we do not know the analytic solution to the homogenous problem $\mathcal{O}_\pm u(\tau)=0$. Nevertheless, it turns out that $\det \mathcal{O}^2_\psi$ and $\det \tilde{\mathcal{O}}^2_\psi$ are related in a simple way as will be explained shortly, and we do know how to compute $\det \tilde{\mathcal{O}}^2_\psi$ analytically\footnote{Notice that $\tilde{\mathcal{O}}_\psi$ is the operator used in \cite{Chu:2009qt,Forini:2010ek,Drukker:2011za} for computing the $q\bar q$-potential partition function, and its generalization. On the other hand in \cite{Kruczenski:2008zk,Kristjansen:2012nz} where the circular Wilson loop was analyzed $\mathcal{O}_\psi$ was used. These operators are related as we shall later see, but give different results.}. In order to simplify $\tilde{\mathcal{O}}_\psi$, we notice that the square of the fermionic operator can be diagonalized and we define the diagonal components, in a similar way to the analysis in \cite{Chu:2009qt}, as
\be
 \tilde{ \mathcal{O}}_\psi^2 = c^{1/2} M \matr{cc}{\tilde{\mathcal{O}}_+ & 0 \\ 0 & \tilde{\mathcal{O}}_-} M^{-1} c^{-1/2} \; ,
\ee
with
\be
\tilde{\mathcal{O}}_\pm = -\partial_\tau^2-\partial_\sigma^2 + \frac{1 \pm \dot{c}}{c^2},
\ee
and a constant matrix $M$ that satisfies $M \Gamma_{01} M^{-1} = \diag(1,-1)$. Thus, as explained in \appref{app:GY}, eq. \eqref{eq:detConj}, the determinant is given by
\be\label{eq:factor}
\det \tilde{\mathcal{O}}^2_{\psi}
\eq
\left(\frac{\sqrt{z/r}|_{\tau=\tau_{\mathrm{max}}}}{\sqrt{z/r}|_{\tau=\tau_{\mathrm{min}}}}\right)^2
\det \tilde{\mathcal{O}}_{+}
\det \tilde{\mathcal{O}}_{-}
=
\left(\frac{r_{\mathrm{min}}}{r_{\mathrm{max}}}\right)
\det \tilde{\mathcal{O}}_{+}
\det \tilde{\mathcal{O}}_{-}.
\ee

We find numerically that the fermionic determinants are related by\footnote{Notice the relation $\det \tilde{\mathcal{O}}^2_\psi=4\frac{r_{\mathrm{min}}}{r_{\mathrm{max}}}\det {\mathcal{O}}^2_\psi$, or $\tilde{\mathcal{Z}}=4^{2\cdot\infty} \bigbrk{ \frac{r_{\mathrm{min}}}{r_{\mathrm{max}}} }^2{\mathcal{Z}}$, which yields a different result for the partition function then the one found in \cite{Chu:2009qt,Forini:2010ek,Drukker:2011za} for the $q\bar q$-potential where $\frac{r_{\mathrm{min}}}{r_{\mathrm{max}}}=1$.
This implies one has to regularize the results for the $q\bar q$-potential in those papers differently.}
\be\label{eq:fermionicOpsDifferensWW}
 \det \mathcal{O}_+ \det \mathcal{O}_-
=
\frac{1}{4}\det \tilde{\mathcal{O}}_+ \det \tilde{\mathcal{O}}_-
 \; .
\ee
Hence, in the following sections, we will work with $\tilde{\mathcal{O}}_\pm$ instead of $\mathcal{O}_\pm$ and then use the relation \eqref{eq:fermionicOpsDifferensWW} in the final answer for the partition function.

\subsection{Evaluating the determinants using Gel'fand Yaglom method}
\label{sec:detsUsingGY}

In this section, we give the analytic solution to \eqref{eqn:partition_functionCWL} in terms of an infinite product, using the Gel'fand Yaglom (GY) method applied to $\mathcal{O}_0$, $\mathcal{O}_2$, $\mathcal{O}_R$, $\tilde{\mathcal{O}}_\pm$. In \appref{app:GY}, we give further details on the method. We start by transforming the operators into Lam\'{e} form, where the analytic solution is well known (see \appref{app:Lame}). Then, we compute the ratio of the determinants using the GY method, and find analytic expression for the partition function.

\begin{figure}%
\begin{center}
\subfloat[$V_2$]{\label{fig:V2}\includegraphics[width=35mm]{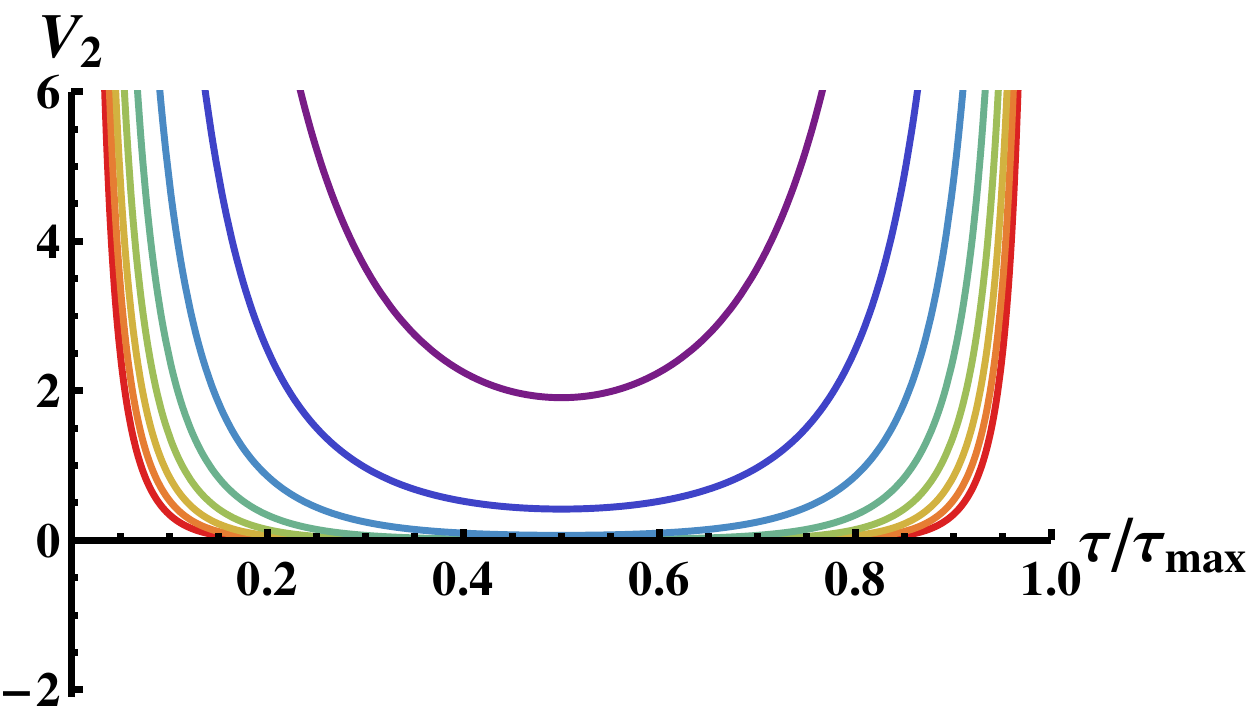}} \hspace{3mm}
\subfloat[$V_R$]{\label{fig:VR}\includegraphics[width=35mm]{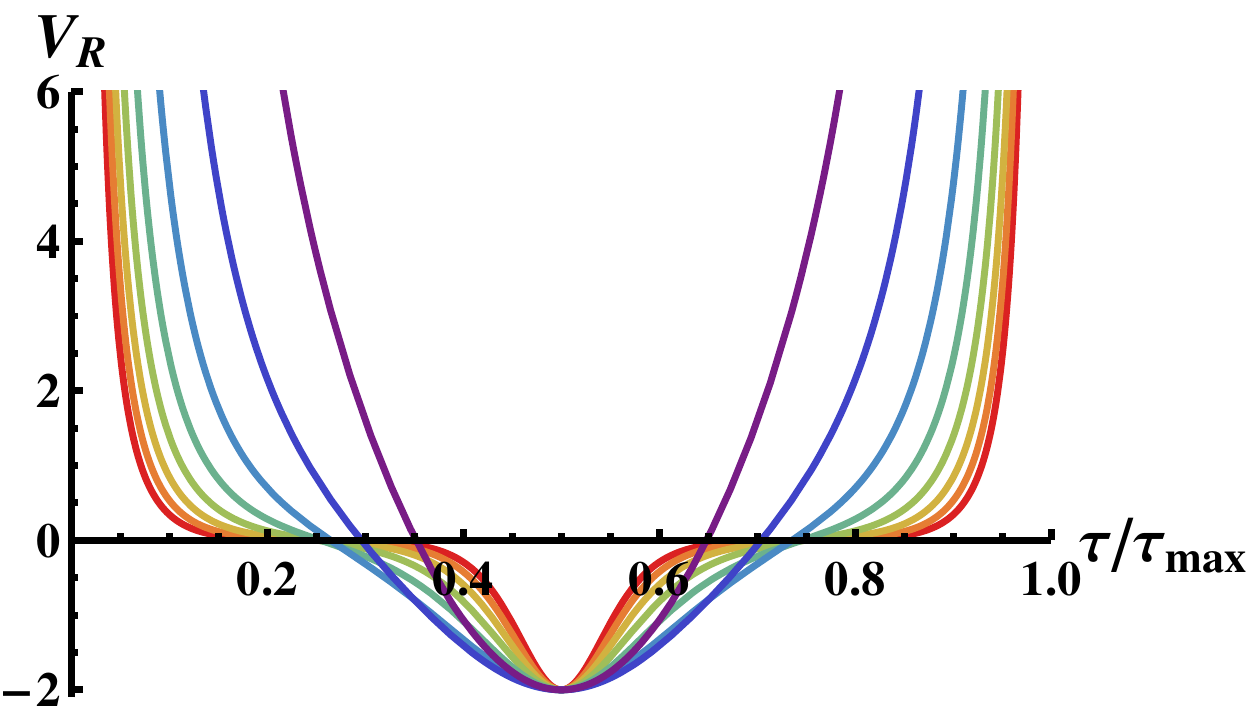}} \hspace{3mm}
\subfloat[$ V_+$]{\label{fig:VP}\includegraphics[width=35mm]{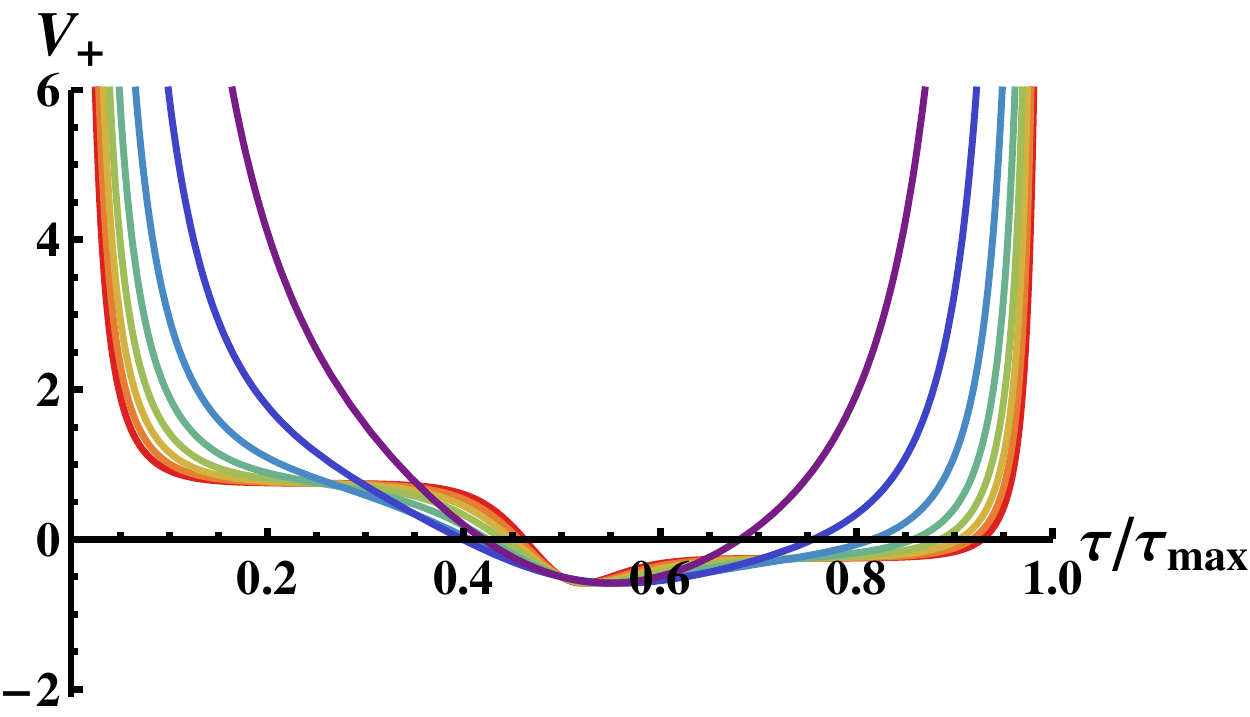}} \hspace{3mm}
\subfloat[$ V_-$]{\label{fig:VM}\includegraphics[width=35mm]{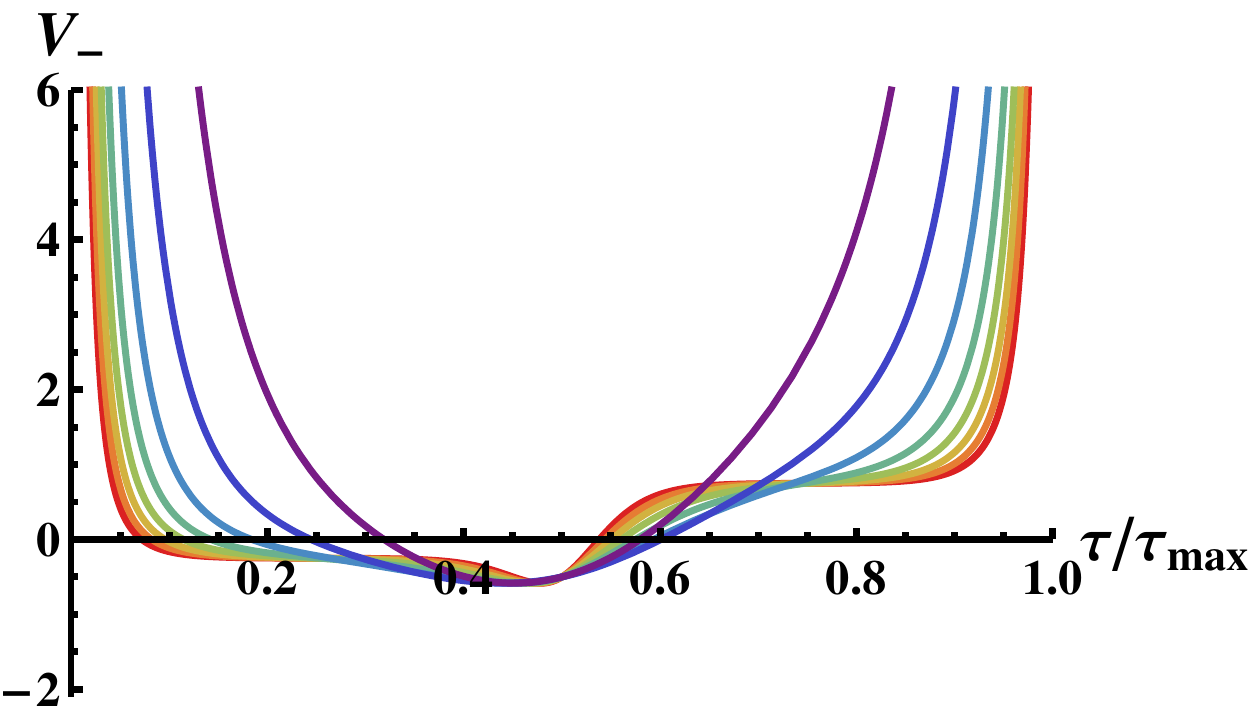}}
\caption{\textbf{Potential for rescaled fluctuations.} The potentials $V_i$ for the rescaled fluctuations $\xi = \frac{1}{z} \zeta$ are plotted as a function of $\tau/\tau_{\mathrm{max}}$ with $j$ as a parameter ranging from $0.002$ (red) to $2.7$ (violett). The formulas for these potentials can be read off from the corresponding operators via $\mathcal{O}_i = - \partial_\tau^2  - \partial_\sigma^2 + V_i(\tau).$ For $V_\pm$ we take $s=\frac{1}{2}$. As can be seen, $V_+$ and $V_-$ are related by reflection.}%
\label{fig:Potentials}%
\end{center}
\end{figure}

\begin{figure}%
\begin{center}
\subfloat[$\tilde V_+$]{\label{fig:tildeVP}\includegraphics[width=35mm]{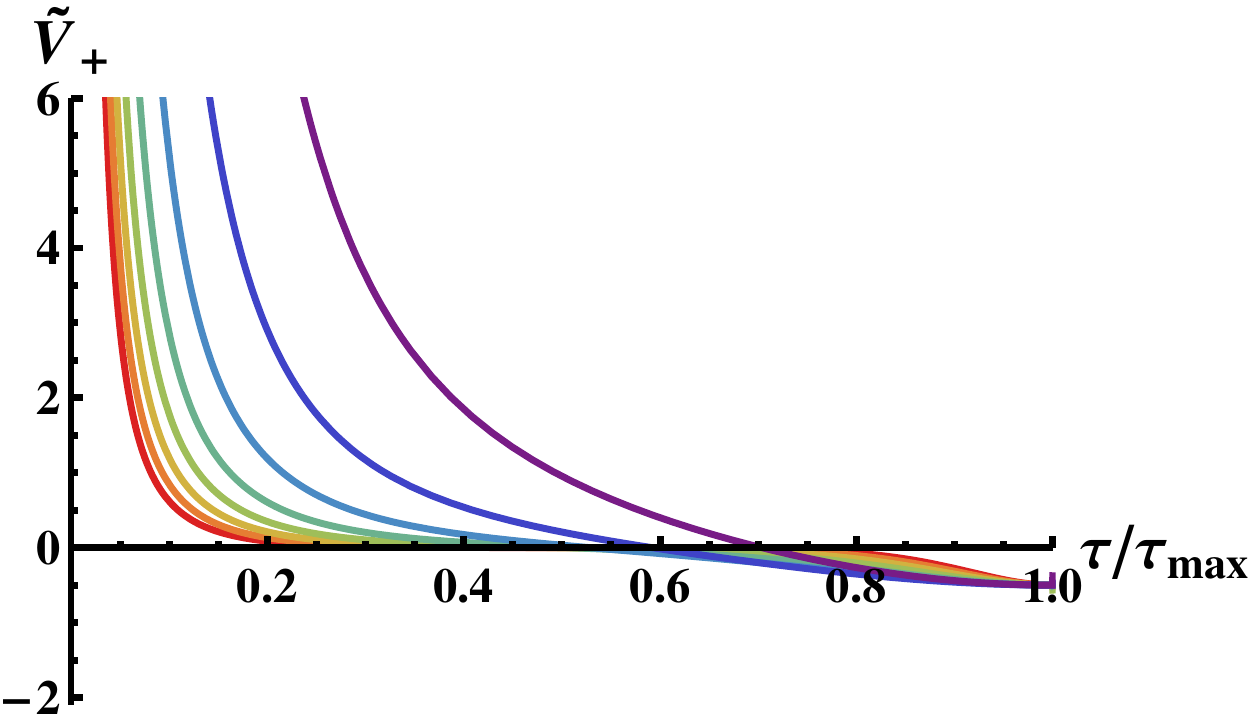}} \hspace{3mm}
\subfloat[$\tilde V_-$]{\label{fig:tildeVM}\includegraphics[width=35mm]{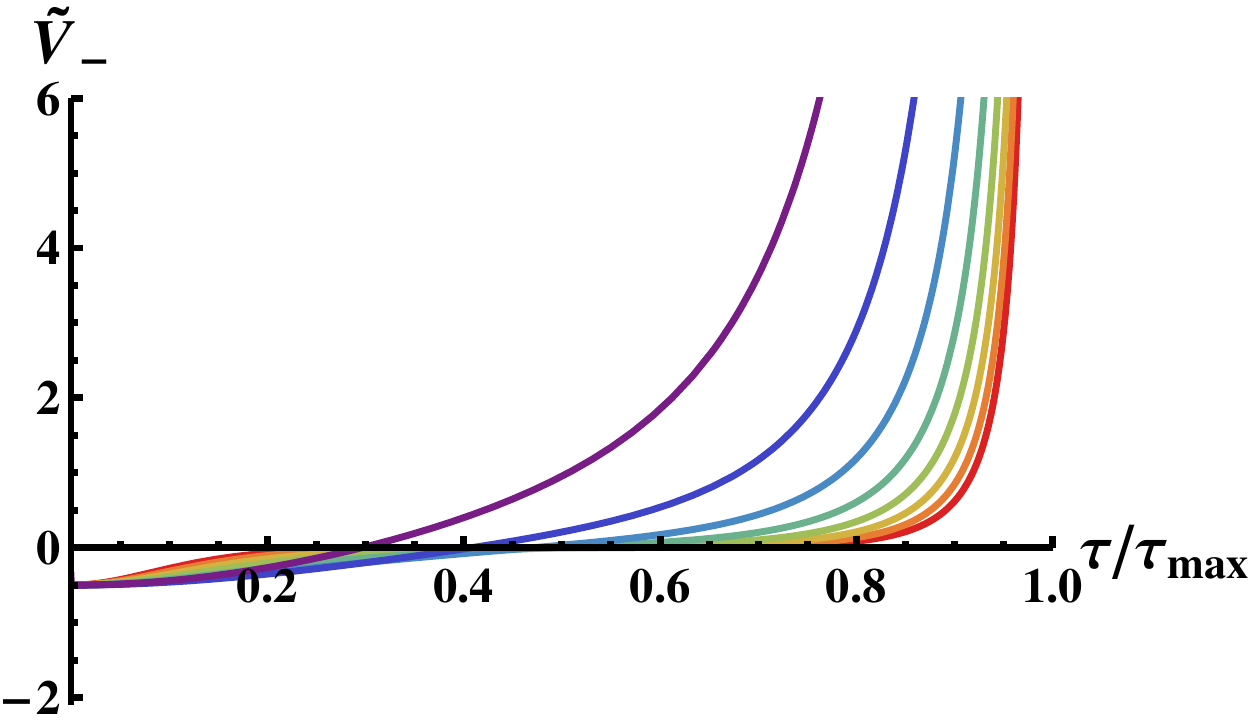}}
\caption{\textbf{Potential for the auxiliary rescaled fluctuations.} The potentials $\tilde V_\p,$  are plotted as a function of $\tau/\tau_{\mathrm{max}}$ with $j$ as a parameter ranging from $0.002$ (red) to $2.7$ (violett). The formulas for these potentials can be read off from the corresponding operators via $\tilde{\mathcal{O}}_\pm = - \partial_\tau^2  - \partial_\sigma^2 + \tilde V_\pm(\tau).$}%
\label{fig:Potentials2}%
\end{center}
\end{figure}

\paragraph{Operators in Lam\'{e} form.} The explicit form of the operators we analyse in this section is given by
\begin{subequations}
\label{eqn:general-operators}
\be
  \label{eqn:general-op0}
  \mathcal{O}_0 \eq - \partial_\tau^2 - \partial_\sigma^2  \; ,\\
  \label{eqn:general-op2}
	\mathcal{O}_2 \eq  - \partial_\tau^2 - \partial_\sigma^2 + 2a \JacobiDS^2(\sqrt{a}\tau|m) \; , \\
  \label{eqn:general-opR}
	\mathcal{O}_R \eq  - \partial_\tau^2 - \partial_\sigma^2 + 2a \JacobiDS^2(\sqrt{a}\tau|m) - 2a m (1-m) \JacobiSD^2(\sqrt{a}\tau|m)\; , \\
  \label{eqn:general-opP}
	\tilde{\mathcal{O}}_\pm  \eq - \partial_\tau^2 - \partial_\sigma^2 + a \JacobiDS^2(\sqrt{a}\tau|m) \pm a \JacobiCS(\sqrt{a}\tau|m) \JacobiNS(\sqrt{a}\tau|m)\; ,
\ee
\end{subequations}
where we took all the operators to have Euclidean world-sheet signature. We remind the reader that
\be \label{eqn:r-over-z}
  \frac{r}{z} = \sqrt{a} \JacobiDS(\sqrt{a}\tau|m)
  \comma
  a = \sqrt{1+j^2}
  \comma
  \Half \le m = \Half + \frac{1}{2a} \le 1,
\ee
and the coordinate range is $0 \le \tau \le \frac{2}{\sqrt{a}} \EllipticK(m)$. In \figref{fig:Potentials} we plot the potentials $V_i(\tau)$ for different values of $j$. As one can see, the fermionic potentials are related to each other by reflection (so we expect the same eigenvalues). In \figref{fig:Potentials2} we plot the potentials $\tilde V_\pm(\tau)$ of the auxiliary operators $\tilde{\mathcal{O}}_\pm$. Notice that, $V_R(\tau)=4 \tilde V_+(2\tau)=4 \tilde V_-(\tau_{\mathrm{max}}-2\tau)$, so the eigenvalues of $\tilde{\mathcal{O}}_\pm$ should be related to the $\mathcal{O}_R$ eigenvalues by $\lambda^{(+)}_{n}=\frac{1}{4}\lambda^{(R)}_{2n}$ where $n=1,2,3,..$ for radially symmetric excitations (otherwise they are still related but a bit differently as will be explained later). Also, notice the difference between $V_\pm$ and $\tilde V_\pm$, where in the first case the potential diverges at both boundaries, while in the other case the potential diverges only at one of the boundaries.


We would like to solve the homogeneous problem for these operators, so we put these operators in the form of a Lam\'{e} operator (see \appref{app:Lame})
\begin{subequations}
\label{eqn:OpsList_in_Lame_form}
\be
  \mathcal{O}_0 \eq \sqrt{1+j^2} \Bigbrk{ -\partial_u^2 + \frac{\ell^2}{\sqrt{1+j^2}} } \; , \\
  \mathcal{O}_2 \eq \sqrt{1+j^2} \Bigbrk{ -\partial_u^2 +\frac{\ell^2-1-\sqrt{1+j^2}}{\sqrt{1+j^2}} + 2\, m \JacobiSN^2(u + \imag \EllipticK'(m)|m) } \; , \\
  \mathcal{O}_R \eq \left(1+\imag j\right) \Bigbrk{ -\partial_v^2 +\frac{\ell^2-2}{1+\imag j} +2\mu\JacobiSN^2\left(v +\imag \EllipticK'(\mu)\big|\mu\right) } \; , \\
  \tilde{\mathcal{O}}_+ \eq \frac{\left(1+\imag j\right)}{4} \Bigbrk{ -\partial_w^2 +\frac{4s^2-2}{1+\imag j} +2\mu\JacobiSN^2\left(w +\imag \EllipticK'(\mu)\big|\mu\right) }  \; , \\
  \tilde{\mathcal{O}}_- \eq \frac{\left(1+\imag j\right)}{4} \Bigbrk{ -\partial_w^2 +\frac{4s^2-2}{1+\imag j} +2\mu\JacobiSN^2\left(w \big|\mu\right) }  \; ,
\ee
\end{subequations}
where we introduced
\be
\mu=\frac{1-\imag j}{1+\imag j} ,\quad
u=(1+j^2)^\frac{1}{4} \, \tau ,\quad
v=\sqrt{1+\imag j} \, \tau ,\quad
w=\frac{1}{2}\sqrt{1+\imag j} \, \tau ,
\ee
and Fourier transformed the $\sigma$ coordinate such that $\partial_\sigma\to \imag\ell$ for bosons and $\partial_\sigma\to \imag s$ fermions. We label the bosonic and fermionic modes differently since in principle they can take different values. The bosons take integer values, while the fermions may take integer or half-integer values. In \cite{Mikhaylov:2010ib} this issue is studied in detail, and it is claimed that for non-simply connected manifolds, such as our world-sheet with the topology of an annulus, one is free to \emph{choose} either integer or half-integer values for the fermionic modes. Here, we present the computation for half-integer fermionic modes. By using the supersymmetric regularization scheme described in \appref{app:Supersymmetric regularization}, an extra factor is created, and we have checked that this extra factor renders the result equal to what one obtains by using integer-moded fermions in the first place.

\paragraph{Determinants.} We apply formula \eqref{eqn:Gelfand-Yaglom}, which follows from the GY method, to the above operators. The expressions are regularized by replacing $L \to L + \frac{\eps}{r_{\mathrm{min}}}$ and $R \to R - \frac{\eps}{r_{\mathrm{max}}}$, where $L$ and $R$ represent the left and right boundaries, respectively. In all terms where it is possible, we send $\eps$ to zero. Recall that \eqref{eqn:Gelfand-Yaglom} gives the determinant up to a normalization by another determinant. Thus, we shell write $\det \mathcal{O}_i\simeq u_i(R)$ where $u_i(R)$ is given in \eqref{eqn:Gelfand-Yaglom}. Eventually, we will normalize all the determinants with the same reference determinant such that this factor will cancel in the expression for the partition function \eqref{eqn:partition_functionCWL}.

The resulting determinants as $\eps\to 0$ are given by
\begin{align}\label{eqn:det_in_GY_general}
  \det \mathcal{O}_0    & \simeq \frac{1}{\ell}\sinh\frac{2\ell \EllipticK(m)}{\sqrt{a}}, \nonumber\\
  \det \mathcal{O}_2    & \simeq -\frac{1}{\eps^2}\frac{2 r_{\mathrm{min}} r_{\mathrm{max}}}{\sqrt{\left(\ell^2-1\right)\bigbrk{4\ell^2\left(\ell^2-1\right)-j^2}}}
  \sinh\left(2Z(\alpha_2(\ell)|m) \EllipticK(m)\right), \nonumber\\
  \det \mathcal{O}_R    & \simeq \frac{1}{\eps^2}\frac{r_{\mathrm{min}} r_{\mathrm{max}}}{\sqrt{\ell ^2 \bigbrk{j^2+\left(\ell^2 - 1\right)^2}}}
  \sinh\left(\frac{\imag \pi \alpha_R(\ell)}{\EllipticK(\mu)}+2Z(\alpha_R(\ell)|\mu)(2\EllipticK(\mu)+\imag \EllipticK'(\mu))\right), \nonumber\\
  \det \tilde{\mathcal{O}}_+    & \simeq \frac{1}{\eps}\frac{4 r_{\mathrm{min}}}{\sqrt{j^2+\left(4 s^2-1\right)^2}}
  \cosh\left(\frac{\imag \pi \alpha_f(s)}{2\EllipticK(\mu)}+Z(\alpha_f(s)|\mu)(2\EllipticK(\mu)+\imag \EllipticK'(\mu))\right), \nonumber\\
  \det \tilde{\mathcal{O}}_-    & \simeq \frac{1}{\eps}\frac{4 r_{\mathrm{max}}}{\sqrt{j^2+\left(4 s^2-1\right)^2}}
  \cosh\left(\frac{\imag \pi \alpha_f(s)}{2\EllipticK(\mu)}+Z(\alpha_f(s)|\mu)(2\EllipticK(\mu)+\imag \EllipticK'(\mu))\right),
\end{align}
with
\be\label{eq:alphaDEF}
\JacobiSN(\alpha_2(\ell)|m)=\sqrt{\frac{1-m+\ell^2/a}{m}},\quad
\JacobiSN(\alpha_R(\ell)|\mu)=\sqrt{\frac{\ell^2}{1-\imag j}},\quad
\JacobiSN(\alpha_f(s)|\mu)=2\sqrt{\frac{s^2}{1-\imag j}},
\ee
where all the function are understood to be functions of $j$. Notice that the overall divergence when all these determinants are multiplied together is $\bigbrk{\frac{1}{\eps^2}}{}^{2-1-1/2}=\frac{1}{\eps}$ as expected.

There are some special cases one should consider separately, namely when $\ell=0,1$. In this case, the general basis solutions to the Lam\'{e} equation, \eqref{eq:LameEqSol}, become linearly dependent (the $\alpha$ parameter vanishes) and two actually independent solutions take a different form (see \appref{app:Lame}). For $\ell=0$ one has
\begin{align}\label{eqn:det_in_GY_0}
  \det \mathcal{O}_0 & \simeq \frac{2\EllipticK(m)}{\sqrt{a}} \; , \nonumber\\
  \det \mathcal{O}_R & \simeq \frac{1}{\eps^2}\frac{4r_{\mathrm{min}} r_{\mathrm{max}}}{a^{3/2}}\left(\EllipticE(m)-\frac{1}{2}\EllipticK(m)\right) \; ,
\end{align}
while $\det \mathcal{O}_2$ is the same as in \eqref{eqn:det_in_GY_general}. For $\ell=1$ one has
\be\label{eqn:det_in_GY_1}
  \det \mathcal{O}_2 \simeq \frac{1}{\eps^2}\frac{8 r_{\mathrm{min}} r_{\mathrm{max}} \sqrt{a}}{j^2}\bigbrk{\EllipticE(m)-(1-m)\EllipticK(m)} \; ,
\ee
while the other determinants take the same form as in \eqref{eqn:det_in_GY_general}. In retrospect, one can see that those special cases can actually be found by carefully taking the limit of the general formula.

We note that $V_\pm$ are related by reflection and diverge at both endpoints of the string, which implies that $\det {\mathcal{O}}_+ = \det {\mathcal{O}}_-$. Thus, using \eqref{eq:fermionicOpsDifferensWW}, we arrive at
\begin{align}
  \det {\mathcal{O}}_\pm & \simeq \frac{1}{\eps}\frac{2 r_{\mathrm{min}}r_{\mathrm{max}}}{\sqrt{j^2+\left(4 s^2-1\right)^2}}
  \cosh\left(\frac{\imag \pi \alpha_f(s)}{2\EllipticK(\mu)}+Z(\alpha_f(s)|\mu)(2\EllipticK(\mu)+\imag \EllipticK'(\mu))\right) \; .
\end{align}

Let us mention that for large Fourier mode numbers $\ell$ and $s$, the determinants take the form
\begin{align}\label{eqn:det_in_GY_general_large_omega}
  \det \mathcal{O}_0    & \simeq \frac{1}{2\ell}\exp\frac{2\ell \EllipticK(m)}{\sqrt{a}} \; , \nonumber\\
  \det \mathcal{O}_2    & \simeq
  \det \mathcal{O}_R     \simeq \frac{1}{\eps^2}\frac{r_{\mathrm{min}} r_{\mathrm{max}}}{2\ell^3}
  \exp\frac{2\ell \EllipticK(m)}{\sqrt{a}} \; , \nonumber\\
  \det \mathcal{O}_+    & \simeq
  \det \mathcal{O}_-     \simeq \frac{1}{\eps}\frac{\sqrt{r_{\mathrm{min}}r_{\mathrm{max}}}}{4s^2}
  \exp\frac{2s \EllipticK(m)}{\sqrt{a}} \; , \nonumber\\
  \det \tilde{\mathcal{O}}_+    & \simeq \frac{1}{\eps}\frac{r_{\mathrm{min}}}{2s^2}
  \exp\frac{2s \EllipticK(m)}{\sqrt{a}} \; , \quad
  \det \tilde{\mathcal{O}}_-    \simeq \frac{1}{\eps}\frac{r_{\mathrm{max}}}{2s^2}
  \exp\frac{2s \EllipticK(m)}{\sqrt{a}} \; .
\end{align}

Putting everything together, we find the partition function is given by
\begin{align}\label{eq:unreg_PF}
\mathcal{Z}
=
\prod_{\ell,s}
\frac{\sqrt{r_{\mathrm{min}} r_{\mathrm{max}} }}{\eps}
\frac{f(\ell,j)}
{g(s,j)^2}
\frac{\cosh^4 x_f}
{
\sinh^{5/2} x_0
\sinh x_2
\sinh^{1/2} x_R
},
\end{align}
where
\begin{align}\label{eq:def_for_PFs_arg}
f(\ell,j)&=\frac{1}{2} \ell^{3}\sqrt{\left(\ell^2-1\right)\left(4\ell^2\left(\ell^2-1\right)-j^2\right)\sqrt{j^2+\left(\ell^2 - 1\right)^2}} \; ,
\nonumber\\
g(s,j)&=\frac{1}{4}\left(j^2+\left(4 s^2-1\right)^2 \right)\; ,
\nonumber\\
x_0(\ell,j)&=\frac{2\ell \EllipticK(m)}{\sqrt{a}} \; ,
\nonumber\\
x_2(\ell,j)&=2Z(\alpha_2(\ell)|m) \EllipticK(m) \; ,
\nonumber\\
x_R(\ell,j)&=\frac{\imag \pi \alpha_R(\ell)}{\EllipticK(\mu)}+2Z(\alpha_R(\ell)|\mu)(2\EllipticK(\mu)+\imag \EllipticK'(\mu)) \; ,
\nonumber\\
x_f(s,j)&=\frac{\imag \pi \alpha_f(s)}{2\EllipticK(\mu)}+Z(\alpha_f(s)|\mu)(2\EllipticK(\mu)+\imag \EllipticK'(\mu)) \; .
\end{align}

At this point we should use the supersymmetric regularization scheme (see \appref{app:Supersymmetric regularization} for details), in order to perform the infinite product over $\ell$ and $s$. The supersymmetric regularization procedure arranges the bosonic and fermionic frequencies in the sum in a symmetric way, see \eqref{eqn:susy-regularization}. However, this arrangement also introduces another contribution to the partition function given by \eqref{eqn:susy-reg-KT}. Using the large $s$ expansion given in \eqref{eqn:det_in_GY_general_large_omega}, the fermionic frequencies are
\be
  \omega^F_s = 2 \ln \lrsbrk{ \frac{r_{\mathrm{min}}r_{\mathrm{max}}}{4^2 \eps^2 s^4} \exp\frac{4s\EllipticK(m)}{\sqrt{a}} } \approx \frac{8s\EllipticK(m)}{\sqrt{a}} \; .
\ee
The terms in \eqref{eqn:susy-reg-KT} evaluate to
\be\label{eq:susyregfactor}
\lim_{\mu\to 0}\left(
-4 \sinh^2\frac{\mu}{4}
\sum_{\ell=1}^{\infty}e^{-\mu \ell} \, \frac{8 \ell \EllipticK(m)}{\sqrt{a}}\right)
= - \frac{2 \EllipticK(m)}{\sqrt{a}} \; .
\ee
Thus, in the supersymmetric regularization the partition function \eqref{eq:unreg_PF} reads
\begin{align}\label{eqn:partition_function_result}
\mathcal{Z}
=
e^{-\frac{2\EllipticK(m)}{\sqrt{a}}}\prod_{\ell\in \mathbb{Z}}
\frac{\sqrt{r_{\mathrm{min}} r_{\mathrm{max}} }}{\eps}
\frac{f(\ell,j)}
{g\left(\ell+\frac{1}{2},j\right)g\left(\ell-\frac{1}{2},j\right)}
\frac{\cosh^2 x_f^+\cosh^2 x_f^-}
{
\sinh^{5/2} x_0
\sinh x_2
\sinh^{1/2} x_R
} \; ,
\end{align}
where $x_f^{\pm}=x_f\left(\ell\pm\frac{1}{2},j\right)$ and all the arguments are defined in \eqref{eq:def_for_PFs_arg}. Let us also define the partition function per Fourier mode as
\be
\mathcal{Z}
\eq
e^{-\frac{2\EllipticK(m)}{\sqrt{a}}}\prod_{\ell\in \mathbb{Z}}
\mathcal{Z}_{\ell}
=
e^{-\frac{2\EllipticK(m)}{\sqrt{a}}}\mathcal{Z}_{0}\mathcal{Z}_{1}^2\prod_{\ell = 2}^{\infty}
\mathcal{Z}_{\ell}^2,
\ee
where the definition of $\mathcal{Z}_{\ell}$ should be obvious by comparing with \eqref{eqn:partition_function_result}. In the second equality we used the fact that $\mathcal{Z}_{\ell}=\mathcal{Z}_{-\ell}$ and separated the $\ell=0,1$ modes from the rest since in these cases we should treat $\mathcal{Z}_{\ell}$ more carefully as was pointed out in \secref{sec:detsUsingGY} (see \eqref{eqn:det_in_GY_0} and \eqref{eqn:det_in_GY_1}). For these Fourier modes, we get
\be\label{eqn:partition_function_result_lis0}
\mathcal{Z}_{0}
\eq
-\frac{\sqrt{r_{\mathrm{min}} r_{\mathrm{max}} }}{\eps}
\frac{
a^2
}
{j^3}
\frac{\cosh^2 x_f^+\cosh^2 x_f^-}
{
\EllipticK^{5/2}(m)
\sinh x_2
\sqrt{2\EllipticE(m)-\EllipticK(m)}
}\Bigg|_{\ell=0},
\ee
and
\be\label{eqn:partition_function_result_lis1}
\mathcal{Z}_{1}
\eq
\frac{\sqrt{r_{\mathrm{min}} r_{\mathrm{max}} }}{\eps}
\frac{2
\left(j^2\right)^{1/4}}
{\sqrt{a}\left(j^2+64\right)}
\frac{\cosh^2 x_f^+\cosh^2 x_f^-}
{\sinh^{5/2} x_0
\left(\EllipticE(m)-(1-m)\EllipticK(m)\right)
\sinh^{1/2} x_R
}\Bigg|_{\ell=1}.
\ee

\subsection{Regularizing the partition function}
\label{sec:REG}
In this section, we regularize the logarithm of the partition function, i.e.\ the effective action
\be
\Gamma = \ln \prod_{\ell}\mathcal{Z}_{\ell}
=\sum_\ell \ln \mathcal{Z}_{\ell}
\equiv
\sum_\ell \Gamma_{\ell} \; .
\ee
In order to perform the regularization, we have to cancel the IR and UV divergences. The IR divergence comes from the $\frac{1}{\epsilon}$ dependence. More precisely, for a circular contour ending on the AdS boundary the IR divergence behaves as $\frac{R}{\epsilon}$, where $R$ is the radius of the circle (as was shown in \secref{sec:classical}). Thus, we refer to the factor $\ln\frac{\sqrt{r_{\mathrm{min}}r_{\mathrm{max}}}}{\epsilon}$ as the IR divergence of $\Gamma_\ell$. The UV divergence stems from the product over Fourier modes. To be more specific, using \eqref{eqn:det_in_GY_general_large_omega}, we see that for large mode number $\ell$, the partition function behaves as
$
Z_\ell \simeq \frac{\sqrt{r_{\mathrm{min}}r_{\mathrm{max}}}}{16 \epsilon\ell} \; .
$

We regulate the IR divergence by subtracting the effective action of a reference solution with the same IR behavior, similar to the treatment at the classical level. Writing the effective action of the reference solution as an expansion in the mode number, this means to subtract some functions from each $\Gamma_\ell$. At the classical level, we subtracted two infinite cylinders, however, the infinite cylinders do not solve the equations of motion, but we prefer to subtract the partition function for the fluctuations around some actual solution. The most natural candidate is the circular Wilson loop solution, ending on the same boundary. As explained in \appref{app:CWL-PF}, the IR behavior of two circular Wilson loops with the same boundary conditions as the two Wilson loops correlator is the same. In principle, one could hope that this subtraction would also cancel the UV divergence since the circular Wilson loop partition function suffers from similar divergences, however, the sub-leading UV behaviors are different and do not cancel.

Nevertheless, we are able to cancel the constant UV divergence at the price of introducing an overall unknown, finite, $j$-independent constant. This means we will calculate the partition function as a function of $j$ for any $j$, up to the same constant which is not important if one is interested in the effective action as a function of $j$, or equivalently as a function of the ratio of the radii $\rho$.

One way to get rid of the UV divergence is to differentiate $\Gamma_\ell$ with respect to $j$, perform the summation, and integrate back. Although this is straight forward in principle, we find it easier to subtract a reference function, e.g.\ $\Gamma_\ell^{\mathrm{ref}}=\ln\frac{\sqrt{r_{\mathrm{min}}r_{\mathrm{max}}}}{16 \epsilon\ell}$ (for $\ell\neq 0$) which yields a finite result. Notice, that apart from $\Gamma_0$, this is the periodic line effective action contribution, thus such a subtraction could be considered as the analogue of the subtraction of the infinite cylinders performed at the classical level (see \appref{app:LINE-PF} for details, also notice that $\Gamma_0$ in not well defined in this case). It is interesting to note that the periodic straight lines regulate the UV behavior while the circular Wilson loops do not.

Since any $j$-independent function is as good as any other, we might as well choose a reference function which makes the regulated sum converge faster. One example is to subtract $\Gamma_\ell^{\mathrm{ref}}=\ln\frac{\sqrt{r_{\mathrm{min}}r_{\mathrm{max}}}}{16 \epsilon\sqrt{\ell^2-1}}$ for $\ell\neq 0,1$. In this case, for $\ell = 0,1$, we may choose $\Gamma_0=\ln\frac{\sqrt{r_{\mathrm{min}}r_{\mathrm{max}}}}{\epsilon}$ and $\Gamma_1=\ln\frac{\sqrt{r_{\mathrm{min}}r_{\mathrm{max}}}}{16 \epsilon}$ for simplicity. Putting this all together, our final expression for evaluating the effective action up to an additive constant is
\begin{align}\label{eqn:partition_function_reg}
\Gamma_{\mathrm{reg}}
& =
 -\frac{2\EllipticK(m)}{\sqrt{a}}
+\ln\left(-
\frac{
a^2
}
{j^3}
\frac{\cosh^2 x_f^+\cosh^2 x_f^-}
{
\EllipticK^{5/2}(m)
\sinh x_2
\sqrt{2\EllipticE(m)-\EllipticK(m)}
}\Bigg|_{\ell=0}\right)
\nonumber\\
& +
2\ln\left(
\frac{
\left(j^2\right)^{1/4}}
{8\sqrt{a}\left(j^2+64\right)}
\frac{\cosh^2 x_f^+\cosh^2 x_f^-}
{\sinh^{5/2} x_0
\left(\EllipticE(m)-(1-m)\EllipticK(m)\right)
\sinh^{1/2} x_R
}\Bigg|_{\ell=1}\right)
\nonumber\\
& +
2\sum_{\ell=2}^\infty\ln\left(
\frac{16\sqrt{\ell^2-1} f(\ell,j)}
{g\left(\ell+\frac{1}{2},j\right)g\left(\ell-\frac{1}{2},j\right)}
\frac{\cosh^2 x_f^+\cosh^2 x_f^-}
{
\sinh^{5/2} x_0
\sinh x_2
\sinh^{1/2} x_R
}\right) \; ,
\end{align}
where $f$ and $g$ are defined in \eqref{eq:def_for_PFs_arg}.

\begin{figure}
\begin{center}
\subfloat[]{\label{fig:PF1}\includegraphics[width=70mm]{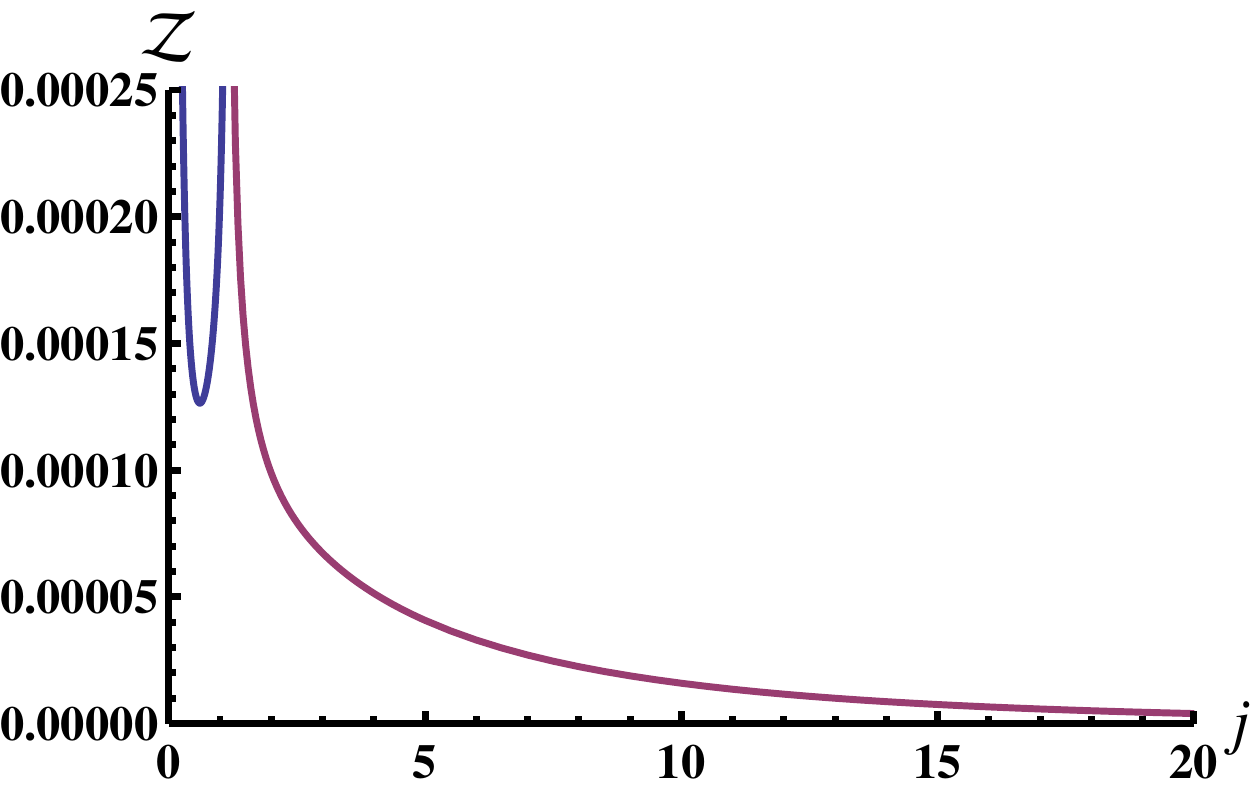}} \hspace{5mm}
\subfloat[]{\label{fig:PF2}\includegraphics[width=70mm]{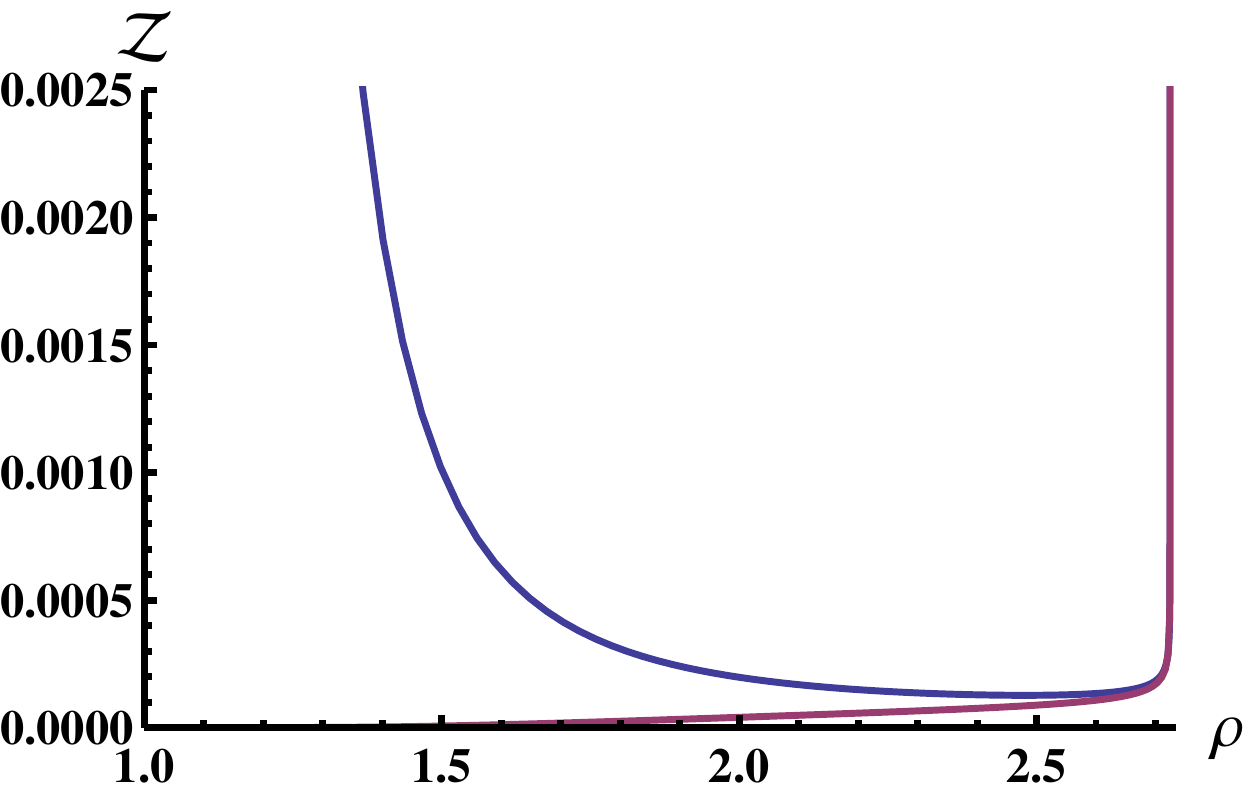}} \\
\subfloat[]{\label{fig:PF3}\includegraphics[width=70mm]{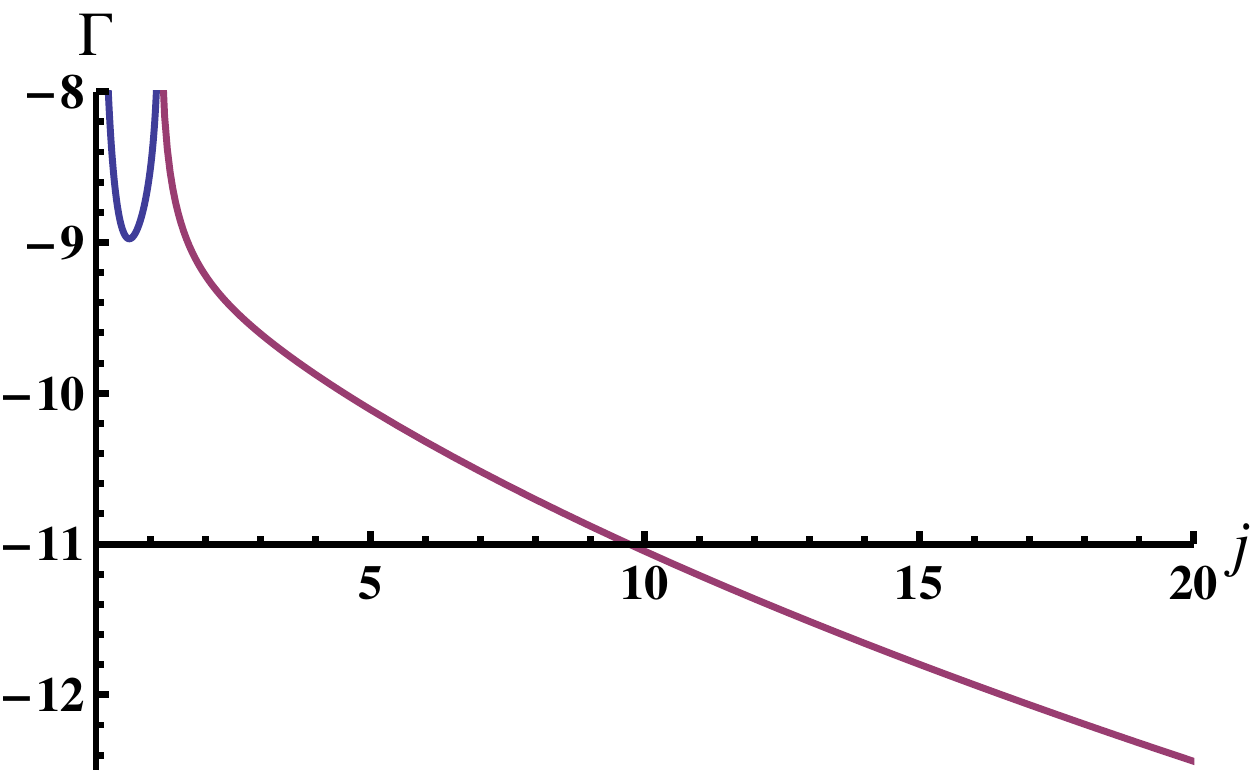}} \hspace{5mm}
\subfloat[]{\label{fig:PF4}\includegraphics[width=70mm]{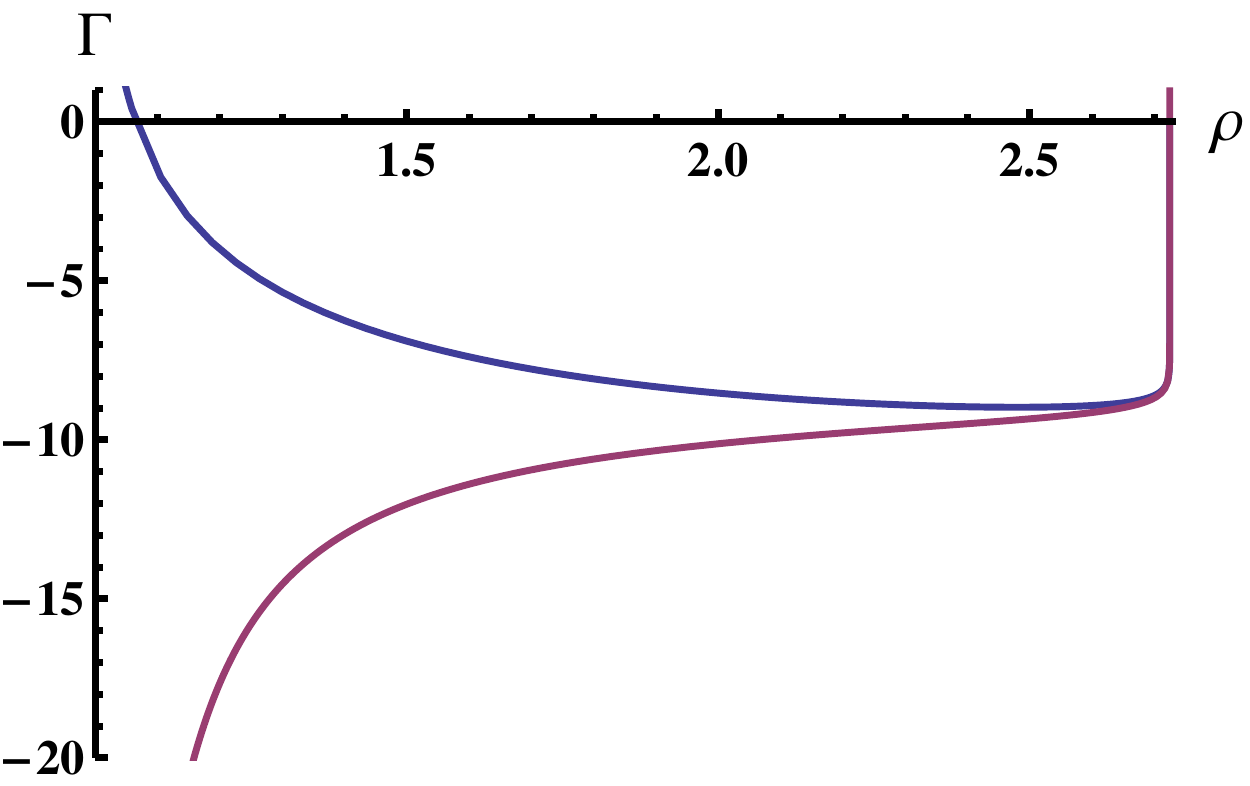}}
\caption{\textbf{Partition function and effective action.}
The purple and blue lines represents the stable and unstable configurations respectively.
In (a) and (b) the values of blue line should be understood as purely \emph{imaginary}, and in (c) and (d) as the real part supplemented by a constant imaginary piece. The partition function and effective action are defined up to a multiplicative and additive unknown constant, respectively. We remind the reader that $\rho = r_{\mathrm{max}}/r_{\mathrm{min}}$.}%
\label{fig:PF}%
\end{center}
\end{figure}

We do not know how to evaluate \eqref{eqn:partition_function_reg} analytically, but we can evaluate it numerically for any $j$ since the series converges quite fast. The value of the partition function is real for the stable configurations where $j>j_c$ and imaginary for unstable ones, where one of the eigenvalues of $\mathcal{O}_R$ is negative. Similarly, the effective action gets an imaginary contribution (namely, $\ln\imag$) for the unstable configurations. In \figref{fig:PF1} and \ref{fig:PF2}, we plot the result for the regularized partition function, and in \figref{fig:PF3} and \ref{fig:PF4}, we plot the effective action as a function of $j$ and $\rho$, respectively. Stable configurations with $j>j_c$ are drawn in purple, while the unstable ones are in blue. Moreover, in \figref{fig:PF1} and \ref{fig:PF2} the value for the unstable configurations should be understood as purely imaginary, and in \figref{fig:PF3} and \ref{fig:PF4} it should be understood that the unstable value is supplemented by $\ln \imag$. Finally, let us point out again that computing the partition function with integer fermionic modes yields the same result.

\section{Fluctuation analysis and stability}
\label{sec:fluctuations}

In this section, we will analyze the fluctuations about the classical solution in more detail. The partition function computed above is actually the product of all fluctuation frequencies, however, by utilizing the Gel'fand-Yaglom method we were able to obtain the partition function without explicitly finding the fluctuation spectrum. Nevertheless, the eigenvalues themselves do contain interesting information which we would like to extract in the following. For one thing, the signs of the eigenvalues indicate whether or not the classical solution is stable. For another, the eigenvalues are not scheme dependent as they can be computed without introducing an IR cutoff (called $\eps$ in the previous sections) and neither does their calculation require the introduction of a UV cutoff or the subtraction of a reference solution as this becomes necessary only when taking their product. Finally, knowing the fluctuation spectrum provides valuable data against which future calculations can be crosschecked.

\paragraph{Eigenvalue equation.} The question about the stability of the classical solution is answered as follows. When the Nambu-Goto action is expanded in powers of a certain perturbation, $\xi(\tau,\sigma)$, about the solution, the quadratic terms,
\be \label{eqn:action-for-fluctuation}
  \Action_2 = \frac{1}{2\pi} \int\!d\tau d\sigma\: \Half \xi \mathcal{O} \xi \; ,
\ee
define a second order differential operator $\mathcal{O}$ whose determinant yields the contribution to the partition function due to this type of fluctuations. Instead, we may solve the eigenvalue problem
\be \label{eqn:eigenvalue-equation}
  \mathcal{O} \xi_{n\ell} = \lambda_{n\ell} \, \frac{\sqrt{g}}{z^2} \, \xi_{n\ell} \; ,
\ee
where $n$ and $\ell$ are the $\tau$- and $\sigma$-quantum numbers, respectively. Then, by writing the perturbation as a superposition of the eigenfunctions, $\xi(\tau,\sigma) = \sum_{n\ell} a_{n\ell} \xi_{n\ell}(\tau,\sigma)$, we see that the action of the classical solution changes by an amount proportional to the eigenvalues:
\be \label{eqn:action-for-fluctuation-modes}
  \Action_2 = \frac{1}{4\pi} \sum_{n\ell} \lambda_{n\ell} \, a_{n\ell}^2 \; .
\ee
If all eigenvalues are positive, then the classical solution is a true minimum and the surface is stable under perturbations. If one or more eigenvalues are negative, then the area decreases under deformations by the corresponding eigenfunction and is thus an unstable saddle point.

The factor $\sqrt{g}$ on the right hand side of \eqref{eqn:eigenvalue-equation} was introduced in order to make the equation and hence the eigenvalues independent of the parametrization of the world-sheet, and the factor $\frac{1}{z^2}$ makes the equation transform covariantly under target-space isometries. It also indicates that we are thinking of $\xi$ as being a fluctuation in $\AdS$ as the latter factor would be absent for fluctuations on the sphere, see \eqref{eqn:natural-norm-fluctuations}.

\paragraph{$\Hyp_3$ fluctuations.} For concreteness, let us consider fluctuations inside $\Hyp_3 \subset \AdS_5$. Since the two-dimensional world-sheet is extended in $\Hyp_3$, there is exactly one direction normal to the classical solution. These fluctuations perturb the solution according to
\be
  r \rightarrow r + \frac{1}{\lambda^{1/4}} \frac{\dot{z}}{\sqrt{\dot{r}^2+\dot{z}^2}} \, \xi_R(\tau,\sigma)
  \comma
  z \rightarrow z - \frac{1}{\lambda^{1/4}} \frac{\dot{r}}{\sqrt{\dot{r}^2+\dot{z}^2}} \, \xi_R(\tau,\sigma)
  \; .
\ee
As introduced in \secref{sec:partition-function}, we continue to denoted this kind of fluctuations by a subscript $R$. The corresponding operator can be read off from \eqref{eqn:action-2B} as
\be
  \mathcal{O}_R =  \sqrt{g} \bigbrk{ - \nabla^2 + \mathcal{R} + 4 } \; ,
\ee
however, since \eqref{eqn:action-2B} is written for the rescaled fluctuations $\zeta_R = \frac{1}{z} \xi_R$, the eigenvalue equation reads
\be \label{eqn:eigenvalue-equation-zeta}
  \mathcal{O}_R \zeta_R = \lambda \, \sqrt{g} \, \zeta_R \; ,
\ee
i.e.\ without the factor $\frac{1}{z^2}$, which is in line with the measure in \eqref{eqn:natural-norm-fluctuations}. Multiplying the equation by $c^2 = \sqrt{g}^{-1} = \frac{z^2}{r^2}$, we arrive at
\be \label{eqn:eigenvalue-equation-R}
  c^2 \mathcal{O}_R \zeta_R = \bigbrk{ - \nabla^2 + \mathcal{R} + 4 } \zeta_R = \lambda \, \zeta_R \; ,
\ee
which identifies the eigenvalues $\lambda$ as those of the \emph{unrescaled} operator given in \eqref{eq:operator-OR}.

\paragraph{Boundary conditions.} The eigenvalue equation for the fluctuations needs to be supplemented by boundary conditions. As the perturbed surface should satisfy the same Dirichlet boundary conditions as the underlying classical solution, we have to require that the flucutations vanish at the boundary of the surface. In the context of the AdS/CFT correspondence, the boundary conditions on the string solution are always imposed at the boundary of the AdS-space as they have the interpretation of Wilson loops in the gauge theory. However, we may temporarily leave this interpretation aside and consider general minimal surfaces of revolution in $\Hyp_3$ (no additional $\Sphere^1$) with boundary conditions in the bulk.

\begin{figure}%
\begin{center}
\subfloat[Profile]{\label{fig:Vert-profile}\includegraphics[width=35mm]{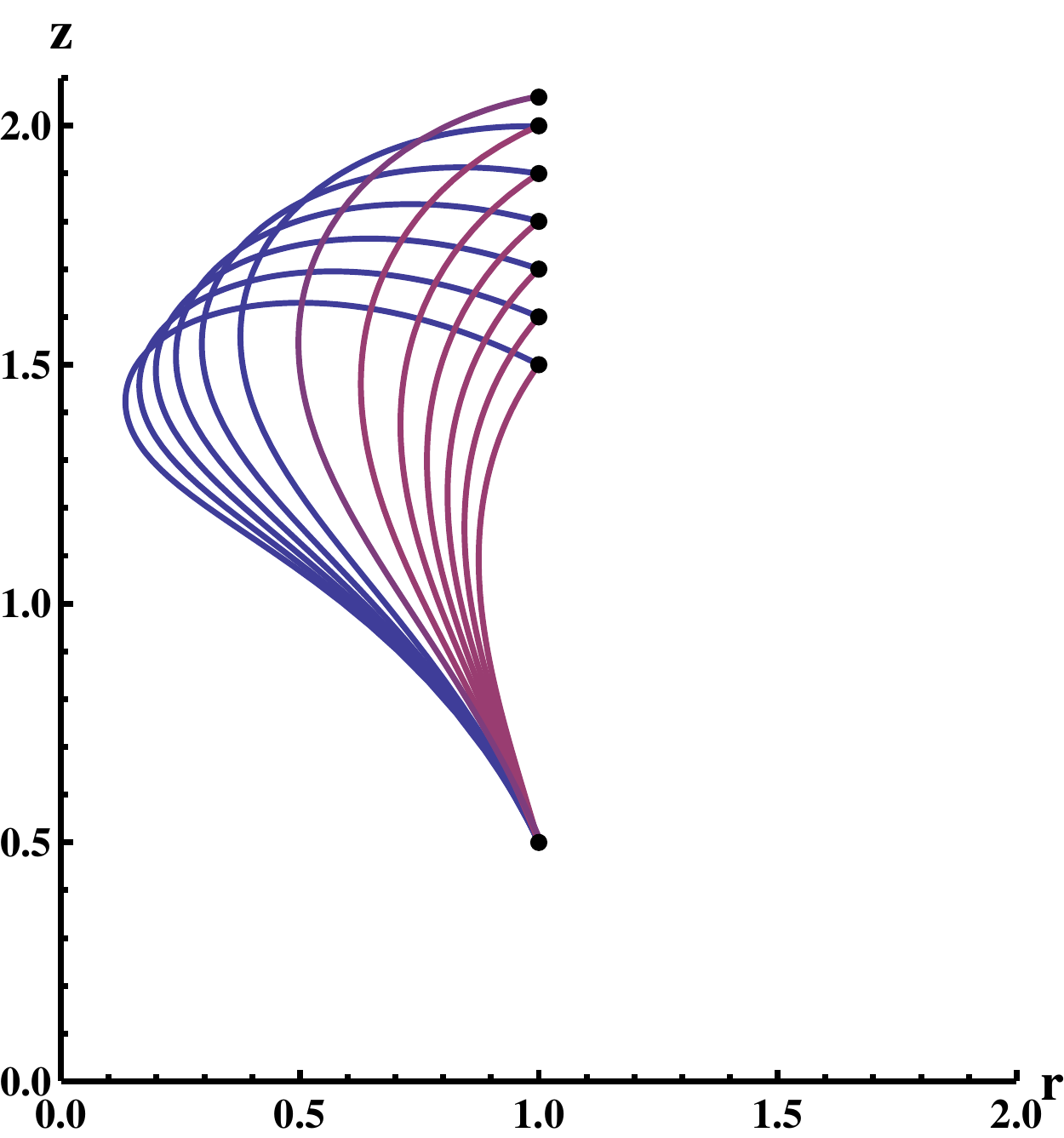}} \hspace{40mm}
\subfloat[Profile]{\label{fig:Horz-profile}\includegraphics[width=50mm]{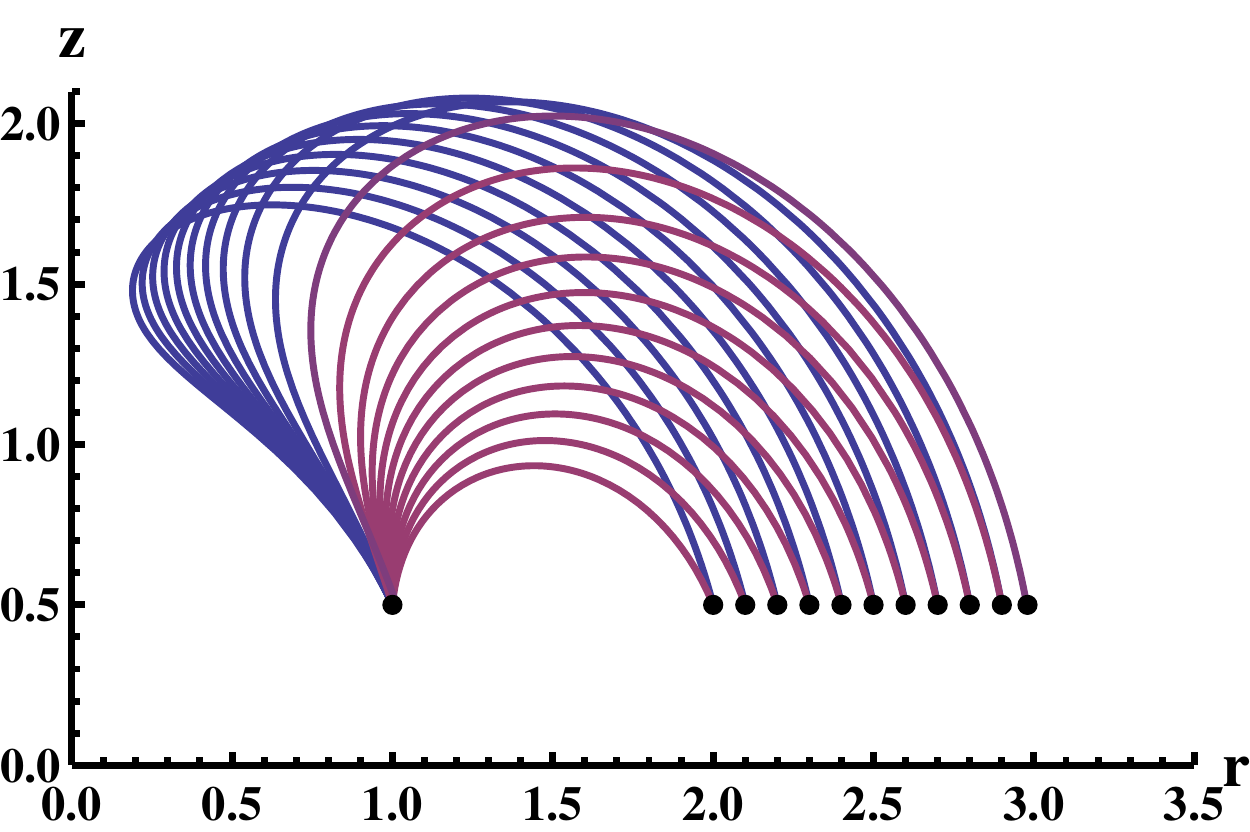}}
\caption{\textbf{Stable and unstable catenoids in hyperbolic space.}
These are the profiles of minimal surfaces in hyperbolic space bounded by two circles. If the circles are close enough together and do not differ too much in size, there exist two solutions, an unstable ``big'' one (blue) and a stable ``small'' one (purple). (a) If the circles are pulled apart from each other, the solutions degenerate and then disappear. For the chosen radius, this happens at $j\approx0.649$. (b) The same happens when the ratio of the circles' radii is increased. In this particular example, the solutions cease to exist at $j\approx1.111$.
}%
\label{fig:Profiles}%
\end{center}
\begin{center}
\subfloat[Eigenvalues]{\label{fig:Vert-eigenvalues}\includegraphics[width=43mm]{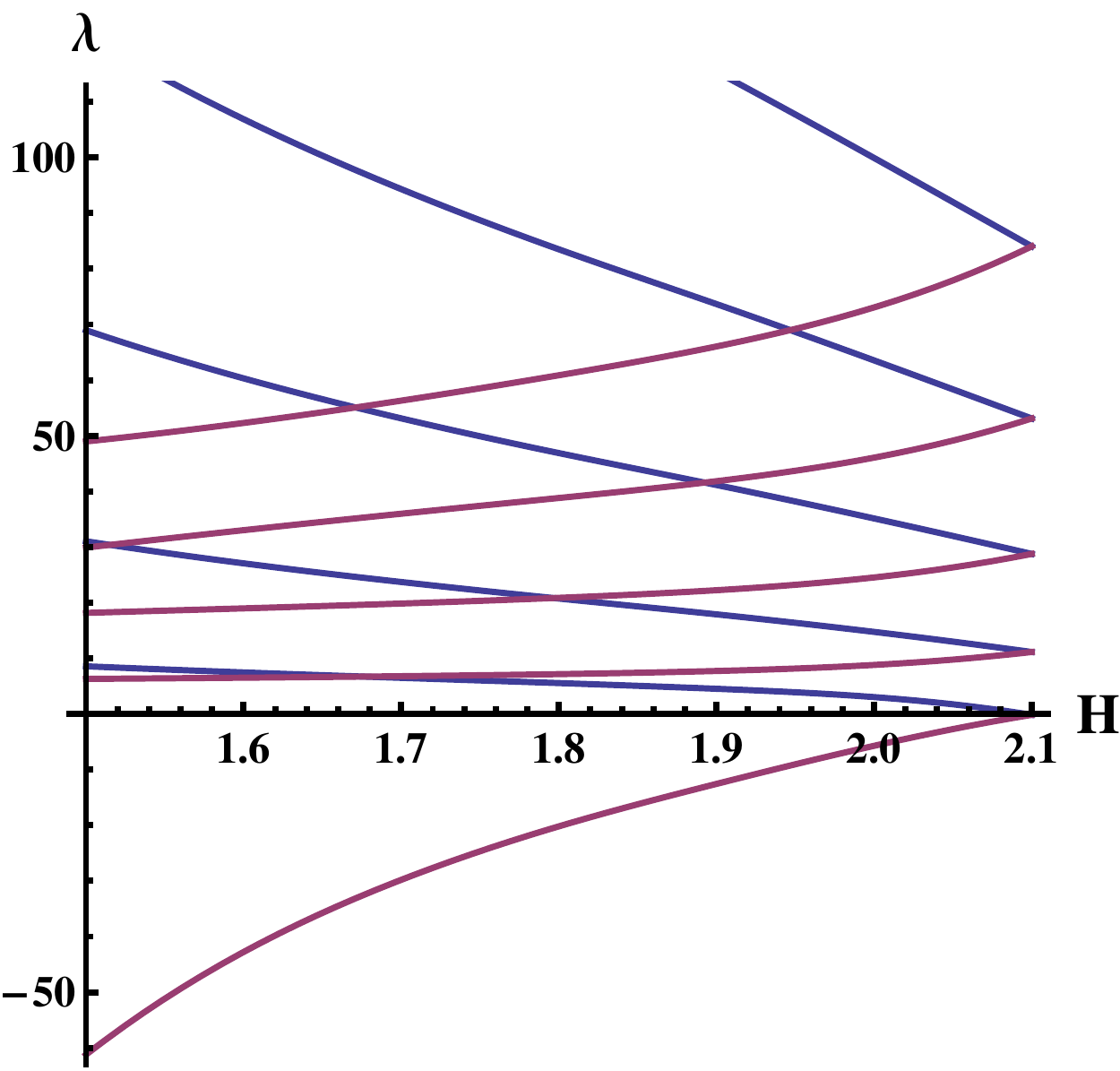}} \hspace{5mm}
\subfloat[``Big'' solution]{\label{fig:Vert-big-fluc}\includegraphics[width=50mm]{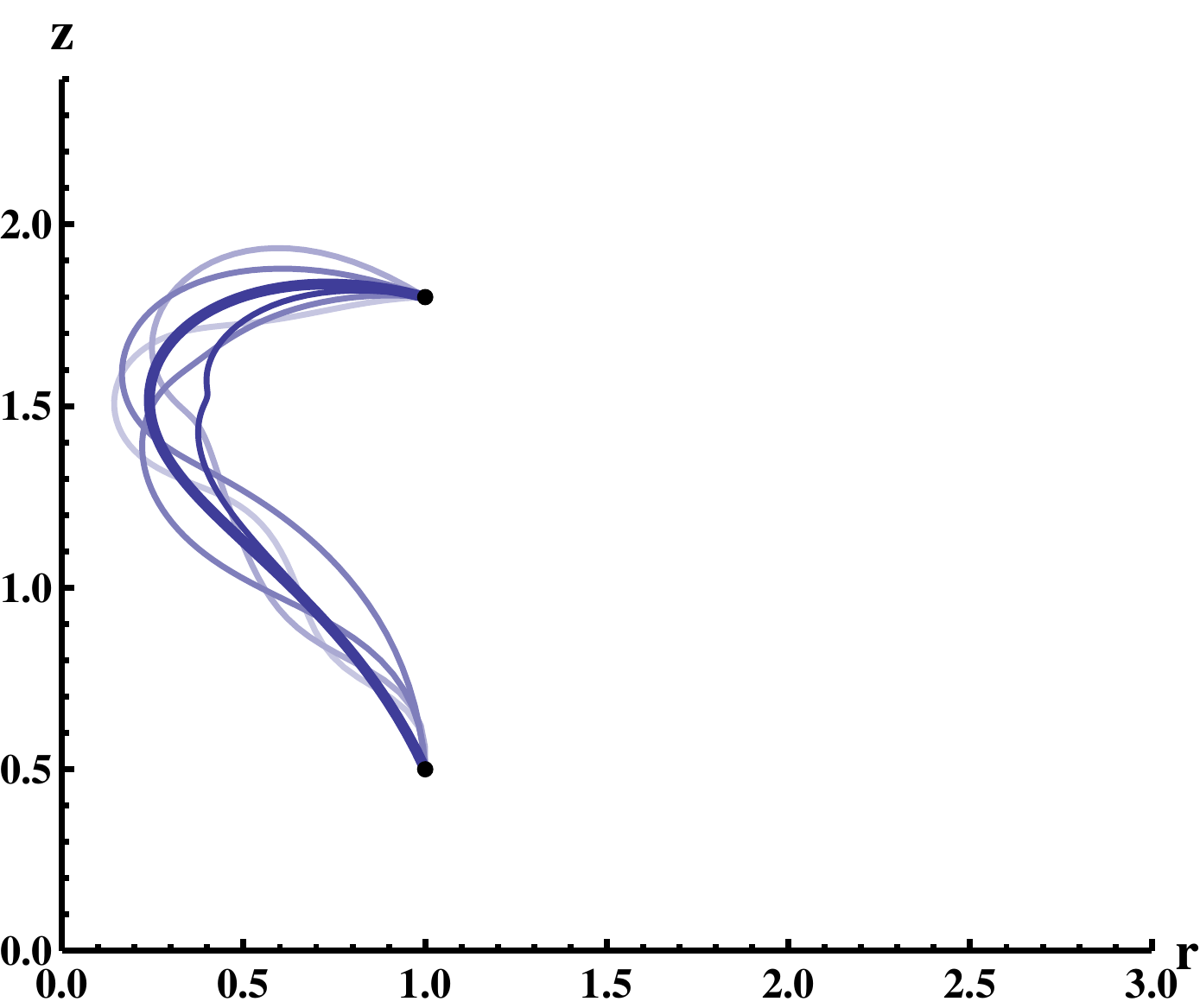}} \hspace{5mm}
\subfloat[``Small'' solution]{\label{fig:Vert-small-fluc}\includegraphics[width=50mm]{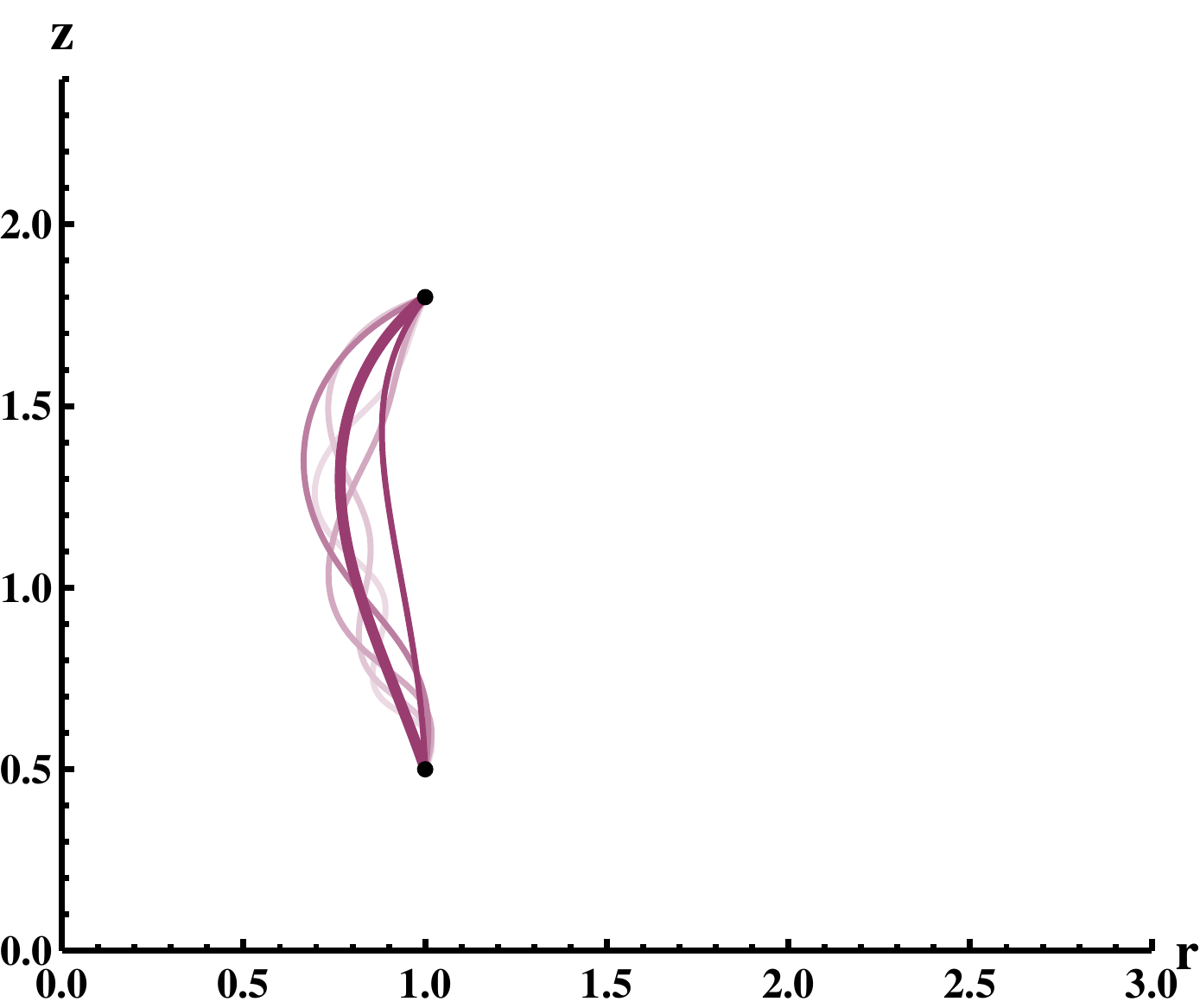}}
\caption{\textbf{Fluctuations of the ``vertical'' configurations.} (a) The eigenvalue spectrum is plotted as a function of the distance $H$. (b,c) The unperturbed solution (bold) for $H=1.8$ and its perturbations due to the first few eigenfunction.}%
\label{fig:Vert-normalfluc}%
\end{center}
\begin{center}
\subfloat[Eigenvalues]{\label{fig:Horz-eigenvalues}\includegraphics[width=43mm]{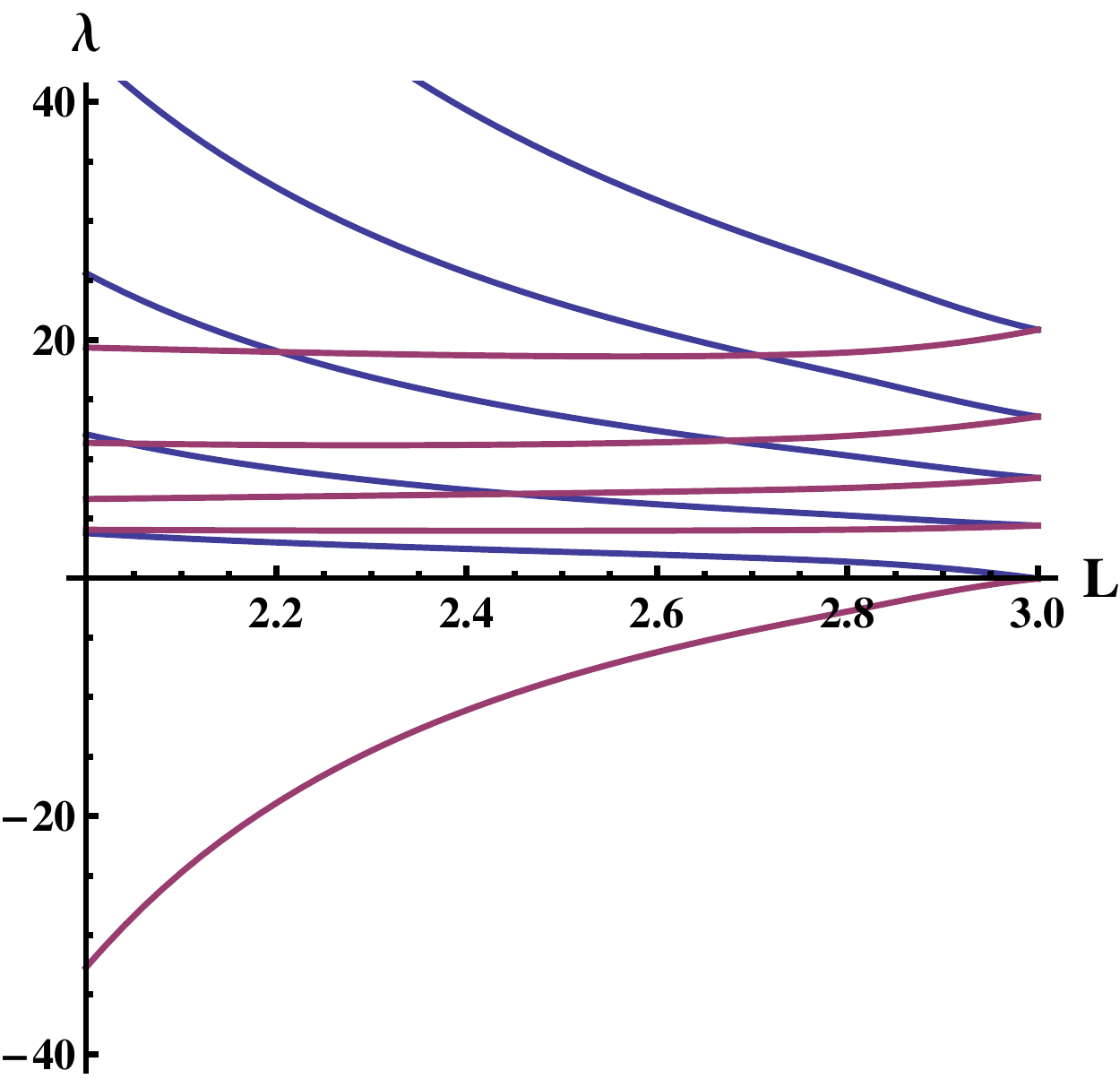}} \hspace{5mm}
\subfloat[``Big'' solution]{\label{fig:Horz-big-fluc}\includegraphics[width=50mm]{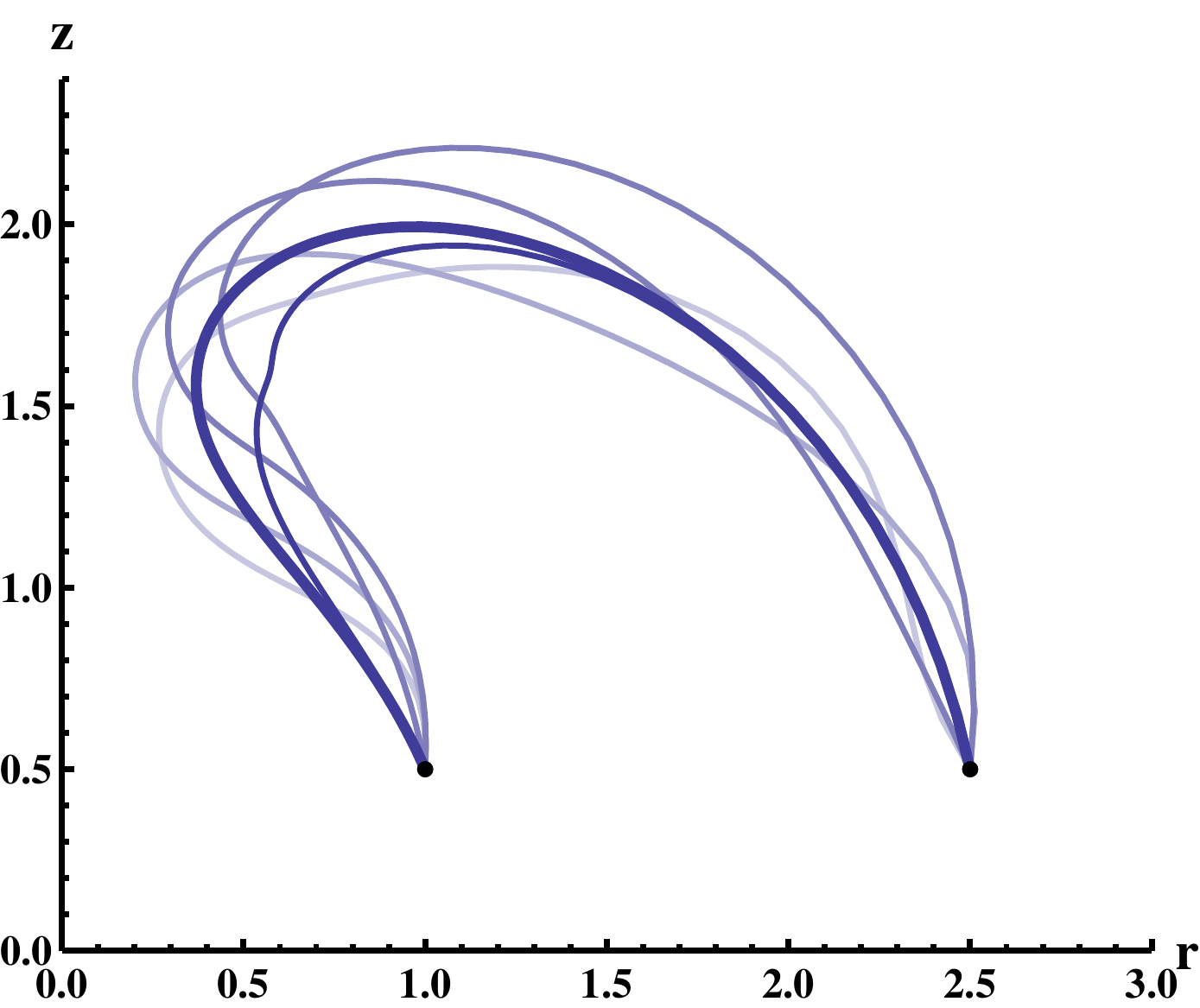}} \hspace{5mm}
\subfloat[``Small'' solution]{\label{fig:Horz-small-fluc}\includegraphics[width=50mm]{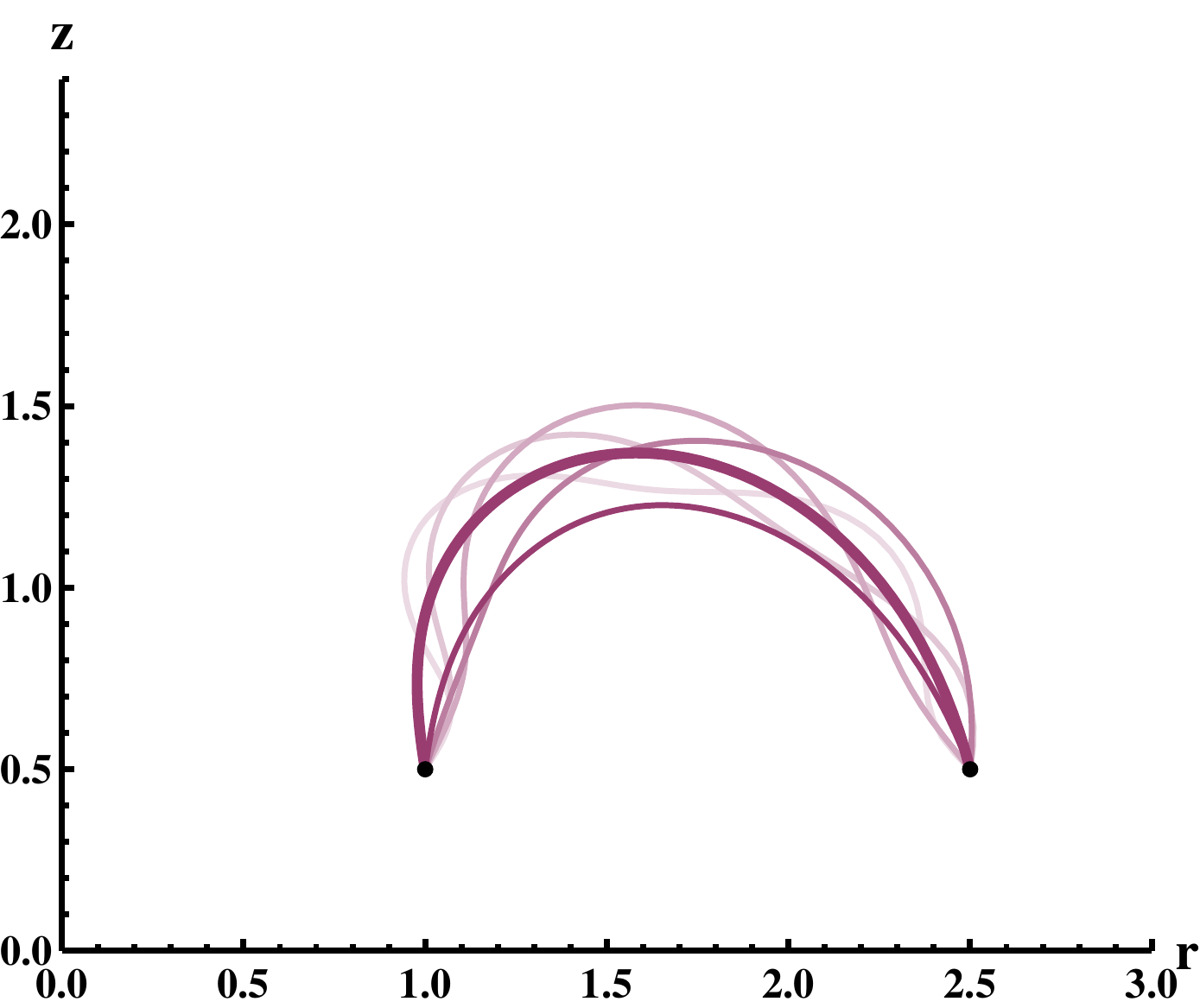}}
\caption{\textbf{Fluctuations of the ``horizontal'' configurations.} (a) The eigenvalue spectrum is plotted as a function of the distance $L$. (b,c) The unperturbed solution (bold) for $L=2.5$ and its perturbations due to the first few eigenfunction.}%
\label{fig:Horz-normalfluc}%
\end{center}
\end{figure}

Hence, we demand the surface to end on two circles of radius $r_0$ and $r_1$ at heights $z_0$ and $z_1$, respectively. From the analysis in \secref{sec:classical} we know that there is either one, two or no solution that ends on these circles. In case there are two solutions, they will have different area and we will call one ``big'' and the other one ``small'' accordingly. The following discussion is to show that the big solution is a saddle point of the Nambu-Goto action and therefore is unstable against perturbations. The small solution is a true minimum and thus stable. The situation in which there is one solution, is the limiting case when the big and small solution coincide.

To give specific examples, we will consider two sets of boundary conditions. We call the first set ``vertical'' and the second ``horizontal''. The vertical configurations are given by the boundary conditions
\be
  r(\tau_{\mathrm{min}}) = 1.0
  \comma
  r(\tau_{\mathrm{max}}) = 1.0
  \comma
  z(\tau_{\mathrm{min}}) = 0.5
  \comma
  z(\tau_{\mathrm{max}}) = H
  \; ,
\ee
with $H$ between $1.5$ and $2.1$. These four equations are conditions on the parameters\footnote{Equivalently, we may trade the parameter $f_0$ for the more physical quantity $r_{\mathrm{min}}$, see \protect\eqref{eqn:r-min-max}.}
\be
  j \comma
  f_0 \comma
  \tau_{\mathrm{min}} \comma
  \tau_{\mathrm{max}} \; .
\ee
The boundary conditions are chosen to lie in a region where there are always two solutions, which essentially coincide for $H\approx2.1$. For larger $H$ there are no solutions. The corresponding profiles are plotted in \figref{fig:Vert-profile}. Now, we solve the eigenvalue equation \eqref{eqn:eigenvalue-equation-zeta} for axially symmetric ($\ell=0$, $\sigma$-independent) fluctuations satisfying
\be
  \zeta_R(\tau_{\mathrm{min}}) = \zeta_R(\tau_{\mathrm{max}}) = 0 \; .
\ee
We find the spectrum numerically using the shooting method. The first few eigenvalues as a function of $H$ are plotted in \figref{fig:Vert-eigenvalues}. The important point is that the lowest eigenvalue is negative for the big (blue) solution and positive for the small solution (purple). The negative eigenvalue increases and the positive eigenvalue decreases with $H$ until they reach zero precisely then when the big and small solution coincide. In \figref{fig:Vert-big-fluc} and \figref{fig:Vert-small-fluc}, we display the corresponding eigenfluctuations on top of the classical solution.

For the horizontal configurations, we choose the boundary conditions
\be
  r(\tau_{\mathrm{min}}) = 1.0
  \comma
  r(\tau_{\mathrm{max}}) = L
  \comma
  z(\tau_{\mathrm{min}}) = 0.5
  \comma
  z(\tau_{\mathrm{max}}) = 0.5
  \; ,
\ee
where $L$ is varied between $2.0$ and $3.0$. The calculation and the conclusions are qualitatively the same as above. The results are drawn in \figref{fig:Horz-profile} and \figref{fig:Horz-normalfluc}.

\paragraph{Critical configurations.} As explained above, depending on the boundary conditions there are either no, one, or two solutions. In case there are two solutions, one of them is stable and the other one is unstable. The case where there is only solution is a limiting case of this situation when the two solution degenerate. We call such a configuration ``critical''.

\begin{figure}%
\begin{center}
\subfloat[``All'' solutions through $(1.0,0.5)$]{\label{fig:allsol}\includegraphics[width=55mm]{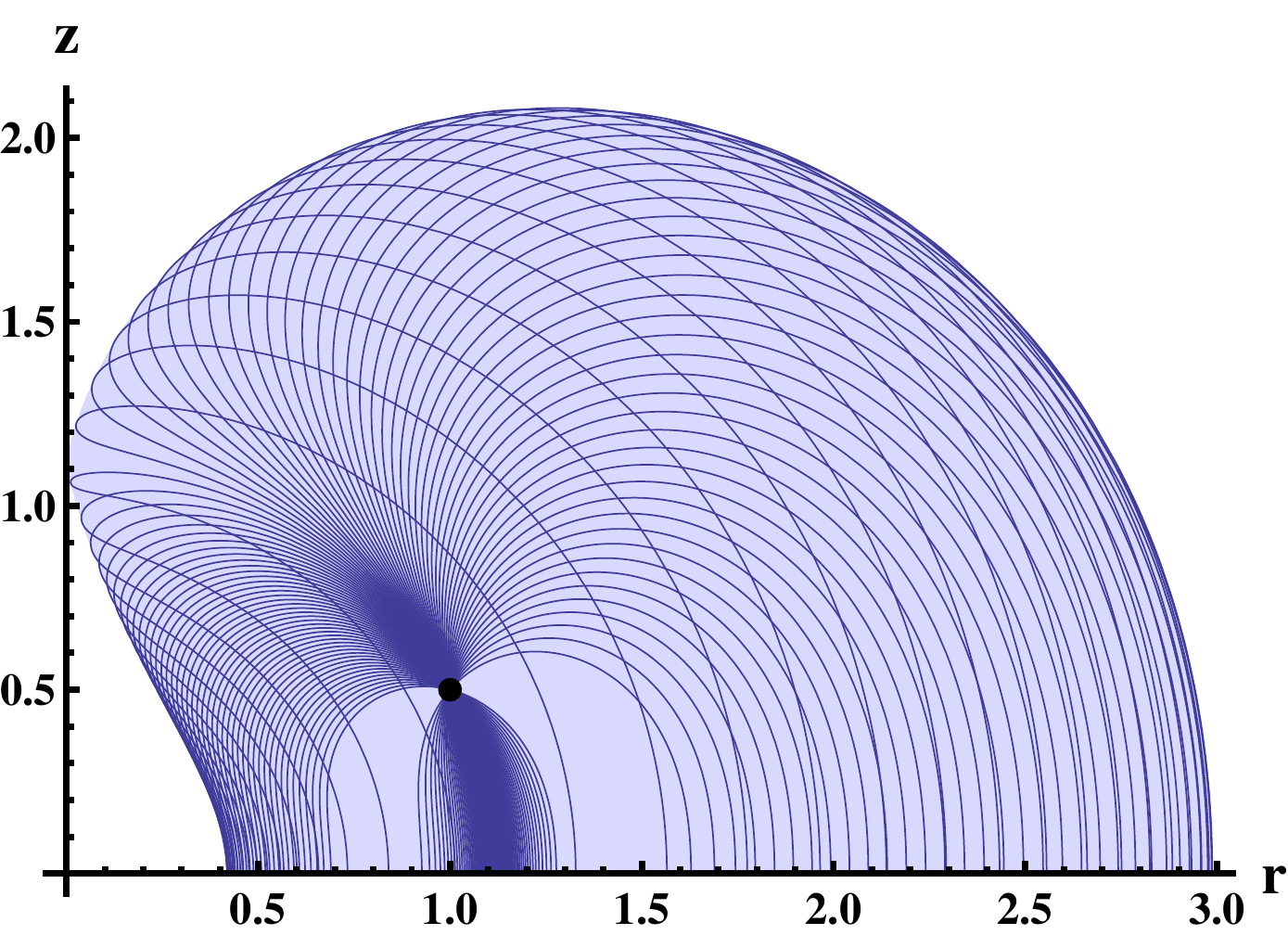}} \hspace{15mm}
\subfloat[Critial solutions through $(1.0,0.5)$]{\label{fig:criticalsol}\includegraphics[width=55mm]{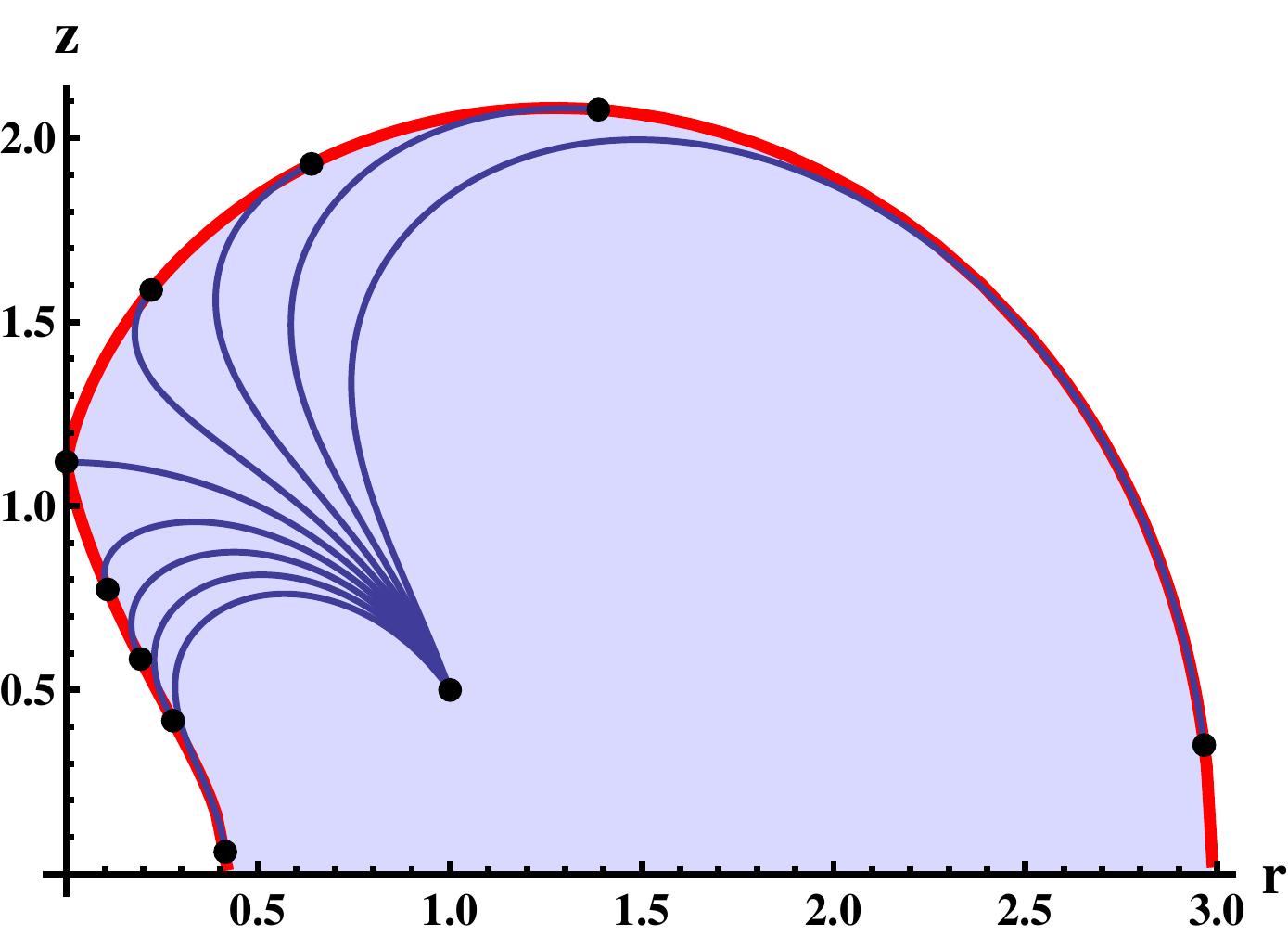}}
\caption{\textbf{Critical solutions.} See the text for explanations.}%
\label{fig:Critical-solutions}%
\end{center}
\end{figure}

Now, we would like to ask how those critical solutions look. Or, what the boundary conditions are that lead to critical solutions. These questions can be answered as follows. The shape of each solution is parametrized by two parameters, namely $j$ and $f_0$. Now, let's say we fix one boundary. As an example we choose
\be
  r(\tau_0) = 1.0 \comma z(\tau_0) = 0.5 \; .
\ee
This is only one condition because $\tau_0$ is arbitrary. We can use this condition to eliminate $f_0$, i.e. to express $f_0$ through $j$ using this condition. By varying $j$ we obtain the set of solutions that go through the point $(1.0,0.5)$, see \figref{fig:allsol}. If we impose as second boundary condition that the surface goes through another point that lies in the interior of the shaded region, then there exist solutions. If we tried to impose that the surface goes through another point outside of the shaded region, then there would not be any solution. Thus, if we place the second boundary point of the surface right on the boundary of the shaded region, we will obtain a critical solution. Some examples are plotted in \figref{fig:criticalsol}.

The boundary of the shaded region is called the \emph{envelope} of the set of curves $\vec{r}(\tau,j) = (r,z)^\trans$. The envelope satisfies the equation
\be
  \partial_\tau \vec{r} \wedge \partial_j \vec{r} \; \equiv \,
	\frac{\partial r}{\partial \tau} \frac{\partial z}{\partial j} - \frac{\partial z}{\partial \tau} \frac{\partial r}{\partial j} = 0 \; .
\ee
This equation gives a relationship between $\tau$ and $j$ with the following meaning. Say, we pick a curve by specifying $j$. Then we can compute $\tau(j)$ as the point on this curve which is also part of the envelope. The red curve in \figref{fig:criticalsol} is thus given by $j\mapsto\vec{r}(\tau(j),j)$.

\paragraph{Critical correlator.} If we go back and impose the boundary condition for the surface on the boundary of AdS,
we can give a compact equation for the parameter $j_c$ of the critical solution. As we have seen, for the critical solution there exists a zero-mode fluctuation. Hence, the product of all eigenvalues, or the determinant, is zero. From \eqref{eqn:det_in_GY_0}, we immediately read off the condition
\be \label{eqn:eqn-for-jc}
  \frac{\EllipticE(m(j_c))}{\EllipticK(m(j_c))}=\frac{1}{2}
  \qquad\text{with}\qquad
  m(j) = \Half + \frac{1}{2\sqrt{1+j^2}} \; .
\ee
This equation can be solved numerically using, e.g., Mathematica to practically arbitrary precision
\be
j_c \eq 1.16220056179001257099525974162879065620254318155768903221387272980593...\qquad\mbox{}
\ee

\begin{figure}%
\begin{center}
\includegraphics[width=60mm]{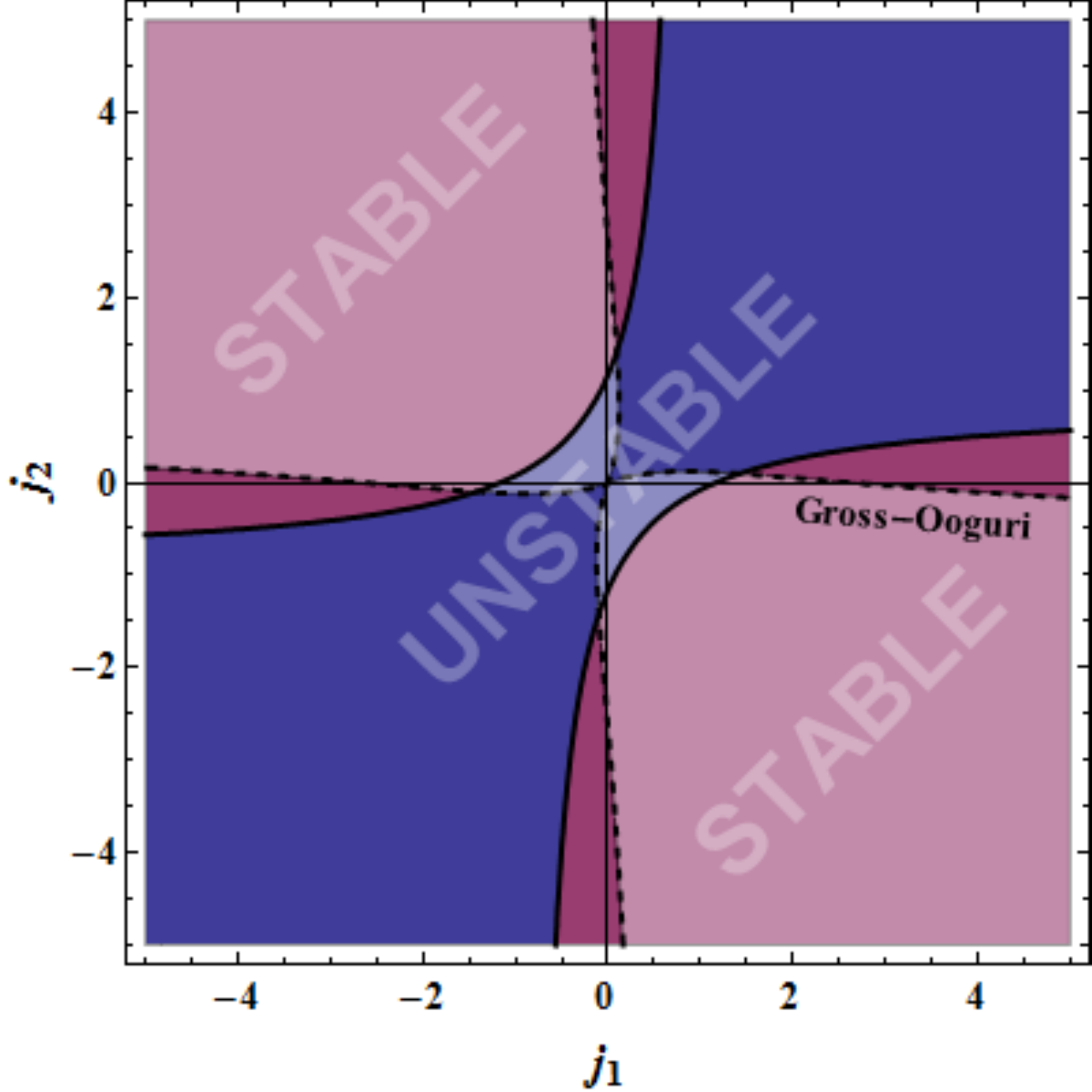}
\caption{\textbf{Stability and phase transition.} For parameters $j_1$ and $j_2$ in the purple region (both light and dark), all fluctuations increase the area which implies that the classical surface is stable. In the complementary region (light and dark blue), one fluctuation mode has a negative eigenvalue and thus renders the surface instable. The dashed line marks the Gross-Ooguri phase transition, i.e.\ the contour of configurations whose area equals that of two disconnected minimal surfaces. In the dark regions, the connected surface dominates in the path integral over the disconnected one, i.e.\ the former has smaller area than the latter.}%
\label{fig:Phases}%
\end{center}
\end{figure}

\paragraph{Generalization to $\Hyp_3 \times \Sphere^1$.} The eigenvalue equation \eqref{eqn:eigenvalue-equation-R} holds, in fact, also for the two-parameter solution, if we use the appropriate scalar curvature. Specializing the general formula for $\mathcal{R}$ given in \eqref{eqn:scalar-curvature} to solutions in $\Hyp_3 \times \Sphere^1$, which satisfy \eqref{eqn:generalized-eom-rz}, we obtain
\be \label{eqn:scalar-curvature-H3S1}
  \mathcal{R} = - 2 - \frac{(j_1-j_2)^2}{2} \frac{z^4}{r^4} \; .
\ee
This generalizes the corresponding expression for solutions in $\Hyp_3$ by the simple replacement $j^2 \mapsto (j_1-j_2)^2$. What is more, when we evaluate \eqref{eqn:scalar-curvature-H3S1} for the explicit solution \eqref{eqn:rz_to_hf} with \eqref{eqn:general-solution-h} and \eqref{eqn:general-solution-f}, we find
\be
  - \nabla^2 + \mathcal{R} + 4
  = c^2 \Bigbrk{ - \partial_\tau^2 - \partial_\sigma^2 + 2a \JacobiDS^2(\sqrt{a}\tau|m) - 2a m (1-m) \JacobiSD^2(\sqrt{a}\tau|m) } \; ,
\ee
i.e.\ the operator is, up to the factor $c^2=\bigbrk{\frac{z}{r}}^2$, formally identical to \eqref{eqn:general-opR} except that now $a$ and $m$ are given by the general formulas \eqref{eqn:parameters-a-m} rather than \eqref{eqn:r-over-z}. If we are interested only in the zero mode of this operator, we may even drop the $c^2$ and end up---again formally---with the same equation to solve. Thus, also the eigenfunction as well as the criticality condition have the same form as functions of $a$ and $m$. The line of critical solutions in the $j_1$-$j_2$-parameter plane is therefore determined by
\be \label{eqn:eqn-for-critical-line}
  \frac{\EllipticE(m(j_1,j_2))}{\EllipticK(m(j_1,j_2))}=\frac{1}{2}
  \qquad\text{with}\qquad
  m(j_1,j_2) = \Half + \frac{1+j_1j_2}{2\sqrt{(1+j_1^2)(1+j_2^2)}} \; .
\ee
The corresponding curve is plotted in \figref{fig:Phases} as the continuous line. Note that this line intersects the axes at $(\pm j_c,0)$ and $(0,\pm j_c)$. In the same plot, we have also added a dashed line for solutions whose regularized area is $-4\pi$, i.e.\ the same area as two hemispheres. This line marks the Gross-Ooguri phase transition.

\section{Discussion}

In this paper, we have studied the correlation function of two concentric circular Wilson loops at strong coupling in $\mathcal{N}=4$ SYM theory using holography. To lowest order at strong coupling, the expectation value of the correlator is related to the area of the classical string solution bounded by the two circles. By taking the circles to be concentric, we have effectively solved the general problem of finding all minimal surfaces of revolution in $\Hyp_3\times\Sphere^1 \subset \AdS_5 \times \Sphere^5$. This set of solutions is characterized by two parameters, $j_1$ and $j_2$, which are related to the $\Hyp_3$ dilatation charge $Q$ and the $\Sphere^1$ angular momentum $J$.

For given boundary configuration there may be several, one, or no solutions. If solutions exist, they may be stable or unstable under small perturbations. We have analyzed their stability by an analysis of the fluctuation spectrum. If negative eigenvalues exist, then the solution is unstable. A zero mode means that the area of the surface is not changed under the corresponding perturbation to quadratic order and we refer to such a solution as being ``critical''. Critical solutions appear in the parameter space where a phase with stable solutions borders on a phase with unstable solutions. We have determined the critical lines as functions of $j_1$ and $j_2$. In addition, we have found the Gross-Ooguri phase transition points in terms of the same parameters. We have also studied the stability of such solutions where the circles that constitute the boundary conditions are located in the bulk of $\Hyp_3$ rather than on its boundary. Finally, we constructed the algebraic curve that encodes the classical solution in terms of $j_1$ and $j_2$, by explicitly calculating the eigenvalues of the monodromy matrix.

After having studied the classical string solution, we computed the one-loop partition function by introducing fluctuations around the classical solution. In this part, we limited ourselves to classical solutions that are confined to $\Hyp_3$, i.e.\ to solutions that do not carry any charge on the sphere. Nonetheless, the fluctuations themselves propagate in all eight directions of $\AdS_5 \times \Sphere^5$ transverse to the classical string as well as along the fermionic coordinates. The resulting object is a formal expression for the one-loop partition function in terms of bosonic and fermionic two-dimensional determinants. Using the azimuthal symmetry, we reduced the problem to an infinite product over one-dimensional non-trivial determinants. Such determinants can be computed using the Gel'fand-Yaglom method, which maps the problem of finding the individual eigenvalues and their product to an initial value problem. Fortunately, in this case the initial value problem can be solve analytically in terms of solutions to the Lam\'{e} equation. Thus, we were able to express the partition function in terms of an infinite product over known function.

In order to compute the partition function, this expression still needs to be regulated for several reasons. First, taking the logarithm of the partition function yields an infinite sum over integer bosonic and half-integer fermionic modes, each of which is divergent separately. These sums should be combined in a ``supersymmetric'' way which amounts to defining the summation purely in terms of bosonic (or fermionic) mode numbers at the cost of producing an extra finite contribution. Although this regularization results in significant cancellations, the sum still suffers from IR and UV divergences, where the IR divergence appears when the solution approaches the AdS boundary, while the UV divergence originates from the infinite summation over Fourier modes. Both divergences can be canceled by subtracting a reference solution. We showed that not any reference solution with the same boundary condition can render the result finite. Eventually, we regulated the (logarithm of the) partition function using a reference function which yields the result up to a constant. The final result is given in \figref{fig:PF}.

As a matter of fact, we could have regulated the result by subtracting an arbitrary function that does not have any physical significance as long as it has the appropriate behavior for large mode numbers and as long as it is independent of the physical parameters (here $j$). The difference in the answer would have been at most an overall factor in the partition function or, equivalently, an additive constant in the effective action. While the $j$-dependence is unique, the overall scale could not be fixed. Trying to determine this scale by, say, taking the limit $j\to0$ and comparing the result to the circular Wilson loop fails because the partition function does not have a smooth limit.

As expected, the partition function diverges at the critical configuration because of the presence of a zero mode. It is real valued for the stable solutions and takes imaginary values for the unstable ones. The partition function decreases for large $j$ (or equivalently as the ratio of the radii becomes smaller). Other than that, the partition function is smooth and does not show any exceptional features. It is interesting to notice that for given boundary, the imaginary part of the effective action for the unstable solution is greater than the real part for the stable one, as is the classical area.

Currently, there is no data available in the literature to which we could compare our results. It would be interesting to re-derive our results using different techniques which, e.g., make use of integrability. A hint that such an approach should be possible comes from the fact that in all computations of this kind \cite{Kruczenski:2008zk,Chu:2009qt,Forini:2010ek,Drukker:2011za,Beccaria:2010ry,Kristjansen:2012nz}, one always ends up with a Lam\'{e} equation of very special type. In order to implement such a program, the semi-classically quantization of closed string solutions using the algebraic curve comes to mind \cite{Gromov:2007aq}. In fact, we were able to construct such a curve, however, the boundary data is missing in this description. It is not clear whether the construction of a curve that does carry information about the boundaries is possible (however, see \cite{Dekel:2011ja}) and whether such a description may be quantized in the spirit of \cite{Gromov:2007aq}. Naively quantizing the curve given in this paper yields trivial results for the spectrum and thus calls for an extension of the current quantization prescription.


In this paper, we did not consider the quantization of the general classical solution in $\Hyp_3\times\Sphere^1$. It would be very intriguing to generalize our calculation to this case, which comprises the interesting case of the correlation function between a circular Wilson loop and a BMN operator.

Recently, the correlation function of two circular Wilson loops in a confining background was studied numerically at strong coupling using classical string theory in \cite{Armoni:2013qda}. It would be nice to generalize our one-loop analysis to their case and see how it gets modified. In order to do that, it would be very helpful to find an analytical solution for the classical problem first.

At strong coupling, Wilson loops in $\mathcal{N}=4$ SYM theory in representations of the gauge group other than the fundamental representation are holographically represented by D3 and D5 branes carrying electric flux rather than by fundamental strings \cite{Drukker:2005kx,Yamaguchi:2006tq,Hartnoll:2006hr,Gomis:2006sb,Gomis:2006im}. The equations for the one-loop correction to the effective action for the case of D5-branes wrapping $\AdS_2\times \Sphere^4$ inside $\AdS_5\times \Sphere^5$, which correspond to Wilson loops in the anti-symmetric representation, were derived for general boundary conditions and then solve for the circular Wilson loop in \cite{Faraggi:2011bb,Faraggi:2011ge}. Furthermore, it was suggested that it should be possible to compute the correlator of two Wilson loops in the anti-symmetric representation using the same techniques. It would be very nice to perform this computation and compare the results with the ones of this paper.

\section*{Acknowledgments}

We would like to thank Nadav Drukker, Valentina Forini, Valentina Giangreco Marotta Puletti, Tristan McLoughlin, Joseph Minahan, Wolfgang M\"{u}ck, Hai-cang Ren and Konstantin Zarembo for valuable discussions.

\appendix

\section{Elliptic functions}\label{app:EllipticFunctions}

Throughout this paper we work with the standard Jacobi elliptic functions with the conventions of \cite{abramowitz1964handbook}:
\be
\JacobiSN(x|m) \; , \quad
\JacobiDN(x|m) \; , \quad
\JacobiCD(x|m) \; , \quad
\mathrm{etc}. \; ,
\ee
where $x$ is the argument and $m$ is the modulus. We further use the incomplete elliptic integrals
\be
\EllipticF(x|m) \eq \int_0^x (1-m\sin^2\theta)^{-\frac{1}{2}} d\theta\; , \\
\EllipticE(x|m) \eq \int_0^x (1-m\sin^2\theta)^{\frac{1}{2}} d\theta\; , \\
\EllipticPi(n;x|m) \eq \int_0^x (1-n\sin^2\theta)^{-1}(1-m\sin^2\theta)^{-\frac{1}{2}} d\theta\; ,
\ee
of the first, second and third kind, respectively, as well as the complete elliptic integrals
\be
\EllipticK(m) = \EllipticF(\tfrac{\pi}{2}|m) \; , \quad
\EllipticE(m) = \EllipticE(\tfrac{\pi}{2}|m) \; .
\ee
We also use the Jacobi amplitude $\JacobiAM(x|m)$ which is the inverse of $\EllipticF(x|m)$. Furthermore, we use the $H$, $\Theta$ and $Z$ Jacobi theta functions which are defined using the standard $\theta$-functions (whose definitions can be found in \cite{abramowitz1964handbook})
\be
H(x|m)      \eq \theta_1\Bigbrk{\frac{\pi x}{2 \EllipticK(m)},q(m)} \; , \\
\Theta(x|m) \eq \theta_4\left(\frac{\pi x}{2 \EllipticK(m)},q(m)\right) \; , \\
Z(x|m)      \eq \frac{\pi}{2 \EllipticK(m)}\frac{\theta_4^{'}\left(\frac{\pi x}{2 \EllipticK(m)},q(m)\right)}{\theta_4\left(\frac{\pi x}{2 \EllipticK(m)},q(m)\right)} \; ,
\ee
where the nome is given by $q(m)=e^{-\pi \EllipticK'(m)/\EllipticK(m)}$, and the prime represents a derivative with respect to $x$. Two useful identities between the various elliptic functions are
\be
\JacobiSN^{-1}(x|m) \eq \EllipticF(\sin^{-1}(x)|m) \; , \\
Z(\EllipticF(x|m)|m) \eq \EllipticE(x|m)-\frac{\EllipticE(m)}{\EllipticK(m)}\EllipticF(x|m) \; .
\ee
In the main body of the paper, we deal with elliptic functions with different moduli, see \eqref{eqn:general-operators} versus \eqref{eqn:OpsList_in_Lame_form}. Some relations between these functions, which can be proved, e.g., using Landen transformations, are
\be\label{eq:modulus-identities}
  \EllipticK(m) \eq \mu^{\Quarter} \, \Bigbrk{ 2\EllipticK(\mu) + \imag \EllipticK'(\mu) } , \; \\
  \EllipticK'(m) \eq \mu^{\Quarter} \, \EllipticK'(\mu), \; \\
  \JacobiSN^2(u|{-1}) \eq  1 - \JacobiSN^2\bigbrk{\sqrt{2} u + \EllipticK(\half)|\half}^2  . \;
\ee

\section{The Lam\'{e} equation}\label{app:Lame}

The Lam\'{e} operator with a general eigenvalue $\Lambda$ is given by
\be \label{eqn:Lame-eq}
  \mathcal{O}_{n,m,\Lambda} = -\partial^2_x + n(n+1) m \JacobiSN^2(x|m) + \Lambda \; .
\ee
In this work we are interested in the special case where $n=1$, which gives
\be \label{eqn:standard-Lame-operator}
  \mathcal{O}_{m,\Lambda} = -\partial^2_x + 2 m \JacobiSN^2(x|m) + \Lambda \; .
\ee
Generally, the basis of solutions to $\mathcal{O}_{m,\Lambda} y(x)=0$ is given by (see e.g.\ \cite{WhWa27})
\be\label{eq:LameEqSol}
y_{\pm}(x)=\frac{H(x\pm \alpha|m)}{\Theta(x|m)}e^{\mp x Z(\alpha|m)}
\comma
\JacobiSN(\alpha|m)=\sqrt{\frac{1+m+\Lambda}{m}} \; ,
\ee
where $H$, $\Theta$ and $Z$ are Jacobi's theta functions, defined in \appref{app:EllipticFunctions}. However, for some special values of $\Lambda$ these solutions degenerate, and one should use a different set of basis solutions. The special cases which are of interest are $\Lambda=-(1+m)$ and $\Lambda=-1$ (another one which we do not need here is $\Lambda=-m$). Next we give these special solutions following \cite{Valent:2005}.

For $\Lambda=-(1+m)$ the Lam\'{e} operator is given by
\be \label{eqn:special-Lame-operator-1}
  \mathcal{O}_{m,\Lambda} = -\partial^2_x + 2 m \JacobiSN^2(x|m) -m-1 \; ,
\ee
and the two independent solutions are given by
\be
  y_1(x) \eq \JacobiSN(x|m) \; , \nln
  y_2(x) \eq \JacobiSN(x|m)\left(\frac{H'(x|m)}{H(x|m)} + \frac{\EllipticE(m)-\EllipticK(m)}{\EllipticK(m)} \, x\right) \; .
\ee
This case is relevant for $\mathcal{O}_R$ with $\ell^2=0$.

For $\Lambda=-1$ the Lam\'{e} operator is given by
\be \label{eqn:special-Lame-operator-2}
  \mathcal{O}_{m,\Lambda} = -\partial^2_x + 2 m \JacobiSN^2(x|m) -1 \; ,
\ee
and the two independent solutions are given by
\be
  y_1(x) \eq \JacobiCN(x|m) \; , \nln
  y_2(x) \eq \JacobiCN(x|m)\left(\frac{H_1'(x|m)}{H_1(x|m)}+\frac{\EllipticE(m)-(1-m)\EllipticK(m)}{\EllipticK(m)} \, x\right) \; .
\ee
This case is relevant for $\mathcal{O}_2$ with $\ell^2=1$.

In the case of the circular Wilson loops correlator the ``eigenvalues'' $\Lambda$ in the Lam\'{e} equation take the following values
\be
\Lambda_2=\frac{\ell^2}{a}-2 m,\quad
\Lambda_R=\frac{\ell^2}{1+\imag j}-(1+\mu),\quad
\Lambda_f=\frac{4 s^2}{1+\imag j}-(1+\mu)
\ee
for $\mathcal{O}_2$, $\mathcal{O}_R$ and $\mathcal{O}_f$ respectively.

\section{Computing determinants using Gel'fand-Yaglom method}\label{app:GY}

In this appendix, we explain how to use the Gel'fand-Yaglom (GY) method for computing one dimensional determinants and how it applies to our case. This discussion follows the one in \cite{McKane:1995vp} and \cite{Kruczenski:2008zk}. In general, the method applies to two one-dimensional second order differential operators on the interval $x\in[a,b]$
\be\label{eq:GYops}
L \eq
 P_0(x)\frac{d^2}{d x^2}
+P_1(x)\frac{d}{d x}
+P_2(x) \\
\hat L \eq
 \hat P_0(x)\frac{d^2}{d x^2}
+\hat P_1(x)\frac{d}{d x}
+\hat P_2(x) \; ,
\ee
with $P_0(x)\equiv\hat P_0(x)$. The GY method and its generalizations give the ratio of the determinants of these operators in terms of the solution of the associated homogeneous problems without direct reference to the individual eigenvalues. The special case that is relevant for us obeys $P_0(x)=\hat P_0(x)=1$ and $P_1(x)=\hat P_1(x)=0$. The determinants clearly depend on the boundary conditions that are imposed on the eigenfunctions. Here we discuss only Dirichlet (D) or Neumann (N) boundary conditions. The procedure involves first of solving the homogeneous problems
\be
L u_i(x) = 0 \; ,\quad
\hat L \hat u_i(x) = 0 \; ,
\ee
where $u_{i=1,2}(x)$ are two independent solutions. Depending on the boundary conditions at the left end of the interval, we define solve the homogeneous equation subject to the initial conditions:
\be
u(a)\eq 0,\quad u'(a)=1\quad\quad \text{for D b.c. at $a$} \; , \\
u(a)\eq 1,\quad u'(a)=0\quad\quad \text{for N b.c. at $a$} \; ,
\ee
where $u(a)$ stands both for $u$ and $\tilde u$. The formula for the ratio of determinants depends on the boundary conditions at the right end of the interval and are given by
\be
\frac{\det L}{\det \hat L} \eq \frac{u(b)}{\hat u(b)} \quad\quad \text{for D b.c. at $b$} \; , \\
\frac{\det L}{\det \hat L} \eq \frac{u'(b)}{\hat u'(b)} \quad\quad \text{for N b.c. at $b$} \; .
\ee
For more general boundary conditions (Robin, periodic or anti-periodic) see, e.g., \cite{Dunne:2007rt}. In the following, we would like to stress some points that are relevant for our cases.

\paragraph{Rescaling.} If one wants to compute the ratio of the determinants of the general operators given in \eqref{eq:GYops}, the result may change under an overall rescaling of the operators. However, if $P_1(x) = \hat P_1(x)$, the the answer does \emph{not} change. Explicitly, we have
\be
\frac
{\det \Bigbrk{ Q(x) \bigbrk{ \frac{d^2}{d x^2} + P_2(x) } } }
{\det \Bigbrk{ Q(x) \bigbrk{ \frac{d^2}{d x^2} + \hat P_2(x) } } }
=
\frac
{\det \bigbrk{\frac{d^2}{d x^2} + P_2(x) } }
{\det \bigbrk{\frac{d^2}{d x^2} + \hat P_2(x) } } \; ,
\ee
i.e.\ the common function $Q(x)$ can be dropped.

However, in case we start with a Dirac operator or some other first order operator
\be
L_1 \eq \sqrt{Q(x)} \Bigbrk{ \frac{d}{d x}+R(x) } \; ,
\ee
then, in general,
\be\label{eq:noncancelation-of-prefactors}
\frac
{\det \Bigbrk{ \sqrt{Q(x)} \bigbrk{\frac{d}{d x}+R(x)} }^2}
{\det \Bigbrk{ Q(x) \bigbrk{ \frac{d^2}{d x^2} + \hat P_2(x) } } }
\neq
\frac
{\det \bigbrk{ \frac{d}{d x} + R(x) }^2}
{\det \bigbrk{ \frac{d^2}{d x^2} + \hat P_2(x) } } \; .
\ee
We felt compelled to stress this fact, because such a relation can sometimes be found in the literature. In this paper, we explicitly show the difference between the left and right hand side for the fermionic determinants of the single circular Wilson loop and the correlator of two circular Wilson loops, see \eqref{eq:fermionicOpsDifferensCWL} and \eqref{eq:fermionicOpsDifferensWW}. In the latter case, except for a factor of $4$, there is also a non-trivial factor given in \eqref{eq:factor}. The factor that relates the left and right hand sides depend on $Q(x)$ and on the boundary.

\paragraph{Conjugation.} Here we show how conjugation of the operator changes the value of the determinant. Let us say we want to compute
\be
\frac{\det \tilde L(x)}{\det \hat L(x)}
\eq
\frac{\det \left(f(x) L(x) f^{-1}(x)\right)}{\det \hat L(x)}
=
\frac{\det \left(f(x) \left(\frac{d^2}{d x^2}+P_2(x)\right) f^{-1}(x)\right)}{\det \hat L(x)} \; .
\ee
In case one was dealing with matrices instead of functions, the answer would be $\frac{\det L(x)}{\det \hat L(x)}$, but here, using the GY method, we show that this is not necessarily the case. According to the general theorem, we have
\be
  \frac{\det \tilde{L}(x)}{\det \hat L(x)} = \frac{\tilde u(b)}{\hat u(b)}
  \comma
  \frac{\det L(x)}{\det \hat L(x)}=\frac{u(b)}{\hat u(b)}
  \; ,
\ee
where the $u$'s satisfies all the conditions given above. For instance, we have $L u(x) = 0$, which implies
\be
\tilde L f(x) u(x) = 0 \; .
\ee
Thus, we can set
\be
  \tilde u(x) = C f(x) u(x)
\ee
for some constant $C$. Taking the $x$-derivative of this relation at $x=a$, we find
\be
  \tilde u'(a) = C f'(a) u(a) + C f(a) u'(a)
\ee
which reduces to $C = 1/f(a)$ upon using the initial conditions. Then
\be\label{eq:detConj}
  \frac{\det \tilde L}{\det \hat L} = \frac{\tilde u(b)}{\hat u(b)} = \frac{f(b)}{f(a)} \frac{\tilde u(b)}{\hat u(b)} = \frac{f(b)}{f(a)} \frac{\det L}{\det \hat L} \; .
\ee

\paragraph{Zero modes.} If one of the operators have a zero mode the determinant vanishes. In case one is interested in the product of the eigenvalues after omitting the zero mode, it is possible to modify the GY method to find the answer, see \cite{Kirsten:2003py}. In our case, there is one configuration with a zero mode, namely the critical configuration for $j\approx 1.16220...$. For this configuration, the zero mode appears only in the spectrum of the bosonic operator $\mathcal{O}_R$, so for this value the partition function diverges. Here we do not extract the finite answer which corresponds to omitting the zero mode.

\paragraph{Application.} In the following, let us give a taste of how the method is applied to the case of the Lam\'{e} operators discussed in this paper. The solution to the ``homogeneous'' Lam\'{e} equation \eqref{eqn:standard-Lame-operator} is given by \eqref{eq:LameEqSol}. Let us denote the left and right boundaries by $L$ and $R$, respectively, and assume Dirichlet boundary conditions at both ends. Then, the solution satisfying the initial conditions
\be
  u(L) = 0 \comma u'(L) = 1
\ee
is formally given by
\be \label{eqn:standard-Lame-solution}
  u(x)=\frac{y_-(L)y_+(x)-y_+(L)y_-(x)}{W(L)} \; ,
\ee
where the denominator is the Wronskian $W(x)=y_-(x)y'_+(x)-y_+(x)y'_-(x)$ evaluated at the left boundary, and $y_\pm(x)$ are defined in \eqref{eq:LameEqSol}. This solution can be written more explicitly by computing the derivative
\be
y'_{\pm}(x) \eq y_{\pm}(x)\left(
\frac{H'(x \pm \alpha)}{H(x \pm \alpha)}
-\frac{\Theta'(x)}{\Theta(x)}+Z(\mp \alpha)\right) \nln
\eq y_{\pm}(x)\left(
   -m \JacobiSN(x\pm\alpha)\JacobiSN(x)\JacobiSN(\alpha)
   +\frac{\JacobiCN(x\pm\alpha)\JacobiDN(x\pm\alpha)}{\JacobiSN(x\pm\alpha)}
   \right) \nln
&\equiv& y_{\pm}(x) f_{\pm}(x) \; ,
\ee
where we introduced a shorthand for the terms in parenthesis. This allows us to write
\be \label{eqn:Gelfand-Yaglom}
u(R) \eq \frac{1}{f_+(L)-f_-(L)}
         \left(\frac{y_+(R)}{y_+(L)}-\frac{y_-(R)}{y_-(L)}\right) \nln[2mm]
     \eq \frac{\JacobiSN^2(\alpha)-\JacobiSN^2(L)}{2\JacobiSN(\alpha)\JacobiCN(\alpha)\JacobiDN(\alpha)}
		     \frac{\Theta(L)}{\Theta(R)}\left(\frac{H(R+\alpha)}{H(L+\alpha)}
				 e^{-Z(\alpha)(R-L)}-\frac{H(R-\alpha)}{H(L-\alpha)}e^{Z(\alpha)(R-L)}\right) \; ,
\ee
which gives the determinant up to a normalization by another determinant. We do not worry about the normalization because in the partition function, we take all the reference operators to be the same, so this factor cancels.

\section{Supersymmetric regularization}\label{app:Supersymmetricregularization}

We use the supersymmetric regularization scheme in order to shift the fermionic Fourier modes such that the summation will be over integer numbers as for the bosons following \cite{Frolov:2004bh}. We will work with the effective action $\Gamma \equiv \ln \mathcal{Z}$ so the product becomes a sum over logarithms. We denote the bosonic contributions as $\omega^B_\ell$ and the fermionic ones by $\omega^F_s$. Thus, we have
\be
\Gamma =
\sum_{s\in \mathbb{Z}+\frac{1}{2}}\omega^F_s
-\sum_{\ell\in \mathbb{Z}}\omega^B_\ell,
\ee
and regularize using a small parameter $\mu$
\be
\Gamma =
\sum_{s\in \mathbb{Z}+\frac{1}{2}}e^{-\mu|s|}\omega^F_s
-\sum_{\ell\in \mathbb{Z}}e^{-\mu|n|}\omega^B_\ell.
\ee
Supersymmetric regularization means that we consider the bosonic frequency $\omega^B_\ell$ together with the fermionic frequencies $\omega^F_{\ell+\frac{1}{2}}$ and $\omega^F_{\ell-\frac{1}{2}}$. Hence, we write
\be
\Gamma \eq
 \sum_{\ell\in \mathbb{Z}} \lrsbrk{
 \frac{1}{2}e^{-\mu|\ell+\frac{1}{2}|}\omega^F_{\ell+\frac{1}{2}}
+\frac{1}{2}e^{-\mu|\ell-\frac{1}{2}|}\omega^F_{\ell-\frac{1}{2}}
           -e^{-\mu|\ell|}            \omega^B_\ell } \; .
\ee
We rearrange the expression by adding and subtracting some terms to obtain
\be \label{eqn:susy-regularization-derivation1}
\Gamma \eq
  \sum_{\ell\in \mathbb{Z}} e^{-\mu|\ell|} \lrbrk{
          \frac{1}{2}\omega^F_{\ell+\frac{1}{2}}
         +\frac{1}{2}\omega^F_{\ell-\frac{1}{2}}
         -\omega^B_\ell
  } \nl
+\frac{1}{2}\sum_{\ell\in \mathbb{Z}}\left(e^{-\mu|\ell+\frac{1}{2}|}-e^{-\mu|\ell|}\right)\omega^F_{\ell+\frac{1}{2}}
+\frac{1}{2}\sum_{\ell\in \mathbb{Z}}\left(e^{-\mu|\ell-\frac{1}{2}|}-e^{-\mu|\ell|}\right)\omega^F_{\ell-\frac{1}{2}} \; .
\ee
We will now simplify the sums in the second line. We split the range of the first sum into $\ell=-\infty..-1$ and $\ell=0..+\infty$, and the range of the second sum into $\ell=-\infty..0$ and $\ell=1..+\infty$. In the sums with negative indices we relabel $\ell\to-\ell$ and thus obtain for the second line in \eqref{eqn:susy-regularization-derivation1}
\be
&&
 \frac{1}{2}\sum_{\ell = 1}^{\infty}\left(e^{-\mu\left(\ell-\frac{1}{2}\right)}-e^{-\mu \ell}\right)\omega^F_{-\ell+\frac{1}{2}}
+\frac{1}{2}\sum_{\ell = 0}^{\infty}\left(e^{-\mu\left(\ell+\frac{1}{2}\right)}-e^{-\mu \ell}\right)\omega^F_{ \ell+\frac{1}{2}} \nln
&&
+\frac{1}{2}\sum_{\ell = 0}^{\infty}\left(e^{-\mu\left(\ell+\frac{1}{2}\right)}-e^{-\mu \ell}\right)\omega^F_{-\ell-\frac{1}{2}}
+\frac{1}{2}\sum_{\ell = 1}^{\infty}\left(e^{-\mu\left(\ell-\frac{1}{2}\right)}-e^{-\mu \ell}\right)\omega^F_{ \ell-\frac{1}{2}} \; .
\ee
At this point, we use\footnote{This assumption is true for the cases we consider.} $\omega^F_{s}=\omega^F_{-s}$ in the third and the first sum so that they become equal to the second and the fourth sum. Then, shifting $\ell$ in the forth sum and combining the frequencies together yields
\be\label{eqn:susy-reg-KT}
-4\sinh^2\frac{\mu}{4}\sum_{\ell=0}^\infty e^{-\mu (\ell+\frac{1}{2})}\omega_{\ell+\frac{1}{2}}^F \; .
\ee
Now, we can write the complete formula as
\be \label{eqn:susy-regularization}
\Gamma^{(1)} \eq
  \sum_{\ell\in \mathbb{Z}} e^{-\mu|\ell|} \lrbrk{
          \frac{1}{2}\omega^F_{\ell+\frac{1}{2}}
         +\frac{1}{2}\omega^F_{\ell-\frac{1}{2}}
         -\omega^B_\ell
  }
  -\lim_{\mu\to 0}4\sinh^2\frac{\mu}{4}\sum_{\ell=0}^\infty e^{-\mu (\ell+\frac{1}{2})}\omega_{\ell+\frac{1}{2}}^F \; .
\ee

It is tempting to expand \eqref{eqn:susy-regularization-derivation1} in $\mu$ after identifying the first and the third sum with the second and third, to give
\be
-\frac{\mu}{2}\omega^F_{\frac{1}{2}}
-\frac{\mu}{2}\sum_{\ell = 1}^{\infty}e^{-\mu \ell }\lrbrk{ \omega^F_{\ell+\frac{1}{2}} - \omega^F_{\ell-\frac{1}{2}} }.
\ee
This result is similar to the one in \cite{Kruczenski:2008zk}, however, it yields a different result in general. For the case at hand, this difference turns out to be a factor of 2.

\section{Partition function for the circular Wilson loop}
\label{app:CWL-PF}

The classical string solution ending on a circle of radius $R$ at the boundary of the Poincar\'{e} patch is given by
\be
  r = R \sech\tau
	\comma
	z = R \tanh\tau
	\comma
	\phi = \sigma
	\comma
	x = 0
	\comma
	y = 0
\ee
where $\sigma = 0..2\pi$ and $\tau = 0..\infty$. The fluctuation analysis around this background \cite{Kruczenski:2008zk} yields the partition function
\be\label{eqn:hemisphere-partition_function}
\mathcal{Z} = \frac{\det^{2}\mathcal{O}_+ \: \det^{2}\mathcal{O}_-}
                   {\det^{5/2}\mathcal{O}_0 \: \det^{3/2}\mathcal{O}_2} \; .
\ee
where the operators have the same formal expression as in \eqref{eqn:general-operators}, except that now $r/z$ is not given by \eqref{eqn:r-over-z} but by
\be
  \frac{r}{z} = \csch\tau \; .
\ee
Note, however, that this is the limit of \eqref{eqn:r-over-z} for $j\to0$ or $m\to1$. So, explicitly we have
\begin{subequations}
\be
  \label{eqn:hemisphere-op0}
  \mathcal{O}_0 \eq - \partial_\tau^2 - \partial_\sigma^2 \; , \\
  \label{eqn:hemisphere-op2}
	\mathcal{O}_2 \eq - \partial_\tau^2 - \partial_\sigma^2 + 2\csch^2\tau \; , \\
  \label{eqn:hemisphere-opPM}
	\mathcal{O}_\pm \eq - \partial_\tau^2 - \partial_\sigma^2 \pm \imag\coth\tau \, \partial_\sigma + \tfrac{1}{4}\lrbrk{ 1 + 3\csch^2\tau } \; , \\
  \label{eqn:hemisphere-opPMtil}
	\tilde{\mathcal{O}}_\pm \eq - \partial_\tau^2 - \partial_\sigma^2 + \lrbrk{ \csch\tau \pm \coth\tau } \csch\tau \; .
\ee
\end{subequations}

The target-space cutoff $\eps$ is related to the world-sheet cutoff $\eps_0$ by $\eps = R \tanh\eps_0$. Furthermore, following \cite{Kruczenski:2008zk}, we introduce an unphysical world-sheet cutoff at large $\tau$, say at $\tau = T$. This cutoff should drop out from the partition function \eqref{eqn:hemisphere-partition_function}. Using GY on the interval $\tau \in [\eps_0,T]$, we obtain for the initial value solutions
\begin{subequations}
\be
  u_0(T) \eq \frac{\sinh T \ell}{\ell} + \order(\eps_0) \simeq \frac{e^{T \ell}}{2\ell} + \order(\eps_0) \; , \\
  u_2(T) \eq \frac{1}{\eps_0} \frac{\ell \cosh T \ell - \coth T \sinh T \ell}
	                                       {\ell(\ell^2-1) } + \order(\eps_0^0) \simeq
\frac{1}{\eps_0} \frac{e^{T \ell}}
	                                       {2\ell(\ell + 1) } + \order(\eps_0^0) \; , \\
  u_+(T) \eq \frac{e^{-s T} \left(1+e^{2 s T} (-\cosh T+2 s \sinh T)\right)}{\sqrt{\eps_0 }\left(4 s^2-1\right) \sqrt{\sinh T}} + \order(\eps_0^0)
  \simeq
  \frac{e^{\frac{1}{2}(2 s+1) T}}{\sqrt{2\eps_0 }\left(2 s + 1\right) } + \order(\eps_0^0)\; , \\
  u_-(T) \eq \frac{e^{ s T} \left(1+e^{-2 s T} (-\cosh T-2 s \sinh T)\right)}{\sqrt{\eps_0 }\left(4 s^2-1\right)  \sqrt{\sinh T}} + \order(\eps_0^0)
\simeq
  \sqrt{\frac{2}{\eps_0}}\frac{e^{\frac{1}{2}(2 s-1) T}}{\left(4 s^2 - 1\right) } + \order(\eps_0^0)  \; , \\
  \tilde u_+(T) \eq \frac{2}{\eps_0} \frac{2 s \cosh T s - \sinh T s \coth\frac{T}{2}}{s(4 s^2-1)} + \order(\eps_0^0)
  \simeq
  \frac{e^{T s}}{\eps_0 s(2 s+1)} + \order(\eps_0^0)\; , \\
  \tilde u_-(T) \eq \frac{2 s \sinh T s - \cosh T s \tanh\frac{T}{2}}{4 s^2 -1} + \order(\eps_0)
  \simeq
  \frac{e^{T s}}{2(2s+1)} + \order(\eps_0)\; ,
\ee
\end{subequations}
where to leading order we can replace $\eps_0$ by $\eps/R$. In the second line, we took the large $T$ limit and assumed $\ell,s>0$.
Notice the following relation
\be\label{eq:fermionicOpsDifferensCWL}
u_+(T)u_-(T)
\simeq
\frac{2 s}{(2 s -1)}\tilde u_+(T)\tilde u_-(T),
\ee
where $\simeq$ means we took the large $T$ limit. This is quite different than the relation we had for the two Wilson loops correlator where the proportionality coefficient was independent of $s$.

Despite the vanishing of the denominators for $\ell=0,\pm1$ and $s=\pm\half$, these formulas have well-defined limits, namely
\begin{subequations}
\be
  \bigeval{u_0(T)}_{\ell\to0} \eq T + \order(\eps_0) \simeq T + \order(\eps_0) \; , \\
  \bigeval{u_2(T)}_{\ell\to0} \eq \frac{T \coth T - 1}{\eps_0} + \order(\eps_0^0) \simeq \frac{T}{\eps_0} + \order(\eps_0^0)\; , \\
	\bigeval{u_2(T)}_{\ell\to\pm1} \eq \frac{\cosh T - T \csch T}{2\eps_0} + \order(\eps_0^0) \simeq \frac{e^T}{4\eps_0} + \order(\eps_0^0) \; , \\
  \bigeval{u_+(T)}_{s\to \frac{1}{2}} \eq \frac{e^{ T  }  T  -\sinh T }{2 \sqrt{e^{ T  } \eps_0 \sinh T }} + \order(\eps_0^0) \simeq \frac{ T}{\sqrt{2\eps_0  }} + \order(\eps_0^0) \; , \\
  \bigeval{u_-(T)}_{s\to \frac{1}{2}} \eq \frac{-1+e^{2 T}-2 T}{4 \sqrt{\eps_0 } \sqrt{e^T \sinh T}} + \order(\eps_0^0) \simeq \sqrt{\frac{2}{\eps_0}}\frac{e^T}{4} + \order(\eps_0^0)\; , \\
  \bigeval{\tilde u_+(T)}_{s\to\pm\frac{1}{2}} \eq - \frac{1}{\eps_0} (T - \sinh T) \csch\frac{T}{2} + \order(\eps_0^0) \simeq \frac{e^{T/2}}{\eps_0} + \order(\eps_0^0) \; , \\
  \bigeval{\tilde u_-(T)}_{s\to\pm\frac{1}{2}} \eq \frac{1}{4} (T + \sinh T) \sech\frac{T}{2} + \order(\eps_0) \simeq \frac{e^{T/2}}{4} + \order(\eps_0)\; .
\ee
\end{subequations}
For the ``zero'' mode ($\ell=0$, $s=\pm\frac{1}{2}$), we actually impose Neumann boundary conditions, so instead of the value of $u$ at $ T$, we need to know the derivatives
\be
  \bigeval{u'_0(T)}_{\ell\to0} \eq 1 + \order(\eps_0) \simeq  1 + \order(\eps_0) \; , \\
  \bigeval{u'_2(T)}_{\ell\to0} \eq \frac{\coth T - T \csch^2 T}{\eps_0}+ \order(\eps_0^0) \simeq  \frac{1}{\eps_0} + \order(\eps_0^0)\; , \\
  \bigeval{u'_+(T)}_{s\to \frac{1}{2}} \eq
	   \frac{e^{-5 T/2} \left(1+2 e^{4 T}-e^{2 T} (3+2 T)\right)}{8 \sqrt{\eps_0 \sinh^3 T}} + \order(\eps_0^0)  \simeq  \frac{1}{\sqrt{2 \eps_0}} + \order(\eps_0^0)\; , \\
  \bigeval{u'_-(T)}_{s\to \frac{1}{2}} \eq
	   \frac{e^{-3 T/2} \left(2+e^{4 T}+e^{2 T} (-3+2 T)\right)}{8 \sqrt{\eps_0 \sinh^3 T}} + \order(\eps_0^0) \simeq  \frac{e^T}{\sqrt{2^3 \eps_0}} + \order(\eps_0^0)\; , \\
  \bigeval{\tilde u'_+(T)}_{s\to\pm\frac{1}{2}} \eq
	   -\frac{1}{2\eps_0} \lrsbrk{ 2 - 2\sinh^2\frac{T}{2} - T \coth\frac{T}{2} } \csch\frac{T}{2} + \order(\eps_0^0) \simeq  \frac{e^{T/2}}{2 \eps_0} + \order(\eps_0^0)\; , \\
  \bigeval{\tilde u'_-(T)}_{s\to\pm\frac{1}{2}} \eq
	    \frac{1}{8}       \lrsbrk{ 2 + 2\cosh^2\frac{T}{2} - T \tanh\frac{T}{2} } \sech\frac{T}{2} + \order(\eps_0) \simeq  \frac{e^{T/2}}{8} + \order(\eps_0)\; .
\ee
Thus, the frequencies (i.e. the product of the eigenvalues of various operators for given angular mode) are
\be
  \omega^B_\ell \eq \omega^B_{-\ell} = \frac{5}{2} \ln \frac{e^{ T \ell}}{2 \ell} + \frac{3}{2} \ln \frac{e^{ T \ell}}{2\eps_0 \ell (\ell+1)}
  =\ln \frac{e^{4 \ell T}}{\eps_0^{3/2}(2\ell)^4(\ell+1)^{3/2}}
	 \; , \\
  \omega^F_s \eq \omega^F_{-s} = 2 \ln \frac{e^{\frac{1}{2}(2 s+1) T}}{\sqrt{2\eps_0 }\left(2 s + 1\right) } + 2 \ln \sqrt{\frac{2}{\eps_0}}\frac{e^{\frac{1}{2}(2 s-1) T}}{\left(4 s^2 - 1\right) }
  =
  2 \ln \frac{e^{2 s T }}{\eps_0\left(2 s + 1\right) \left(4 s^2 - 1\right)}
\ee
for $\ell = 1,2,...$ and $s=\frac{1}{2},\frac{3}{2},...$ (Dirichlet boundary conditions), while in the special cases $\ell=0$ and $s = \pm\frac{1}{2}$ (Neumann boundary conditions), we have
\be
  \omega^B_0 = \frac{3}{2} \ln \frac{1}{\eps_0}
  \comma
  \omega^F_{\pm\frac{1}{2}} = 2 T + 2\ln \frac{1}{4\eps_0}
  \; .
\ee
We are now ready to sum these frequencies according to the supersymmetric regularization formulas \eqref{eqn:susy-regularization}. We use\footnote{Notice that we use the supersymmetric regularization differently than it was done in \cite{Kruczenski:2008zk}. There, one considers only the regularization of the $\mathcal{O}_+$ frequencies, and eventually uses this result in the partition function, while here we treat the fermionic frequencies as a whole and thus treat $\mathcal{O}_+$ and $\mathcal{O}_-$ on an equal footing. This does not change the UV behaviour but does give a slightly different final answer.}
\be\label{eq:CWLfrequency}
   \frac{1}{2}\omega^F_{\ell+\frac{1}{2}}
   +\frac{1}{2}\omega^F_{\ell-\frac{1}{2}}
   -\omega^B_\ell
  = \ln \frac{\ell}{4\sqrt{\eps_0} (\ell-1)\sqrt{\ell+1}}
\ee
for $\ell\not=0$, and
\be
   \frac{1}{2}\omega^F_{\frac{1}{2}}
   +\frac{1}{2}\omega^F_{-\frac{1}{2}}
   -\omega^B_0
  = 2 T + \ln \frac{1}{16\sqrt{\eps_0}} \; ,
\ee
as well as
\be
\omega^F_{\ell+\frac{1}{2}} - \omega^F_{\ell-\frac{1}{2}} = 2\ln\frac{e^{2  T }\ell\left(\ell-1\right) }{\left(\ell+1\right)^2} \; .
\ee
For the sums, we obtain
\be
-\frac{\mu}{2}\sum_{\ell = 1}^{\infty} e^{-\mu \ell } \left(\omega^F_{\ell+\frac{1}{2}} - \omega^F_{\ell-\frac{1}{2}} \right) = -2 T +\order(\mu) \; .
\ee
The $2 T$ cancels with the one coming from the zero mode.

\section{Partition function for periodic straight line}
\label{app:LINE-PF}

The partition function for a periodically identified straight line was also worked out in \cite{Kruczenski:2008zk}. Here we just quote the form of the rescaled operators
\begin{subequations}
\be
  \mathcal{O}_0   = - \partial_\tau^2 - \partial_\sigma^2 \; , \quad
  \mathcal{O}_2   = - \partial_\tau^2 - \partial_\sigma^2 + \frac{2}{\tau^2} \; , \quad
  \mathcal{O}_\pm = - \partial_\tau^2 - \partial_\sigma^2 \pm \imag\frac{1}{\tau}\partial_\sigma + \frac{3}{4\tau^2} \; ,
\ee
\end{subequations}
where $\eps < \tau < T$ with $\eps_0$ is small and $T$ is large. In order to evaluate the determinants using the GY method we need
\be
  u_0(T) \simeq \frac{e^{T \ell}}{2\ell} \; , \quad
  u_2(T) \simeq \frac{e^{T \ell}}{2\ell^2\eps_0}  \; , \quad
  u_+(T) \simeq \frac{e^{T \ell}\sqrt{T}}{2\ell\sqrt{\eps_0}} \; , \quad
  u_-(T) \simeq \frac{e^{T \ell}}{4\ell^2\sqrt{\eps_0 T}} \; ,
\ee
where we expanded in $\eps_0$ and $T$ and assumed $\ell>0$. Putting all of this together, we get
\be
  \mathcal{Z}_\ell = \frac{1}{\sqrt{16\ell \eps_0}}.
\ee
This equation is not valid for $\ell=0$, where the solutions to the homogeneous equations are quite different. In this case, we have
\be
  u_0(T)   \simeq T \; , \quad
  u_2(T)   \simeq \frac{T^2}{3\eps_0}  \; , \quad
  u_\pm(T) \simeq \frac{T^{3/2}}{2\sqrt{\eps_0}} \; ,
\ee
so we get $\mathcal{Z}_0=\sqrt{\frac{27 T}{256 \eps_0}}$, which diverges when $T$ is taken to infinity. This does not cause a problem in \cite{Kruczenski:2008zk}, because the partition function is eventually being integrated over $\ell$ and not summed, so this term is effectively ignored.

\section{Frequencies}
\label{app:frequencies}

In the main text, we have considered the evaluation of the product of all eigenvalues, but did not discussed the individual excitations. In this appendix, we address this second question. Specifically, we derive equations for the fluctuation frequencies that can both be easily evaluated numerically as well as used to obtain a quadratic approximation for their values. We will consider here only the rescaled operators given in \eqref{eqn:general-operators}. The unrescaled operators have a different spectrum.

We define the frequencies $\lambda_i$ of $\mathcal{O}_i$ by the following eigenvalue equation
\be
\mathcal{O}_i \psi_n = \lambda_n \psi_n \; .
\ee
We can re-use the solutions \eqref{eqn:det_in_GY_general} derived before, if we shift $\ell^2\rightarrow \ell^2+\lambda_n$ and then find the eigenvalues by requiring that the expression for the determinant vanishes.

For $\mathcal{O}_2$ the condition translates to
\be
\sinh(2 Z(\alpha_2)\EllipticK(m))=0 \; ,
\ee
where $\alpha_2 = \alpha_2(\ell^2+\lambda_n)$ and we assume that $\ell\neq \pm 1$. This implies
\be
2 Z(\alpha_2(\ell^2+\lambda_n))\EllipticK(m)=\imag \pi n,
\ee
for $n\in \mathbb{Z}$. We note that $n=0$ should be dismissed because the Jacobi functions which multiply the $\sinh$ diverge (see \appref{app:GY}), and there is no solution. There are also no solutions for negative $n$'s, so $n>0$, hence, $n=1,2,3,..$ and $n=1$ is the lowest eigenvalue. Rewriting the Jacobi Zeta function in terms of elliptic integrals, this equation becomes
\be
\EllipticE\left(\sin^{-1}\sqrt{\frac{1-m-\frac{\lambda_n}{a}}{m}}\bigg|m\right)
-\frac{\EllipticE(m)}{\EllipticK(m)} \EllipticF\left(\sin^{-1}\sqrt{\frac{1-m-\frac{\lambda_n}{a}}{m}}\bigg|m\right)
=\frac{\imag \pi n}{2 \EllipticK(m)} \; ,
\ee
which can be expanded for $\lambda_n/a \gg n $ such that
\be
\imag\sqrt{\frac{\lambda_n}{a}}
-\frac{\imag\pi}{2 \EllipticK(m)}
-\imag\sqrt{\frac{a}{\lambda_n}}\left(1-m-\frac{\EllipticE(m)}{\EllipticK(m)}\right)+\mathcal{O}(\lambda_n^{-3/2})
=\frac{\imag \pi n}{2 \EllipticK(m)} \; ,
\ee
so to leading order we have
\be \label{eqn:asymp-eigenvalues-O2}
\lambda_n=a\left(\left(\frac{\pi(n+1)}{2\EllipticK(m)}\right)^2+2 \left(1-m-\frac{\EllipticE(m)}{\EllipticK(m)}\right)\right)+\mathcal{O}(n^{-1}),\quad
n=1,2,3...
\ee
Note that the leading term is identical to the $\mathcal{O}_0$ eigenvalues,
\be
\lambda_n^{(0)}=\left(\frac{\pi n}{2\EllipticK(m)/\sqrt{a}}\right)^2+\mathcal{O}(n).
\ee
A similar analysis for $\mathcal{O}_R$ yields
\be \label{eqn:asymp-eigenvalues-OR}
\lambda_n^{(R)}=a\lrbrk{\lrbrk{\frac{\pi(n+1)}{2\EllipticK(m)}}^2 + 4 \lrbrk{1-m-\frac{\EllipticE(m)}{\EllipticK(m)} } } + \order(n^{-1}),
\ee
where $n=1,2,...$. We notice that the $n^2$ and $n$ coefficient coincide with those of $\mathcal{O}_2$. For $\tilde{\mathcal{O}}_\pm$ eigenvalues we should replace $n$ by $2n$ and rescale by $1/4$. To summarize we have
\begin{align}
\lambda^{(0)}_n & = \left(\frac{\pi n}{2\EllipticK(m)/\sqrt{a}}\right)^2 \; , \nonumber\\
\lambda^{(2)}_n & = \left(\frac{\pi (n+1)}{2\EllipticK(m)/\sqrt{a}}\right)^2+2 a \lrbrk{1-m-\frac{\EllipticE(m)}{\EllipticK(m)} }+ \order(n^{-1}) \; , \nonumber\\
\lambda^{(R)}_n & = \left(\frac{\pi (n+1)}{2\EllipticK(m)/\sqrt{a}}\right)^2+4 a \lrbrk{1-m-\frac{\EllipticE(m)}{\EllipticK(m)} }+ \order(n^{-1}) \; , \nonumber\\
\tilde \lambda^{(\pm)}_n & = \left(\frac{\pi (n+\frac{1}{2})}{2\EllipticK(m)/\sqrt{a}}\right)^2+a \lrbrk{1-m-\frac{\EllipticE(m)}{\EllipticK(m)} }+ \order(n^{-1}) \; ,
\end{align}
with $n=1,2,3...$, so for large $n$ they all coincide. This can also be summarized as
\be
  \lambda^{(i)}_n = \left(\frac{\pi (n+x_i)}{2\EllipticK(m)/\sqrt{a}}\right)^2+y_i a \lrbrk{1-m-\frac{\EllipticE(m)}{\EllipticK(m)} } + \order(n^{-1}) \; ,
\ee
where $(x^{(0)},x^{(2)},x^{(R)},x^{(\pm)})=(0,1,1,\frac{1}{2})$ and $(y^{(0)},y^{(2)},y^{(R)},y^{(\pm)})=(0,2,4,1)$.

\section{Large \texorpdfstring{$j$}{j} limit and the \texorpdfstring{$q\bar q$}{qqbar}-potential}\label{sec:qqbar}

We expect our analysis to reduce to the $q\bar q$-potential, when the radii of the two Wilson loops go to infinity, while their difference is kept fixed. Remember that $r_{\mathrm{max}}=r_{\mathrm{min}} \rho = e^{f_0} \rho $ and that $\rho =e^{2 \EllipticK(m) Z(\alpha|m)}$. Now, $\Delta r=r_{\mathrm{max}}-r_{\mathrm{min}}=e^{f_0}(\rho-1)$ is finite while $e^{f_0}\to \infty$, so $\rho=1+\rho_0 e^{-f_0}+...$. This implies that for stable configurations $j$ is large since $\rho\to 1+\mathcal{O}(j^{-1/2})$ for $j\to \infty$. In this limit $\rho\simeq 1+2 \EllipticK(m) Z(\alpha|m)$ so\footnote{The numerical constant is $d=-4\EllipticK\left(\frac{1}{2}\right)\frac{\Theta''\left(\EllipticK\left(\frac{1}{2}\right),\frac{1}{2}\right)}{\Theta\left(\EllipticK\left(\frac{1}{2}\right),\frac{1}{2}\right)}\simeq 1.69443$.} $\Delta r=2 \EllipticK(m) Z(\alpha|m) e^{f_0}=d\frac{1}{\sqrt{j}}e^{f_0}\equiv L=\mathrm{finite}$. This means that $r_{\mathrm{min}}\simeq r_{\mathrm{max}}\simeq \frac{L}{d}\sqrt{j}$.

By sending $j\to \infty$, the range of world-sheet coordinate $\tau$ becomes zero. In order to keep the range finite, and to make contact with the results in \cite{Forini:2010ek}, we rescale the world-sheet coordinates $\tau$ and $\sigma$ by $a(j)$. Thus, in the large $j$ limit we have to replace $\omega^2\to j \omega^2$ in \eqref{eqn:partition_function_result}, and then send $j\to \infty$. In this way, the $\tau$-range remains finite, $\tau\in [0,2\EllipticK(\frac{1}{2})]$, while the range of $\sigma$ becomes infinite. Finally, the square of the prefactor in \eqref{eqn:partition_function_result} becomes
\be
\frac{8^2 L^2 \omega ^8 \sqrt{1+\omega ^4} \left(4 \omega ^4-1\right)}{d^2 \eps \left(1+16 \omega ^4\right)^4}.
\ee
Up to an irrelevant constant which cancels in the regularization, we get the same prefactor as in \cite{Forini:2010ek}.
Other than that, all the operators in \eqref{eqn:OpsList_in_Lame_form} assume the same form as the operators given in \cite{Forini:2010ek}, which can be proved using \eqref{eq:modulus-identities}, so the results coincide\footnote{We remind the reader that we use different fermionic operators then the ones used in \cite{Forini:2010ek}, and their determinant is related to the determinant of the operators used in \cite{Forini:2010ek} by a constant, see \eqref{eq:fermionicOpsDifferensWW}.}. Notice that the supersymmetric regularization factor \eqref{eq:susyregfactor} vanishes in the limit $j\to \infty$, so it should not contribute to the $q\bar q$-potential computation.

\bibliographystyle{nb}
\bibliography{Revolution}

\end{document}